\documentclass[12pt,a4paper]{article}
\usepackage{geometry} 
\usepackage[T1]{fontenc}
\usepackage[latin9]{inputenc}
\usepackage{authblk}

\usepackage{mathrsfs}
\usepackage{amssymb}
\usepackage{amsmath}
\usepackage{mathtools}
\usepackage{ascmac}
\usepackage{amsthm}
\usepackage{graphicx}
\usepackage{natbib}
\usepackage{setspace}
\usepackage{bm}
\usepackage{comment}

\usepackage{xcolor}
\usepackage{ulem}

\usepackage{multirow}
\usepackage{here}
\usepackage{url}

\usepackage[unicode=true,bookmarks=true,bookmarksnumbered=false,bookmarksopen=false,breaklinks=false,pdfborder={0 0 1},backref=page,colorlinks=true]{hyperref}
\hypersetup{citecolor=blue, linkcolor=blue}

\makeatletter

\newtheorem{df}{Definition}
\newtheorem{thm}{Theorem}
\newtheorem{lem}{Lemma}
\newtheorem{cor}{Corollary}
\newtheorem{prop}{Proposition}

\theoremstyle{definition}
\newtheorem{rem}{Remark}

\newtheorem{assd}{Assumption}

\newtheorem{assk}{Assumption}

\newtheorem{assm}{Assumption}

\newtheorem{assdd}{Assumption}

\DeclareMathOperator*{\argmin}{argmin}

\makeatother

\allowdisplaybreaks

\begin{document}

\title{\bf Fully Data-driven Normalized and Exponentiated Kernel Density Estimator \\ with Hyv\"arinen Score}
\author[1]{Shunsuke Imai}
\author[2]{Takuya Koriyama}
\author[3]{Shouto Yonekura}
\author[4]{Shonosuke Sugasawa}
\author[1]{Yoshihiko Nishiyama}

\affil[1]{Institute of Economic Research, Kyoto University}
\affil[2]{Department of Statistics, Rutgers University}
\affil[3]{Kodansha Ltd.}
\affil[4]{Center for Spatial Information Science, The University of Tokyo}

\maketitle

\begin{abstract}
    Recently, \cite{jewson2021general} proposed a new approach for kernel density estimation using an exponentiated form of kernel density estimators. 
    The density estimator contained two hyperparameters that flexibly controls the smoothness of the resulting density.
    We tune them in a data-driven manner by minimizing an objective function based on the Hyv\"arinen score to avoid the optimization involving the intractable normalizing constant caused by the exponentiation. 
    We show the asymptotic properties of the proposed estimator and emphasize the importance of including the two hyperparameters for flexible density estimation. 
    Our simulation studies and application to income data show that the proposed density estimator is promising when the underlying density is multi-modal or when observations contain outliers.  
\end{abstract}

\noindent%
\vfill

\newpage
\section{Introduction} \label{sec:intro}
Nonparametric density estimation, particularly kernel density estimation (KDE), is routinely used in a variety of fields, including economics \citep[e.g.][]{dinardo1995labor}.
For independent observations, $X_1,\ldots,X_n$, the KDE at a given value $x$ is defined as 
\begin{equation}\label{KDE}
\hat{f}_{h}(x)=\frac1{nh}\sum_{i=1}^nK\left(\frac{x-X_i}{h}\right),
\end{equation}
where $K(\cdot)$ is a kernel function and $h$ is the bandwidth parameter. 
To apply the KDE (\ref{KDE}), one needs to specify the kernel function $K(\cdot)$ as well as the bandwidth parameter $h$ controlling the smoothness of $\hat{f}$.
The selection of $h$ is known to be more significantly related to the accuracy of the density estimation than that of kernel function $K(\cdot)$ \citep[e.g.][]{fan1996local,wasserman2006all}.
If $h$ is smaller than necessary, the estimated density will be more serrated than required; moreover, if $h$ is larger than necessary, the estimated density will be over-smoothed such that it fails to capture the essential features of the true density. 
Hence, many studies have attempted to develop appropriate selection methods for $h$.
Popular techniques include several types of cross validation (CV) \citep{rudemo1982empirical,scott1987biased,hall1992scv} and plug-in methods \citep{sheather1991reliable,Hall1991gpi}; these methods have been reviewed in \cite{jones1996brief}, and \cite{sheather2004density}.

However, the performance of the KDE (\ref{KDE}) with the single tuning parameter $h$ selected by the existing methods may not be satisfactory in some cases. 
This may be because the KDE includes only a single tuning parameter to control the smoothness. 
For example, when observations contain a small number of outliers, the resulting value of $h$ tends to be smaller than necessary, leading to serrated estimation results.

Recently, in the framework of general Bayesian inference \citep{bissiri2016general},  \cite{jewson2021general} consider the exponential of a weighted log-likelihood for KDE, $\exp\{w \sum_{i=1}^n \log \hat{f}_h(X_i)\} = \sum_{i=1}^n \hat{f}_h(X_i)^w$,  as an unnormalized model, where $w>0$ denotes another tuning parameter (learning rate). 
Consequently, they propose an extended  KDE by using the normalized version of the unnormalized density estimation, defined as the power of (\ref{KDE}), which is given by:
\begin{equation}\label{KDE2}
\hat{f}_{w, h}(x) \equiv  \left\{ \int \hat{f}_{h}(t)^{w} dt \right\}^{-1} \hat{f}_{h}(x)^{w} \propto \hat{f}_{h}(x)^{w} = \left\{\frac1{nh}\sum_{i=1}^nK\left(\frac{x-X_i}{h}\right)\right\}^{w}.
\end{equation}
Parameter $w$ can provide additional flexibility to the KDE. 
In particular, a representative role of $w$ is to adjust the smoothness of the KDE, i.e., when $h$ is smaller than necessary, $w$ can be too small to smooth the density estimator. 
In other words, $h$ acts on the kernel for each observations so provides local smoothness, and $w$ acts on the combination of all the kernels so provides global smoothness of the density.
To optimize the two parameters $(w, h)$, \cite{jewson2021general} employ techniques of fitting feasible unnormalized model $\exp\{w \sum_{i=1}^n \log \hat{f}_h(X_i)\}$ with the Fisher divergence \citep{hyvarinen2005estimation}, such that the estimator of $(w, h)$ can be defined as the minimizer of the Hyv{\"a}rinen score. 
In this study, we then theoretically and numerically investigate the new fully data-driven tuning method and the final density estimator.
Our main contributions can be summarized as follows:
\begin{enumerate} 
    \item We theoretically characterize asymptotically optimal $w$ and $h$  in terms of the Fisher divergence (defined in Section \ref{sec:characterisation} as $w_n^*$ and $h_n^*$, respectively).
    In particular, our results show that optimal $w_n^*$ is not necessarily 1, which suggests the theoretical relevance of using $\hat{f}_{w,h}$ in the context of KDE.
    In addition, the results of numerical experiments also suggest cases in which $\hat{f}_{w,h}$ is more effective in comparison to other methods.
    \item We also demonstrate sharp asymptotic results including consistency of the estimated tuning parameters. 
    Notably, our results do not assume the independence of each term in the Hyv\"arinen Score, unlike previous study \citep{jewson2021general}. 
    The assumption of independence is not valid because KDE refers to all observations. 
    To the best of our knowledge, this is the first time that such statements have been proven without assuming independence.
\end{enumerate}

In a closely related study, \cite{jewson2021general} also conduct a theoretical analysis of their method within a general framework.
However, it differs from our work in several respects. 
First, their work is limited to Gaussian kernels, whereas we analyze general symmetric kernels. 
In addition, as we shall point out later, their theoretical work assumes that the generalized likelihood is an additive form of each observation; however, this assumption does not hold true for KDE. Our theoretical study does not rely on this assumption. 
Finally, and most importantly, we provide theoretical characterizations of the parameters $(w_n^*,h_n^*)$ and give an empirical explanation of their roles. 
We believe that this study goes beyond the case of KDE and serves as the first step in guaranteeing the validity of \cite{jewson2021general}'s methodology for many subjects that were not fully addressed by \cite{jewson2021general}, and in revealing its unknown properties.

The remainder of this paper is organized as follows.
In Section~\ref{sec:prob}, we describe the background of the method and propose the estimator of the two tuning parameters in the unnormalized KDE (\ref{KDE2}) with the Hyv{\"a}rinen score.
Section~\ref{sec:theory} presents the primary theoretical results.
We use the proposed method in Section \ref{sec:numerical} to present the numerical results and comparisons, and conclude the paper in Section \ref{sec:remark}.
The supplementary material contains additional technical results and proofs.

\section{Methodology} \label{sec:prob}

\subsection{Settings and Fisher divergence} 
\label{sec:motivation}

Let $\mathcal{X}\subset\mathbb{R}$ be a data domain.
Suppose that we are given independent and identically distributed samples $X_{1:n}=(X_1,...,X_n)$ defined on $\mathcal{X}$ with an unknown distribution $F$ and density function $f$.
Then, $K$ is said to be an $L$-th order kernel, for positive integer $L$ if
\begin{equation}
    \int u^lK(u)du = 
    \begin{cases}
    1 & (l=0) \\
    0 & (1\le l\le L-1)\\
    C \neq 0,<\infty& (l=L), \nonumber
    \end{cases}
\end{equation}
see \citet[pp.55]{wasserman2006all} or \citet[pp.3]{tsybakov2009nonpara} for typical second-order kernels and \citet[pp.32-35]{wand1996smoothing} for higher-order kernels.

Under the model (\ref{KDE2}), the joint (unnormalized) density $g(X_{1:n}; w, h)$ for the observed sample $X_{1:n}=(X_1,\ldots,X_n)$ is given by 
\begin{equation}\label{joint-model}
g(X_{1:n}; w, h) \equiv \prod_{i=1}^n \hat{f}_{w, h}(X_i|X_{1:n}) \propto 
\exp\left\{w\sum_{i=1}^n \log \hat{f}_{h}(X_i|X_{1:n})\right\},
\end{equation}
where $\hat{f}_{h}(X_i|X_{1:n})$ and $\hat{f}_{w, h}(X_i|X_{1:n})$ are the same as (\ref{KDE}) and (\ref{KDE2}), respectively, but the dependence on the entire sample $X_{1:n}$ is addressed in these notations. 
The proper likelihood of $(w,h)$ can be obtained by appropriately normalizing the joint density (\ref{joint-model}).
However, as noted in \cite{jewson2021general}, $\hat{f}_{h}(X_i|X_{1:n})$ is bounded below by a positive constant as
\begin{align*}
\hat{f}_{h}(X_i|X_{1:n})=\frac{K(0)}{nh} + \frac{1}{nh}\sum_{j\neq i}^n K\left(\frac{X_i-X_j}{h}\right)
\geq  \frac{K(0)}{nh},
\end{align*}
such that the joint model (\ref{joint-model}) has an infinite normalizing constant.
Models with diverging normalizing constants cannot be  interpreted directly in terms of probability; however, the model can be interpreted as describing relative probabilities \citep{jewson2021general}.

As an alternative construction of the joint probability model for $X_{1:n}$, we consider the leave-one-out (LOO) form
$$
\hat{f}_h^{(-i)}(X_i|X_{1:n}) \equiv \frac1{nh}\sum_{ j\neq i}^nK\left(\frac{X_i-X_j}{h}\right),
$$
and define an alternative joint model as follows:
\begin{equation}\label{joint-model2}
g_{\mathtt{LOO}}(X_{1:n}; w, h) \propto  \exp\left\{w\sum_{i=1}^n \log \hat{f}_h^{(-i)}(X_i|X_{1:n})\right\}.
\end{equation}
Although the normalizing constant of the joint model (\ref{joint-model2}) may not diverge unlike that in (\ref{joint-model}), it requires $n$-dimensional intractable integration.


\cite{jewson2021general} considered the joint model of $X_{1:n}$ as a generalized statistical model with diverging normalizing constant and employed the Hyv\"arinen score (hereafter referred to as the ``H-score") to the joint model to obtain a proper objective function for $(w, h)$.
Note that the parameter estimation using the H-score is called as \textit{score matching} \citep{hyvarinen2005estimation}, and the H-score is an empirical version of the Fisher divergence between the joint model $g(X_{1:n})$ and true density $f(X_{1:n}) = \prod_{i=1}^n f(X_i)$, which is defined as 
\begin{align}
    J(f||g) \equiv \frac{1}{2}\int_{\mathcal{X}^n} \left\|  \nabla_{X_{1:n}}\log g(X_{1:n}) - \nabla_{X_{1:n}}\log f(X_{1:n})\right\|^2 f(X_{1:n})dX_{1:n},
\end{align}
where $\nabla_{X_{1:n}}$ denotes the $n$-dimensional gradient vector (see \citealp[pp.697]{hyvarinen2005estimation}). This divergence is suitable for the inference of unnormalized models because the gradient of the logarithm does not depend on the normalizing constants. 

Although $g_{\mathtt{LOO}}$ is normalizable, it involves an unrealistic $n$-dimensional integral, as mentioned earlier. \cite{jewson2021general} considered the Fisher-divergence for non-normalizable models, but \cite{hyvarinen2005estimation} is designed for cases where the normalizing constants are difficult to compute and is suitable for coping with our problem.

\begin{rem}
    In the following section, we consider two types of Fisher divergences.
    The first version is the Fisher-divergence between the joint unnormalized model $g$ or $g_{\mathtt{LOO}}$, which are defined as (\ref{joint-model}) or (\ref{joint-model2}), respectively, and true density  $f$. We denote them as $J_{X_{1:n}}(f||g)$ and $J_{X_{1:n}}(f||g_{\mathtt{LOO}})$, respectively. 
    This type of Fisher-divergence was first introduced by \cite{jewson2021general}. In their concept, the exponential of a loss function is considered as an unnormalized probability model. 
    In the context of bandwidth selection for KDE, the feasible loss function based on the negative log-likelihood is given by $- \sum_{i=1}^n \log \hat{f}_h(X_i)$ or $- \sum_{i=1}^n \log \hat{f}_h^{(-i)}(X_i)$ and then the resulting unnormalized model is $g$ or $g_{\mathtt{LOO}}$.
    
    The second version represents the Fisher divergence between the estimator for future observations $\hat{f}_{w,h}$ and true density $f$ (denoted as $J_x(f||\hat{f}_{w,h})$). 
    Because the estimator to be used is $\hat{f}_{w,h}$, the tuning parameter $(w,h)$ should be selected such that $J_x(f||\hat{f}_{w,h})$ is minimized. In Section \ref{sec:consistency}, we confirm the asymptotic equivalence between the H-score based on  $J_{X_{1:n}}(f||g)$ or $J_{X_{1:n}}(f||g_{\mathtt{LOO}})$ and Fisher divergence $J_x(f||\hat{f}_{w,h})$. 
    \begin{align*}
        & J_{X_{1:n}}(f||g) \equiv  \frac{1}{2} \int_{\mathcal{X}^n}  \left\| \nabla_{X_{1:n}} \log  \hat{f}_h(X_{1:n})^w  - \nabla_{X_{1:n}}\log f(X_{1:n})\}  \right\|^2 f(X_{1:n})dX_{1:n},  \\
        & J_x(f||\hat{f}_{w,h}) \equiv  \frac{1}{2} \int_{\mathcal{X}}  \left( \frac{\partial}{\partial x}\log \hat{f}_{w,h}(x) - \frac{\partial}{\partial x} \log f(x)\right)^2 f(x)dx.
    \end{align*}
    From a frequentist standpoint, it is standard to start from $J_x(f||\hat{f}_{w,h})$ and make its  empirical counterpart with leave-one-out technique (LHS defined later); however in the generalized Bayesian method, it is standard to consider $J_{X_{1:n}}(f||g)$.
    Although we consider $J_{X_{1:n}}(f||g) $ as a starting point for constructing the H-score, which is applicable to both frequentistic and generalized Bayesian methods, the theorems we present bellow are equally valid, even if we start from $J_{x}(f||\hat{f}_{w,h})$.
\end{rem}

\subsection{Estimation of parameters in the exponentiated KDE}
\label{sec:methodology}

We introduce two objective functions to estimate $(w, h)$ using the H-scores of the in-sample joint density (\ref{joint-model}) (denoted by IHS) and the LOO-based joint density (\ref{joint-model2}) (denoted by LHS).
Note that IHS has already been adopted in \cite{jewson2021general}; however, LHS is an alternative that follows the standard approach to model evaluation in the context of kernel smoothing.
As discussed latter (Remark \ref{rem:LHS_IHS}), IHS and LHS are asymptotically equivalent. 
From the Fisher divergence $J_{X_{1:n}}(f||g)$, the $i$-th component of IHS is given by:
\begin{align}
    \mathcal{H}(X_i;g)  &\equiv 2\frac{\partial^2}{\partial X_i^2}\log \hat{f}_{h}(X_i)^w + \left\{\frac{\partial}{\partial X_i}\log \hat{f}_{h}(X_i)^w \right\}^2 \nonumber\\
    & = 2w\left\{\frac{1}{h^2}\sum_{j\neq i}^nK''_{ij}\right\}\left\{K(0)+\sum_{j\neq i}^nK_{ij}\right\}^{-1} \nonumber\\
    & \quad + (w^2-2w)\left\{\frac{1}{h}\sum_{j\neq i}^nK'_{ij}\right\}^2\left\{K(0) +\sum_{j\neq i}^nK_{ij}\right\}^{-2},
    \label{eq:H-score}
\end{align}
where $K_{ij}\equiv K\{(X_i-X_j)/h\}$, $K'_{ij}\equiv dK(u)/du|_{u=(X_i-X_j)/h}$ and $K''_{ij}\equiv d^2K(u)/du^2|_{u=(X_i-X_j)/h}$.
Similarly, LHS is derived from the unnormalized model $g_{\mathtt{LOO}}$, given by:
\begin{align*}
    \mathcal{H}(X_i;g_{\mathtt{LOO}})  \equiv  2w\left\{\frac{1}{h^2}\sum_{j\neq i}^nK''_{ij}\right\}\left\{\sum_{j\neq i}^nK_{ij}\right\}^{-1} + (w^2-2w)\left\{\frac{1}{h}\sum_{j\neq i}^nK'_{ij}\right\}^2\left\{\sum_{j\neq i}^nK_{ij}\right\}^{-2}.
\end{align*}
Now, we have the objective functions:
\begin{align}
    & \mathcal{H}(w,h) \equiv n^{-1}\sum_{i=1}^n \mathcal{H}(X_i;g) \label{eq:IHS}\\
    & \mathcal{H}_{\mathtt{LOO}}(w,h) \equiv n^{-1}\sum_{i=1}^n \mathcal{H}(X_i;g_{\mathtt{LOO}}) \label{eq:LHS}
\end{align} 
We then estimate $w$ and $h$ as follows: 
\begin{equation}
    (\hat{w}_n,\hat{h}_n) \equiv \argmin_{w,h} \mathcal{H}(w,h), ~~ \text{or} ~~ (\hat{w}_n,\hat{h}_n) \equiv \argmin_{w,h} \mathcal{H}_{\mathtt{LOO}}(w,h) \label{eq:hw-hat}
\end{equation}
or setting $w$ to be some constant (e.g., $w=1$),
\begin{equation}
    \hat{h}_n \equiv \argmin_{h} \mathcal{H}(1,h) ~~ \text{or} ~~ \hat{h}_n \equiv \argmin_{h} \mathcal{H}_{\mathtt{LOO}}(1,h). \label{eq:h-hat}
\end{equation}
Because the objective functions in (\ref{eq:hw-hat}) and (\ref{eq:h-hat}) are smooth functions of $h$ and $w$, the optimization problems can be solved easily using iteration algorithms. 

Given the values of $(w,h)$, the unnormalized density estimator can be obtained from (\ref{KDE2}). Thus, the normalized version is given by:
$$
\hat{f}_{w,h}(x)=\left\{\int \hat{f}_h(t)^w dt\right\}^{-1}\hat{f}_h(x)^w,  \ \  \ x\in \mathcal{X}
$$
such that $\int \hat{f}_{w,h}(x) dx=1$.

\begin{rem}
As we mentioned previously, \cite{jewson2021general} also consider a similar estimator for the density estimation problem.
 However, unlike our study, they restrict the problem to the Gaussian kernel case only.
 Furthermore, although they treated (\ref{eq:H-score}) as an independent term and performed theoretical analysis, their argument was not rigorous in this sense, evident from the form of the H-score during KDE, which depends on all observations.
 Therefore, contrary to their claims, their theoretical considerations would not provide any guarantee that \eqref{eq:IHS} and \eqref{eq:LHS} converge to the Fisher divergence.
\end{rem}


\section{Theoretical Results} \label{sec:theory}
Here, we present the asymptotic properties of the proposed exponentiated KDE. 
In particular, we derive the following four properties under suitable regularity conditions, as stated later:
\begin{enumerate}
\item 
Decomposition of the expected Fisher divergence into two terms (bias and variance terms), given in Section~\ref{sec:deco}.
\item 
Characterization of the theoretically optimal $w$ and $h$, given in Section~\ref{sec:characterisation}.
\item 
Consistency of the parameters estimated by the Hyv\"arinen score, given in Section~\ref{sec:consistency}.
\item 
Properties of the expected Fisher divergence and Hyv\"arinen score when $w$ is fixed, given in Section~\ref{sec:w_fix}.
\end{enumerate}

First, we state the regularity conditions to establish the theoretical properties.

\medskip
\begin{assd}[Data Generating Process] \label{ass:datageneratingprocess}\mbox{}
\begin{enumerate} 
    \renewcommand{\theenumi}{D(\alph{enumi})}
    \renewcommand{\labelenumi}{\theenumi:}
    \item  $X_{1:n}$ are random samples with an absolutely continuous distribution with Lebesgue density $f$. \label{ass:DGP}
    \item $x$ is an interior point in the support of $X$. \label{ass:interiorpoint}
    \item  In a neighborhood of $x$, $f$ is $(L+2)$-times continuously differentiable and its first $(L+2)$-derivatives are bounded. \label{ass:fdiffrentiability1}
    \item The following conditions hold:
    \begin{center}
        $ \lim_{x\rightarrow\pm\infty}f'(x)=0, \quad \lim_{x\rightarrow\pm\infty}f^{(L+1)}(x)=0 , \quad \lim_{x\rightarrow\pm\infty}f'(x)f^{(L)}(x)f(x)^{-1}=0, 
        \newline
        \lim_{x\rightarrow\pm\infty}f^{(L)}(x)f^{(L+1)}(x)f(x)^{-1}=0, \quad \lim_{x\rightarrow\pm\infty}f'(x)f^{(L)}(x)^2f(x)^{-2}=0.
        $
    \end{center}  \label{ass:hyvarinen1}
    \item For the density function $f$, there are no constant $C$ (independent of $x$) such that the following equality holds: 
     $Cf'(x) = f^{(L+1)}(x) - f'(x)f^{(L)}(x)f(x)^{-1}, \quad \forall x\in\mathcal{X}$.
    \label{ass:csequality}
\end{enumerate}
\end{assd}

\begin{assk}[Class of kernel functions]\label{ass:kernelfunction} \mbox{} 
\begin{enumerate}
    \renewcommand{\theenumi}{K(\alph{enumi})}
    \renewcommand{\labelenumi}{\theenumi:}
    \item Kernel function $K$ is bounded, even function, of order $L\ge 2$, $\int  K(u)du=1$ and twice continuously differentiable almost everywhere on the support of $K$.  \label{ass:Kernel}
    \item 
        $\lim_{u\rightarrow\pm\infty}|K(u)|=0, \quad \lim_{u\rightarrow\pm\infty}|K'(u)|=0$ and $\quad \lim_{u\rightarrow\pm\infty}|K''(u)|=0 $ \label{ass:integrationbyparts}
    \item $K''K$, $K'K'$, $K''KKK$ and $K'K'KK$ are H\"older continuous.  \label{ass:kernelholder}
\end{enumerate}
\end{assk}

In the following, we assume that the bandwidth and the learning rate depend on the sample size $n$, denoted by $h_n$ and $w_n$ respectively.
In addition, following the standard asymptotic framework in the literature on kernel smoothing, we assume that the sequence $h_n$ tends toward $0$ as $n\to\infty$ (for example  
\citealt[pp.20]{wand1996smoothing}, \citealt[pp.73]{wasserman2006all} or \citealt[pp.3]{tsybakov2009nonpara}). 
As an analogy of this assumption on the bandwidth, we impose the assumption $w_n \to 1$ on the learning rate. 
In addition, we perform the theoretical analysis without this assumption for $w_n$ in Section  \ref{sec:w_fix}.

\medskip
\begin{assm}[Asymptotic framework]\label{ass:model} \mbox{}
\begin{enumerate}
    \renewcommand{\theenumi}{M(\alph{enumi})}
    \renewcommand{\labelenumi}{\theenumi:}
    \item The parameter space of the bandwidth is defined as $\mathscr{H}_n$ and limited as $\mathscr{H}_n = [Cn^{-1/3+\epsilon_h}, Cn^{-\epsilon_h}]$ for arbitrarily small $\epsilon_h\in\mathbb{R}_{++}$ and some constant $C>0$. \label{ass:nhinfty}
    \item The parameter space of the learning rate is defined as $\mathscr{W}_n$ and limited as  $\mathscr{W}_n \equiv[\delta \vee 1-Cn^{-\epsilon_w},1+Cn^{-\epsilon_w}]$ for arbitrarily small $\epsilon_w\in\mathbb{R}_{++}$ and $\delta \in \mathbb{R}_{++}$ and some constant $C>0$. \label{ass:wlocalise}
\end{enumerate}
\end{assm}

\medskip
Assumption~\ref{ass:interiorpoint} excludes the theoretical analysis at the boundary of the support of $X$. 
It is well known that simple kernel-based estimators, such as KDE and Nadaraya-Watson estimator suffer from a bias at the boundary of the support of the observations without an ingenuity \citep{fan1996local} (see \cite{cattaneo2020polynomialdensity} and references therein for details and the recent progress on this problem in density estimation). 
We use Assumption \ref{ass:fdiffrentiability1} for the expansions in Lemma \ref{lem:k'k_x}-\ref{lem:k'k'kk}.
We need Assumption \ref{ass:hyvarinen1} to show the equivalency of the expected Hyv\"arinen score to the expected Fisher divergence, as shown in  Theorem \ref{thm:eh} using integration by parts. 
Our operation can be understood as the inverse of the transformation of Fisher divergence into an empirically estimable form proposed by \cite{hyvarinen2005estimation}. 
Assumption \ref{ass:csequality} is a high-level condition that guarantees the existence of a theoretically optimal bandwidth $h_n^*$ defined as Corollary \ref{cor:opt_h}.
 Even if this assumption does not hold true, we can derive theoretically optimal bandwidth that differs from Corollary \ref{cor:opt_h}; see Remark \ref{rem:csequality} for further details.

Assumption \ref{ass:Kernel} is a standard one in the context of KDE. 
We use Assumption \ref{ass:integrationbyparts} for the integration by parts in the proof of lemmas.  
This assumption is not restrictive and the Gaussian Kernel and many other kernels satisfy this condition.
Finally, Assumption \ref{ass:kernelholder} is required for Theorem~\ref{thm:uniform}. 
Assumption \ref{ass:kernelholder} is not essential because that Theorem~\ref{thm:uniform} could be proved without it if one adopt the other strategies for the proof.

Under  Assumption \ref{ass:nhinfty} or  both \ref{ass:nhinfty} and \ref{ass:wlocalise}, the Fisher-divergence and Hyv\"arinen Score are expanded. 
Note that Assumption \ref{ass:model} limits the parameter space of the bandwidth sequence $h_n$ such that $h_n \to 0$ and $nh_n^3 \to \infty$ as $n\to\infty$ holds true, and that of the learning rate $w_n$ such that $w_n \to 1$ as $n \to \infty$ and $w_n>0$ hold true.

\begin{rem} \label{rem:LHS_IHS}
\cite{jewson2021general} constructed the H-score without the LOO technique. 
According to \cite{jewson2021general}'s and our numerical studies, it is evident that estimator $\hat{f}_{w,h}(x)$ with parameters selected based on such Hyvärinen score does not overfit (overfitting indicates that the estimated bandwidth is significantly close to $0$).

This is theoretically explained below. 
The definition of the Fisher-divergence contains the ratio of the first-order density derivative estimator $\partial\hat{f}_{h_n}(x)/\partial x$ to density estimator $\hat{f}_{h}(x)$. 
The order of the variance of $\partial\hat{f}_{h_n}(x)/\partial x$ is $O\{(nh_n^3)^{-1}\}$, which dominates  that of $\hat{f}_{h_n}(x)$ (See \citealp{jones1994kernel}). 
However, when deriving the H-score, the term with the order, which occurs because of not performing LOO and can cause overfitting, is eliminated by the derivative at the evaluation point $X_i$. 
It can be observed that $K''(0)/h_n^2$ and $K'(0)/h_n$ do not appear in the numerator of (\ref{eq:H-score}). 
To rephrase, a part of the density derivative is substantially estimated in the LOO manner.
Although the term $K(0)/nh_n$ remains without performing LOO, its influence is asymptotically negligible, because the density derivative estimator, whose variance rate is $O\{(nh_n^3)^{-1}\}$, is dominant as explained. 
Therefore, parameter selection based on the IHS (\ref{eq:H-score}) does not result in overfitting.
In addition, our theoretical results are asymptotically invariant irrespective of H-score being constructed in LOO manner.
\end{rem}

\subsection{Decomposition of expected Fisher divergence} \label{sec:deco}
We first consider the decomposition of the expected Fisher divergence into two terms. 
These terms can be regarded as bias and variance like terms, similar to the decomposition of Mean Integrated Squared Erorr (MISE) in the standard nonparametric estimation. 
The results are presented in the following proposition.

\begin{prop}[Decomposition of the expected Fisher divergence] \label{prop:bv_deco} 
For a given density estimator $\hat{f}$ for density $f$, possibly different  from $\hat{f}_h$ and $\hat{f}_{w,h}$, it holds that 
\begin{align}
    \mathbb{E}_{X_{1:n}}[J_x(f||\hat{f})] = \frac{1}{2}(V_J + B_J) \nonumber,
\end{align}
provided that $V_J$ and $B_J$ exist, where $\mathbb{E}_{X_{1:n}}$ is the expectation operator over the whole observation $X_{1:n}$. The definitions of $V_J$ and $B_J$ are given below.
\begin{align*}
    & V_J \equiv \int \mathbb{E}_{X_{1:n}}\left[\left\{\frac{\partial}{\partial x} \log\hat{f}(x)-\mathbb{E}_{X_{1:n}}\left[\frac{\partial}{\partial x} \log\hat{f}(x)\right]\right\}^2\right]f(x)dx,\\
    & B_J \equiv \int \mathbb{E}_{X_{1:n}}\left[\frac{\partial}{\partial x} \log\hat{f}(x)-\frac{\partial}{\partial x} \log f(x)\right]^2f(x)dx .
\end{align*}
\begin{proof}
See Section~\ref{sec:bv_deco} in Supplementary Material. 
\end{proof}
\end{prop}

This decomposition provides us clear insights into the expected Fisher divergence. 
Specifically, we understand that the expected Fisher divergence is the weighted MISE of the estimator of $\frac{\partial}{\partial x}\log f(x)$ and balances its squared bias and the variance.

\subsection{Characterization of theoretically optimal parameters} \label{sec:characterisation}
Here, we provide the asymptotic results for characterizing theoretically optimal $w$ and $h$. 
Accordingly, we provide a fundamental result showing the asymptotic representation of the expected Fisher divergence as follows:

\begin{thm}[Asymptotic representation of the expected Fisher divergence] \label{thm:ej} 
Under Assumptions \ref{ass:DGP}-\ref{ass:fdiffrentiability1}, \ref{ass:Kernel},  \ref{ass:integrationbyparts}, and \ref{ass:model}, the expected Fisher divergence is expanded as follows:
\begin{align*}
    \mathbb{E}_{X_{1:n}}[J_x(f||\hat{f}_{w_n,h_n})] 
    &= C_{(B,2L)}h_n^{2L} + (w_n-1)^2C_{(B,1)} + (w_n-1)C_{(B,L)}h_n^L  +  \frac{C_V}{nh_n^3} \\
    & \quad + o\left\{h_n^{2L}+(w_n-1)h_n^L+(w_n-1)^2+\frac{1}{nh_n^3}\right\} 
\end{align*}
where
    \begin{align*}
        & C_{(B,2L)} \equiv \kappa_L^2  \mathbb{E}_x\Bigl[\{f^{(L+1)}(x)f(x)^{-1}-f'(x)f^{(L)}(x)f(x)^{-2}\}^2\Bigl] \\
        & C_{(B,L)} \equiv 2\kappa_L \mathbb{E}_x\Bigl[f'(x)f^{(L+1)}(x)f(x)^{-2}-f'(x)^2f^{(L)}(x)f(x)^{-3}\Bigl] \\
        & C_{(B,1)} \equiv \mathbb{E}_x[f'(x)^2f(x)^{-2}], \quad
        C_V \equiv R(K')\mathbb{E}_x[f(x)^{-1}]
    \end{align*}
where $\kappa_l \equiv \frac{1}{l!}\int u^lK(u)du$ and $R(K')\equiv \int K'(u)^2du$. 
In addition, for a function $g: \mathcal{X} \mapsto \mathbb{R}$, we define operator  $\mathbb{E}_x$ as $\mathbb{E}_x[g(x)] \equiv \int g(x)f(x)dx$.
\begin{proof}
See Section \ref{section:outline_ej} in Supplementary Material for the outline of proof and Sections \ref{section:bias} and \ref{section:variance} for the details of the computation. 
\end{proof}
\end{thm}

Among the leading terms of the expected Fisher-divergence, the first three terms represent the asymptotic bias of the score function of KDE and the last term represent the asymptotic variance of it. 
So, asymptotically, the minimization of the Fisher-divergence can be interpreted as selecting the $(w_n,h_n)$ that balances these bias and variance terms.

\begin{rem}
    When the support of the underlying density is non-compact, moment $\mathbb{E}_x[f(x)^{-1}]$ does not exist. 
    It is well known that a similar phenomenon occurs in the least-squares CV method for regression estimators such as the Nadaraya-Watson estimator, where the inverse of density emerges in the variance term; thus, trimming significantly improves the accuracy (see \citealp{racine2004nadarayacv,hall2007nadarayacv}). 
    Because the data must lie within a finite range, trimming procedures will not be necessary in our case. 
    Although the DGPs in our numerical studies have non-compact support, the selected bandwidths and learning rates do not behave in a problematic manner.
\end{rem}

Based on Theorem \ref{thm:ej}, we define the asymptotically optimal bandwidth $h_n^*$ and the asymptotically optimal learning rate $w_n^*$. 
\begin{df}[Asymptotically Optimal Parameters] \label{def:opt}
Asymptotically optimal parameters $(w_n^*, h_n^*)$ are defined as the minimizer of the leading terms of the expected Fisher divergence.
\begin{align*}
    (w_n^*, h_n^*) \equiv \argmin_{(w_n,h_n) \in \mathscr{W}_n \times \mathscr{H}_n} \left( C_{(B,2L)}h_n^{2L} + (w_n-1)^2C_{(B,1)} + (w_n-1)C_{(B,L)}h_n^L  +  \frac{C_V}{nh_n^3} \right)
\end{align*}
\end{df}

\begin{rem}
    In Definition \ref{def:opt}, the parameter space is limited to $\mathscr{W}_n \times \mathscr{H}_n$. 
    This implies that the asymptotically optimal bandwidth $h_n^*$ is optimal in bandwidths that satisfy the conditions $h_n \to 0$ and $nh_n^3 \to \infty$. 
    This kind of definition is standard in the literature on kernel smoothing. 
    For example, the MISE optimal bandwidth is defined in the same manner, i.e., its leading terms are minimized under $h_n \to 0, nh_n \to \infty$; see \citet[equation 2.13]{wand1996smoothing}, \citet[equation 6.30]{wasserman2006all}, and \citet[pp.15]{tsybakov2009nonpara}.
\end{rem}

The optimal learning rate $w_n^*$ is explicitly given by the following corollary:  
\begin{cor}[Theoretically optimal $w_n$] \label{cor:opt_w} Under the same assumptions as those in Theorem \ref{thm:ej} and Assumption \ref{ass:csequality},  theoretically optimal $w_n$ can be given by:
\begin{align}
    w_n^* &= 1 - \frac{C_{(B,L)}}{2C_{(B,1)}}{h_n^*}^L. \nonumber
\end{align}
The defintion of $h_n^*$ is given below.
In particular, the optimal $w$ is larger than 1 if $L=2$.
\begin{proof}
The first claim follows immediately from Theorem \ref{thm:ej} and the first-order condition for minimization. 
Note that integration by parts yields
    \begin{align}
        \mathbb{E}_x[f'(x)f^{(L+1)}(x)f(x)^{-2}-f'(x)f^{(L)}(x)f(x)^{-3}] = -\int f''(x)f^{(L)}(x)f(x)^{-1}dx. \nonumber
    \end{align}
Hence, when $L=2$, $C_{(B,L)}=-2\kappa_L\int f''(x)^2f(x)^{-1}dx$.
Because $\kappa_L, f''(x)^2$ and $f(x)^{-1}$ are always strictly positive, $C_{(B,L)}$ is always negative when $L=2$. This proves the second claim.
\end{proof}
\end{cor}
Although we show that $w_n^* \ge 1$ only when $L=2$, this will be adequate for the following reasons; $(1)$ $L=2$ is the most common case, $(2)$ $L>2$ is unsuitable for the proposed method because a higher-order kernel makes $\hat{f}_{h_n}(x)$ negative at some points and thus, $\hat{f}_{w_n,h_n}(x) \propto \hat{f}_{h_n}(x)^{w_n}$ does not take any value in $\mathbb{R}_{+}$.

We can also characterize the theoretically optimal $w_n$ and $h_n$ as $n$-dependent sequences. 
The following two corollaries state that $h_n^*$ converges to $0$ at the rate of $n^{-1/(2L+3)}$ and $w_n^*$ converges to $1$ at the rate of $n^{-L/(2L+3)}$.

\begin{cor}[Theoretically optimal bandwidth] \label{cor:opt_h} Under the same assumptions as those in Theorem \ref{thm:ej} and Assumption \ref{ass:csequality}, 
\begin{align}
    h_n^* = \left(\frac{6C_{(B,1)}C_V}{L[4C_{(B,1)}C_{(B,2L)}-C_{(B,L)}^2]}\right)^{1/(2L+3)}n^{-1/(2L+3)} \nonumber
\end{align}
\begin{proof}
See Section \ref{sec:proof_opt_h} of Supplementary Materials.
\end{proof}
\end{cor}

\begin{rem} \label{rem:csequality}
For $h_n^*$ to exist, $4C_{(B,1)}C_{(B,2L)}-C_{(B,L)}^2$ must be strictly positive. 
The non-negativity is proven in Section \ref{sec:proof_opt_h} by the Cauchy-Schwarz inequality. 
Also, the equality $4C_{(B,1)}C_{(B,2L)}-C_{(B,L)}^2=0$ does not hold true for many densities. For example, $h_n^*$ has the form of Corollary \ref{cor:opt_h} for all DGPs in our numerical studies.  
In addition, even when the equality is valid, higher-order expansion of the expected Fisher-divergence yields a theoretically optimal bandwidth. This expansion can be obtained by using Assumption \ref{ass:wlocalise} and rearranging Theorem \ref{thm:ej_fix} with respect to $(w-1)$.
\end{rem}

By inserting the explicit form of $h_n^*$ in Proposition \ref{cor:opt_h} into $h_n^{*}$ in Proposition \ref{cor:opt_w}, we can characterize the optimal $w_n^*-1$ as $n$-dependent sequence. 
Specifically, the following corollary explicitly states that $w_n^*$ converges to $1$ at the rate of $n^{-L/(2L+3)}$. 

\begin{cor}[Theoretically optimal $w_n-1$ as $n$-dependent sequence] \label{cor:opr_wn} Under the same assumptions as those in Theorem  \ref{thm:ej} and Assumption \ref{ass:csequality},
\begin{align}
    w_n^*-1 = \frac{-C_{(B,L)}}{2C_{(B,1)}}\left(\frac{6C_{(B,1)}C_V}{L[4C_{(B,1)}C_{(B,2L)}-C_{(B,L)^2}]}\right)^{L/(2L+3)}n^{-L/(2L+3)} \nonumber
\end{align}
\end{cor}

Finally, the following theorem guarantees, at least asymptotically, that the Fisher divergence is convex and the vector of such theoretically optimal parameters satisfies the second-order condition for minimization; that is, we can select the parameters empirically if the sample size is large.

\begin{thm} \label{thm:convex}
Under the same assumptions as those in Theorem~\ref{thm:ej}, the expected Fisher divergence is asymptotically a convex function with respect to $(w_n,h_n)$; and thus, $(w_n^*,h_n^*)$  satisfies the sufficient condition for minimizing of the expected Fisher divergence. 
\begin{proof}
See Section~\ref{sec:proof_hessian} of Supplementary Material.
\end{proof}
\end{thm}

\begin{rem}
In the Supplementary Material, we derive the asymptotic expansion of MISE as Theorem \ref{thm:MISE}: $MISE \left[ \hat{f}_{w_n, h_n}(x) \right] = M_{(B,2L)}h_n^{2L} +  (w_n-1)M_{(B,L)}h_n^L + (w_n-1)^2M_{(B,1)}  + \frac{M_V}{nh_n} + o\left\{\frac{1}{nh_n}+ (w_n-1)^2 + (w_n-1)h_n^L + h_n^{2L}\right\}$, where $M_{(B,2L)}, M_{(B,L)}, M_{(B,1)}$ and $M_{V}$ are some constants. 
In view of this expansion and Corollary \ref{cor:opt_h} and \ref{cor:opr_wn}, the proposed KDE with Fisher-divergence optimal parameters, $\hat{f}_{w_n^*,h_n^*}$, does not achieve the optimal convergence rate (for example, $n^{-2/5}$ for twice continuously differentiable densities).
\end{rem}

\subsection{Consistency} \label{sec:consistency}
In this section, we provide propositions for the consistency of $(\hat{w}_n,\hat{h}_n)$ with $(w_n^*,h_n^*)$. 
Accordingly, we first provide an asymptotic representation of the expected H-score.
\begin{thm}\label{thm:eh}  
Under Assumptions \ref{ass:DGP}-\ref{ass:hyvarinen1}, \ref{ass:Kernel},  \ref{ass:integrationbyparts}, and \ref{ass:model}, the expected IHS can be expanded as follows:
\begin{align}
    &\mathbb{E}_{X_{1:n}}[\mathcal{H}(w_n,h_n)] + \mathbb{E}_x[f'(x)^2f(x)^{-1}] \nonumber\\
    & = 2\mathbb{E}_{X_{1:n}}[J_x(f||\hat{f}_{w_n,h_n})] + o\left\{\frac{1}{nh_n^3}+ (w_n-1)^2 + (w_n-1)h_n^L + h_n^{2L}\right\}. \nonumber
\end{align}
In addition, the leading term on the expected LHS is the same as that on IHS.
\begin{proof}
See Section \ref{sec:proof_eh} in Supplementary Material for an outline of the proof and Section~\ref{sec:the_expectation_of_Hyvarinen_score} for detailed computation.
\end{proof}
\end{thm}

The above theorem gives asymptotic equivalency of the expected Hyv\"arinen score to expected Fisher divergence and thus their minimizers. 

\begin{cor} \label{cor:equivalency}
Under the same assumptions as those in Theorem \ref{thm:eh}, the minimizer of the expected H-score (both IHS and LHS) and that of the expected Fisher divergence is equivalent. 
\begin{proof}
This can be immediately deduced from Theorem \ref{thm:eh}.
\end{proof}
\end{cor}

The following theorem states that the components that depend on $w_n$ and $h_n$ in the stochastic process  $\mathcal{H}(w_n,h_n)-\mathbb{E}_{X_{1:n}}[\mathcal{H}(w_n,h_n)]$ converges (at the rate required for consistency of tuning parameters) to 0 uniformly in $h_n$ and $w_n$ on the given parameter space. 

\begin{thm}\label{thm:uniform} 
Under Assumptions \ref{ass:DGP} - \ref{ass:fdiffrentiability1}, \ref{ass:kernelfunction} and \ref{ass:model},  the process $\mathcal{H}(w_n,h_n) - \mathbb{E}_{X_{1:n}}[\mathcal{H}(w_n,h_n)]$ is given by:
\begin{align}
    \mathcal{H}(w_n,h_n) - \mathbb{E}_{X_{1:n}}[\mathcal{H}(w_n,h_n)] &= Q_n + R_n(w_n,h_n),  \nonumber
\end{align}
where $Q_n$ is defined as:
\begin{align*}
    Q_n & \equiv \frac{1}{n}\sum_{i=1}^n \Bigl(\{2f''(X_i)f(X_i)^{-1}-f'(X_i)^2f(X_i)^{-2}\} -\mathbb{E}_{X_i}\left[\{2f''(X_i)f(X_i)^{-1}-f'(X_i)^2f(X_i)^{-2}\}\right]\Bigl),
\end{align*}
and does not depend on $h_n$ or $w_n$. $\mathbb{E}_{X_i}$ is the expectation operator over $i$-th observation. Additionally, $R_n(w_n,h_n)$ is a stochastic process which satisfies 
\begin{align}
    \sup_{(w_n,h_n)\in\mathscr{W}_n\times\mathscr{H}_n} \left\{h_n^{2L}+(w_n-1)h_n^L+(w_n-1)^2+\frac{1}{nh_n^3}\right\}^{-1} R_n(w_n,h_n) \xrightarrow{p} 0 \nonumber.
\end{align}
In addition, this statement holds true for the case of LHS.
\begin{proof}
See Section~\ref{sec:proof_uniform} of Supplementary Material for an outline of the proof.
\end{proof}
\end{thm}
Corollary \ref{cor:equivalency} states that the minimizer of the expected Fisher-divergence and the expected H-score is equivalent, and Theorem \ref{thm:uniform} guarantees that we can get the minimizer of the expected H-score by minimizing the empirical H-score. Consequently, Theorem \ref{thm:uniform} and Corollary \ref{cor:equivalency} imply the consistency of $(\hat{w}_n,\hat{h}_n)$ with $(w_n^*,h_n^*)$ in probability.
\begin{cor} \label{cor:consistency}
Under the same assumptions as those in Theorem \ref{thm:uniform}, we obtain
\begin{align*}
    (\hat{h}_n - h_n^*) / h_n^* \to 0, \quad \{(\hat{w}_n - 1) - (w_n^* - 1)\} / (w_n^* -1) \to 0.
\end{align*}
\begin{proof}
    See Section \ref{sec:proof_consistency} in Supplementary Material for the proof.
\end{proof}
\end{cor}

Note that we need the rescaling factors $h^*$ and $(w_n^*-1)$ in the denominators because these parameters converge to $0$ as $n\to\infty$. 
Therefore, we must consider this property to guarantee the consistency of the parameters.
\subsection{Properties under fixed $w$} \label{sec:w_fix}
In this section, we provide the asymptotic representation of the expected Fisher divergence and expected H-score with fixed $w$. 
It should be noted that $w$ does not depend on $n$; therefore, we drop the subscript $n$ in this section. First, for the higher-order expansion of the expected Fisher divergence and the expected H-score, we strengthen the assumptions on the data generating process.

\begin{assdd}[Data generating process] \label{ass:datageneratingprocessfix}
\mbox{}
\begin{enumerate} 
    \renewcommand{\theenumi}{D'(\alph{enumi})}
    \renewcommand{\labelenumi}{\theenumi:}
    \item In a neighborhood of $x$, $f$ is $(2L+2)$-times continuously differentiable and its first $(2L+2)$-derivatives are bounded. \label{ass:fdiffrentiability2}
    \item For any integer $L\le l \le 2L$, the following holds true:
\begin{center}
    $\lim_{x\rightarrow\pm\infty}f'(x)=0, \quad \lim_{x\rightarrow\pm\infty}f^{(l+1)}(x)=0 , \quad \lim_{x\rightarrow\pm\infty}f'(x)f^{(l)}(x)f(x)^{-1}=0, \newline
    \lim_{x\rightarrow\pm\infty}f^{(L)}(x)f^{(L+1)}(x)f(x)^{-1}=0, \quad \lim_{x\rightarrow\pm\infty}f'(x)f^{(L)}(x)^2f(x)^{-2}=0.$
\end{center}  \label{ass:hyvarinen2}
\end{enumerate}
\end{assdd}
Note that we need, among Assumption \ref{ass:model}, only Assumption \ref{ass:nhinfty} and  not \ref{ass:wlocalise} for the following two theorems.
The main result is the following theorem.
\begin{thm}\label{thm:ej_fix} Under Assumptions \ref{ass:DGP}, \ref{ass:interiorpoint}, \ref{ass:fdiffrentiability2}, \ref{ass:Kernel}, \ref{ass:integrationbyparts}, and \ref{ass:nhinfty}, the expected Fisher divergence can be expanded as follows.
\begin{align*}
    \mathbb{E}_{X_{1:n}}[J_x(f||\hat{f}_{w,h_n})] &= w^2C_{(B,2L)}h_n^{2L} + (w-1)^2C_{(B,1)} + (w-1)C_{(B,L)}h_n^L  +  w^2\frac{C_V}{nh_n^3} \\
    & \quad + 2w(w-1)\sum_{l=L+1}^{2L}C'_{(B,l)}h_n^l+ 2w(w-1)C'_{(B,2L)}h_n^{2L} + o\left\{h_n^{2L}+\frac{1}{nh_n^3}\right\} 
\end{align*}
where
    \begin{align*}
        & C'_{(B,l)} \equiv 2\kappa_l\mathbb{E}_x[f'(x)f^{(l+1)}(x)f(x)^{-2}-f'(x)^2f^{(l)}(x)f(x)^{-3}] \\
        & C'_{(B,2L)} \equiv \kappa_L^2\mathbb{E}_x[f'(x)^2f^{(L)}(x)^2f(x)^{-4}-f'(x)f^{(L)}(x)f^{(L+1)}(x)f(x)^{-3}]
    \end{align*}
\begin{proof}
See Section \ref{section:outline_ej} in Supplementary Material for the outline of the proof and Sections \ref{section:bias} and Section \ref{section:variance} for the details of the computation. 
\end{proof}
\end{thm}
If $w$ is fixed as $w\neq 1$, $\mathbb{E}_{X_{1:n}}[J_x(f||\hat{f}_{w,h_n})]$ does not converge to $0$ as $n\rightarrow\infty$ because $(w-1)^2C_{(B,1)}>0$. 

Furthermore, we can establish asymptotic equivalency of the expected Hyv{\"a}rinen score to the expected Fisher divergence when $w$ is fixed.

\begin{thm}\label{thm:eh_fix} 
Under Assumptions \ref{ass:DGP}, \ref{ass:interiorpoint}, \ref{ass:datageneratingprocessfix}, \ref{ass:Kernel}, \ref{ass:integrationbyparts}, and \ref{ass:nhinfty}, the expected Hyv{\"a}rinen score (both IHS and LHS) is asymptotically equivalent to the expected Fisher divergence. 
\begin{align}
    \mathbb{E}_{X_{1:n}}[\mathcal{H}(w,h_n)] + \mathbb{E}_x[f'(x)^2f(x)^{-1}] = 2\mathbb{E}_{X_{1:n}}[J_x(f||\hat{f}_{w,h_n})] + o\left\{\frac{1}{nh_n^3} + h_n^{2L}\right\} \nonumber
\end{align}
\end{thm}

Even when $w$ is fixed, the expected H-score is asymptotically equivalent to the expected Fisher divergence. Combined with Theorem \ref{thm:ej_fix}, we can see that the Hyv{\"a}rinen Score selects the bandwidth that minimizes an unreasonable criterion when $w$ is fixed to some value that is not $1$, so one should learn $w$ and $h_n$ simultaneously as long as consistency is one's concern.

\section{Numerical Examples} \label{sec:numerical}

\subsection{Simulation studies}\label{sec:sim}
We compare the numerical performance of the proposed exponentiated KDE with the standard KDE methods through simulation studies. 
Following the work by \citet{marron1992exact,jewson2021general}, we generated the synthetic data from the following Gaussian mixtures: $g(y) = \sum_{j=1}^Jm_j\mathcal{N}(y;\mu_j,\sigma_j^2)$,
where $\mu_j$ and $\sigma_j^2$ are the mean and variance of the normal distribution, respectively, and $m_j$ is the mixing proportions that satisfies $\sum_{j=1}^J m_j=1$.
Specifically, we considered the following five scenarios: 
\begin{enumerate}
    \item Bimodal: $J=2,\mu_1=-1.5,\mu_2=1.5,\sigma_1=\sigma_2=1/2$ and, $m_1=m_2=0.5$.
    \item Trimodal: $J=3,\mu_1=-1.2,\mu_2=0,\mu_3=1.2,\sigma_1=\sigma_3=0.4,\sigma_2=0.2,m_1=m_3=0.45$, and $m_2=0.1$
    \item Claw: $J=6,\mu_1=0,\sigma_1=1$, and $m_1=0.5$ then $\mu_j=-2+j/2,\sigma_j=0.1,m_j=0.1$ for $j=2,...,6$.
    \item Skewed: $J=5$, and $\mu_j=3(\sigma_j-1),\sigma_j=(2/3)^{j-1},m_j=1/J$ for $j=1,...,J$.
    \item Outlier: $J=2$, and $\mu_1=\mu_2=0,\sigma_1=1,\sigma_2=\sqrt{1/10},m_1=1/10,m_2=9/10$.
\end{enumerate}

For the simulated dataset, we applied the exponentiated KDE with $h_n$ and $w_n$ estimated using the LOO H-score (LHS) and in-sample H-score (IHS). 
We also applied the standard KDE using the bandwidth selected by the following two methods: 
\begin{itemize}
\item[-] 
{\bf Unbiased cross-validation (CV)}: Bandwidth $\hat{h}_{\rm CV}$ is determined by minimizing 
$
CV(h_n)=\int\hat{f}_{h_n}^2(X_i)-\frac{2}{n}\sum_{i=1}^n\hat{f}_{{h_n}, -i}^2(X_i), \ \ {\rm with} \ \ 
\hat{f}_{{h_n}, -i}^2(X_i)=\frac{1}{(n-1)h_n}\sum_{j\neq i}K\left(\frac{X_i-X_j}{h_n}\right).
$
Note that $CV(h_n)+\int f(x)^2dx$ is an unbiased estimator of MISE.

\item[-]
{\bf  MISE optimal plug-in (PI)}: Bandwidth is set to the asymptotic mean integrated squared error (AMISE) optimal bandwidth given by:\\
$
h_{\rm AMISE} = \left\{{R(K)}/{(2LC_L^2I_L)}\right\}^{1/(2L+1)}n^{-1/(2L+1)},
$
where $C_L = (L!)^{-1}\int u^LK(u)du$ and $I_L = \int \{f^{(L)}(x)\}^2dx$.
For $I_L$, we adopt the following estimator given in \cite{Hall1987derivative}:
$
\hat{I}_L = \int \{\hat{f}^{(L)}(x)\}^2 dx, \text{ with } \hat{f}^{(L)}(x) = \frac{1}{nb_n^{L+1}}\sum_{i=1}^n H^{(L)}\left(\frac{x-X_i}{b_n}\right),
$
where the kernel function $H$ and bandwidth $b_n$ may be different from $K$ and $h_n$, respectively.
\end{itemize}

\begin{figure}[htbp!]
\centering
\includegraphics[width=13cm,clip]{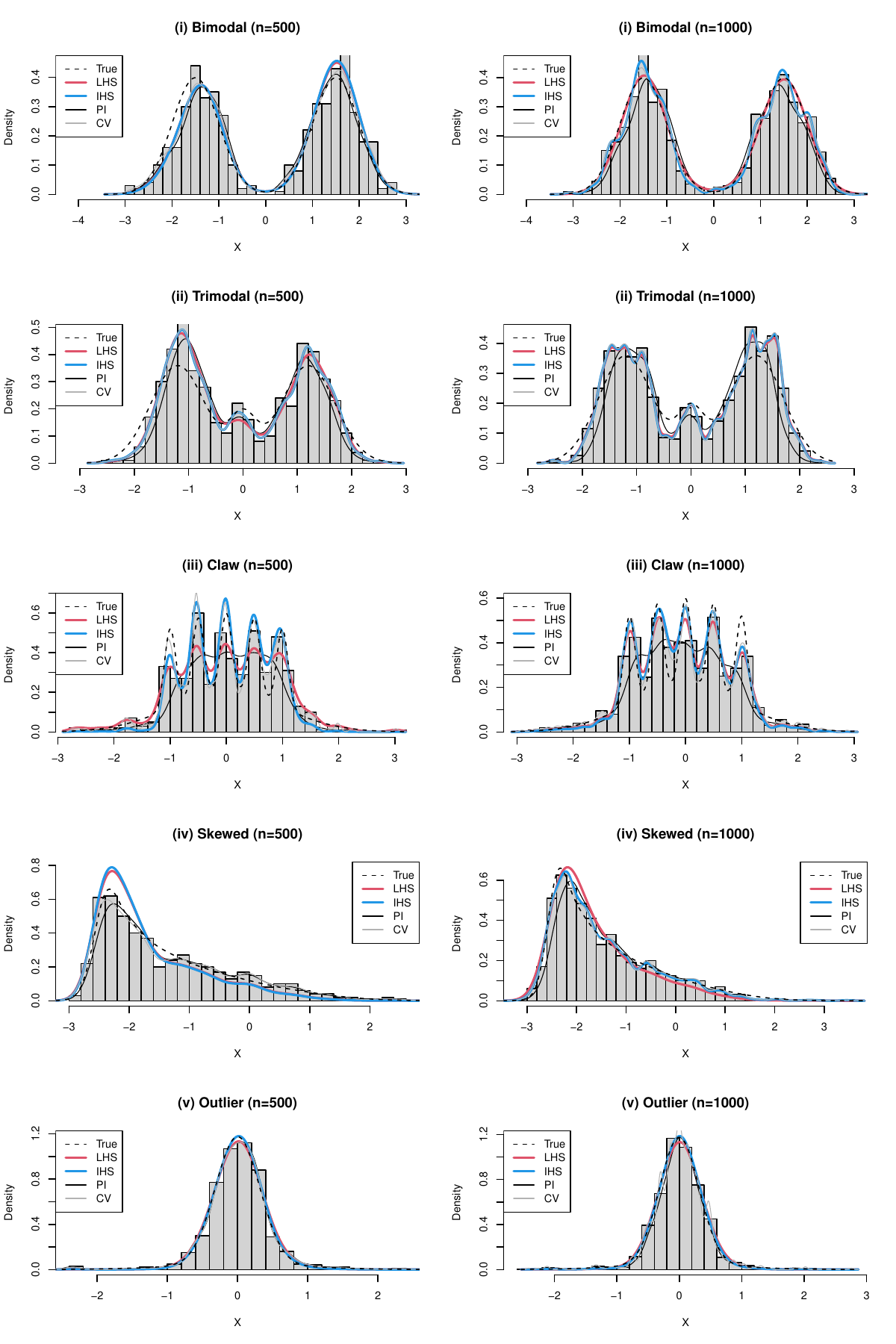}
\caption{Histograms of simulated data,  true density function, and estimated density functions based on a single Monte Carlo sample.}
\label{fig:oneshot}
\end{figure}

Throughout the experiment, we adopted a Gaussian kernel for all the methods.
Furthermore, we set the minimum values of $h_n$ for LHS and IHS to $0.5\min(\hat{h}_{\rm CV}, \hat{h}_{\rm AMISE})$ to avoid numerical instability.
Under the same five scenarios of the true density forms, we generated $n=500$ and $n=1000$ samples and estimated the density functions based on the generated data. 
In Figure \ref{fig:oneshot}, we first present the density functions estimated using the IHS, LHS, CV, and PI methods with the true density under a single simulated dataset.

To quantitatively compare the performances of the four density estimations, we computed the integrated squared error (ISE) and Kullback-Leibler divergence (KL), defined as follows:
$
{\rm ISE}_f=\int \{\hat{f}(x)-f(x)\}^2dx, ~~
{\rm KL}=\int f(x)\log\left(\frac{f(x)}{\hat{f}(x)}\right)dx.
$
Furthermore, to measure the estimation accuracy of the smoothness of the true density, we also computed the ISE of the curvature, defined as:
$
{\rm ISE}_C=\int \{\hat{C}(x)-C(x)\}^2dx, ~~C(x)=\frac{f''(x)}{\{1+f'(x)^2\}^{3/2}}.
$
Note that the above values were approximated by a set of grid points, $x_1<\cdots<x_K$, in our study.
Based on 1000 Monte Carlo replications, we computed 1000 values of ISE$_f$, KL, and ISE$_C$, and we obtained $25\%$, $50\%$, and $75\%$ quantiles of 1000 replications. 
The results of ISE$_f$ are summarized in Table \ref{tab:sim-MISE}.
It is reasonable that PI exhibits the best performance in most scenarios since the bandwidth of PI is selected to minimize AMISE. 
However, it is interesting that the proposed HS methods provide better results in some scenarios. 
Since IHS and LHS are not ISE-oriented objective functions, we suspect that this improvement in ISE is caused by the introduction of $w_n$.
To examine these conjectures, we theoretically derive the asymptotic MISE of $\hat{f}_{w_n,h_n}$ and MISE optimal parameters in Sections \ref{sec:MISE} and \ref{sec:MISE_opt} of Supplementary Material. 
Based on the theoretical results, we compare the asymptotic MISE of $\hat{f}_{w_n,h_n}$ and $\hat{f}_{h_n}$, and provide an additional discussion on the numerical results of ISE in Section \ref{sec:MISE_relative}.
Further, the results of KL and ISE$_C$ are presented in Table \ref{tab:sim-KL} and Table \ref{tab:sim-dMISE}, respectively.
It is observed that the results of KL are similar to those of MISE in Table \ref{tab:sim-MISE}.
Regarding the estimation accuracy of the curvature, we can see that either IHS or LHS attains the minimum values among the four methods, showing a benefit of introducing an additional smoothness parameter $w$. 
The performance differences can be observed between IHS and LHS in the finite sample, but the differences are smaller when $n=1000$ compared to the $n=500$ case since these two methods are asymptotically equivalent.

\begin{table}[htb!]
{\small 
\begin{centering}
\begin{tabular}{cccccccccccc}
\hline 
&&& \multicolumn{4}{c}{$n=500$} && \multicolumn{4}{c}{$n=1000$} \\
Scenario  & Quantiles &  & LHS & IHS & CV & PI &  & LHS & IHS & CV & PI \\
\hline
 & $25\%$ &  & 0.16 & 0.16 & 0.30 & 0.26 &  & 0.07 & 0.07 & 0.17 & 0.15 \\
(i) Bimodal & $50\%$ &  & 0.28 & 0.29 & 0.43 & 0.37 &  & 0.13 & 0.13 & 0.24 & 0.21 \\
 & $75\%$ &  & 0.49 & 0.51 & 0.66 & 0.50 &  & 0.23 & 0.24 & 0.35 & 0.28 \\
 \hline
 & $25\%$ &  & 0.87 & 1.17 & 0.62 & 0.60 &  & 0.94 & 1.10 & 0.66 & 0.65 \\
(ii) Trimodal & $50\%$ &  & 1.24 & 1.53 & 0.89 & 0.86 &  & 1.17 & 1.36 & 0.85 & 0.83 \\
 & $75\%$ &  & 1.62 & 1.95 & 1.26 & 1.21 &  & 1.45 & 1.64 & 1.06 & 1.04 \\
  \hline
 & $25\%$ &  & 1.70 & 1.28 & 5.41 & 0.94 &  & 1.15 & 0.80 & 4.42 & 0.55 \\
(iii) Claw & $50\%$ &  & 2.27 & 1.64 & 5.66 & 1.17 &  & 1.45 & 1.03 & 4.70 & 0.69 \\
 & $75\%$ &  & 3.20 & 2.11 & 6.02 & 1.44 &  & 1.88 & 1.33 & 4.98 & 0.84 \\
  \hline
 & $25\%$ &  & 0.80 & 0.84 & 0.72 & 0.39 &  & 0.62 & 0.59 & 0.41 & 0.23 \\
(iv) Skewed & $50\%$ &  & 1.18 & 1.23 & 1.10 & 0.54 &  & 0.94 & 0.90 & 0.64 & 0.30 \\
 & $75\%$ &  & 1.61 & 1.74 & 1.82 & 0.71 &  & 1.30 & 1.32 & 1.06 & 0.41 \\
  \hline
 & $25\%$ &  & 0.27 & 0.18 & 0.38 & 0.35 &  & 0.16 & 0.12 & 0.22 & 0.21 \\
(v) Outlier & $50\%$ &  & 0.48 & 0.32 & 0.57 & 0.59 &  & 0.27 & 0.19 & 0.33 & 0.33 \\
 & $75\%$ &  & 0.82 & 0.63 & 0.87 & 0.91 &  & 0.43 & 0.33 & 0.48 & 0.49 \\
\hline 
\end{tabular}
\par\end{centering}
\centering{}
\caption{$25\%$, $50\%$, and $75\%$ quantiles of 1000 Monte Carlo replications of ISE$_f$ under $n=500$ and $n=1000$.
The values are multiplied by 100. 
}
\label{tab:sim-MISE}
}
\end{table}

\begin{table}[htb!]
{\small 
\begin{centering}
\begin{tabular}{cccccccccccc}
\hline 
&&& \multicolumn{4}{c}{$n=500$} && \multicolumn{4}{c}{$n=1000$} \\
Scenario  & Quantiles &  & LHS & IHS & CV & PI &  & LHS & IHS & CV & PI \\
\hline
 & $25\%$ &  & 0.52 & 0.53 & 1.61 & 0.97 &  & 0.26 & 0.26 & 0.82 & 0.58 \\
(i) Bimodal & $50\%$ &  & 0.81 & 0.83 & 2.99 & 1.29 &  & 0.39 & 0.39 & 1.89 & 0.75 \\
 & $75\%$ &  & 1.31 & 1.34 & 5.80 & 1.65 &  & 0.61 & 0.62 & 3.92 & 0.96 \\
 \hline
 & $25\%$ &  & 3.31 & 4.34 & 5.99 & 2.39 &  & 3.51 & 4.19 & 5.23 & 2.62 \\
(ii) Trimodal & $50\%$ &  & 4.38 & 5.55 & 8.73 & 3.26 &  & 4.30 & 4.98 & 7.20 & 3.22 \\
 & $75\%$ &  & 5.67 & 6.96 & 13.42 & 4.39 &  & 5.26 & 6.01 & 10.06 & 3.93 \\
 \hline
 & $25\%$ &  & 3.54 & 4.78 & 25.53 & 2.96 &  & 3.00 & 3.68 & 19.22 & 1.93 \\
(iii) Claw & $50\%$ &  & 4.72 & 6.44 & 27.89 & 4.00 &  & 3.78 & 5.17 & 20.82 & 2.68 \\
 & $75\%$ &  & 7.08 & 8.65 & 30.52 & 7.32 &  & 4.78 & 6.61 & 22.57 & 4.46 \\
 \hline
 & $25\%$ &  & 2.85 & 3.12 & 3.94 & 1.39 &  & 2.26 & 2.08 & 2.68 & 0.84 \\
(iv) Skewed & $50\%$ &  & 4.11 & 4.33 & 5.98 & 1.73 &  & 3.25 & 3.08 & 3.91 & 1.03 \\
 & $75\%$ &  & 5.64 & 5.96 & 9.29 & 2.19 &  & 4.39 & 4.25 & 6.08 & 1.29 \\
 \hline
 & $25\%$ &  & 1.39 & 1.34 & 0.66 & 1.24 &  & 1.11 & 0.98 & 0.48 & 0.82 \\
(v) Outlier & $50\%$ &  & 2.12 & 1.90 & 1.50 & 1.76 &  & 1.69 & 1.36 & 1.03 & 1.11 \\
 & $75\%$ &  & 3.11 & 2.85 & 4.27 & 3.12 &  & 2.34 & 1.94 & 2.65 & 2.07 \\
\hline 
\end{tabular}
\par\end{centering}
\centering{}
\caption{$25\%$, $50\%$, and $75\%$ quantiles of 1000 Monte Carlo replications of KL under $n=500$ and $n=1000$.
The values are multiplied by 100. 
}
\label{tab:sim-KL}
}
\end{table}

\begin{table}[htb!]
{\small 
\begin{centering}
\begin{tabular}{cccccccccccc}
\hline 
&&& \multicolumn{4}{c}{$n=500$} && \multicolumn{4}{c}{$n=1000$} \\
Scenario  & Quantiles &  & LHS & IHS & CV & PI &  & LHS & IHS & CV & PI \\
\hline
 & $25\%$ &  & 0.09 & 0.09 & 0.34 & 0.38 &  & 0.05 & 0.04 & 0.25 & 0.27 \\
(i) Bimodal & $50\%$ &  & 0.16 & 0.17 & 0.47 & 0.55 &  & 0.08 & 0.08 & 0.35 & 0.40 \\
 & $75\%$ &  & 0.30 & 0.32 & 0.67 & 0.89 &  & 0.14 & 0.14 & 0.47 & 0.64 \\
 \hline 
 & $25\%$ &  & 1.25 & 1.56 & 0.81 & 1.01 &  & 1.33 & 1.59 & 0.99 & 1.18 \\
(ii) Trimodal & $50\%$ &  & 1.80 & 2.24 & 1.19 & 1.64 &  & 1.78 & 2.09 & 1.29 & 1.64 \\
 & $75\%$ &  & 2.63 & 3.58 & 1.79 & 2.80 &  & 2.33 & 2.81 & 1.67 & 2.30 \\
 \hline 
 & $25\%$ &  & 2.00 & 1.13 & 6.52 & 1.49 &  & 1.14 & 0.66 & 5.65 & 1.10 \\
(iii) Claw & $50\%$ &  & 2.95 & 1.61 & 6.65 & 1.90 &  & 1.75 & 0.91 & 5.93 & 1.37 \\
 & $75\%$ &  & 4.22 & 2.45 & 6.75 & 2.47 &  & 2.57 & 1.28 & 6.14 & 1.70 \\
 \hline 
 & $25\%$ &  & 0.14 & 0.14 & 0.21 & 0.15 &  & 0.10 & 0.10 & 0.14 & 0.11 \\
(iv) Skewed & $50\%$ &  & 0.20 & 0.21 & 0.28 & 0.21 &  & 0.15 & 0.15 & 0.19 & 0.15 \\
 & $75\%$ &  & 0.28 & 0.30 & 0.40 & 0.29 &  & 0.20 & 0.20 & 0.27 & 0.21 \\
 \hline 
 & $25\%$ &  & 0.07 & 0.04 & 0.20 & 0.17 &  & 0.04 & 0.03 & 0.15 & 0.12 \\
(v) Outlier & $50\%$ &  & 0.13 & 0.09 & 0.31 & 0.30 &  & 0.07 & 0.06 & 0.21 & 0.20 \\
 & $75\%$ &  & 0.21 & 0.22 & 0.45 & 0.60 &  & 0.12 & 0.12 & 0.29 & 0.38 \\
\hline 
\end{tabular}
\par\end{centering}
\centering{}
\caption{$25\%$, $50\%$, and $75\%$ quantiles of 1000 Monte Carlo replications of ISE$_C$ under $n=500$ and $n=1000$. 
The values under Scenarios (i) and (ii) are multiplied by 10.
}
\label{tab:sim-dMISE}
}
\end{table}

\begin{figure}[htb!]
\centering
\includegraphics[scale=0.35]{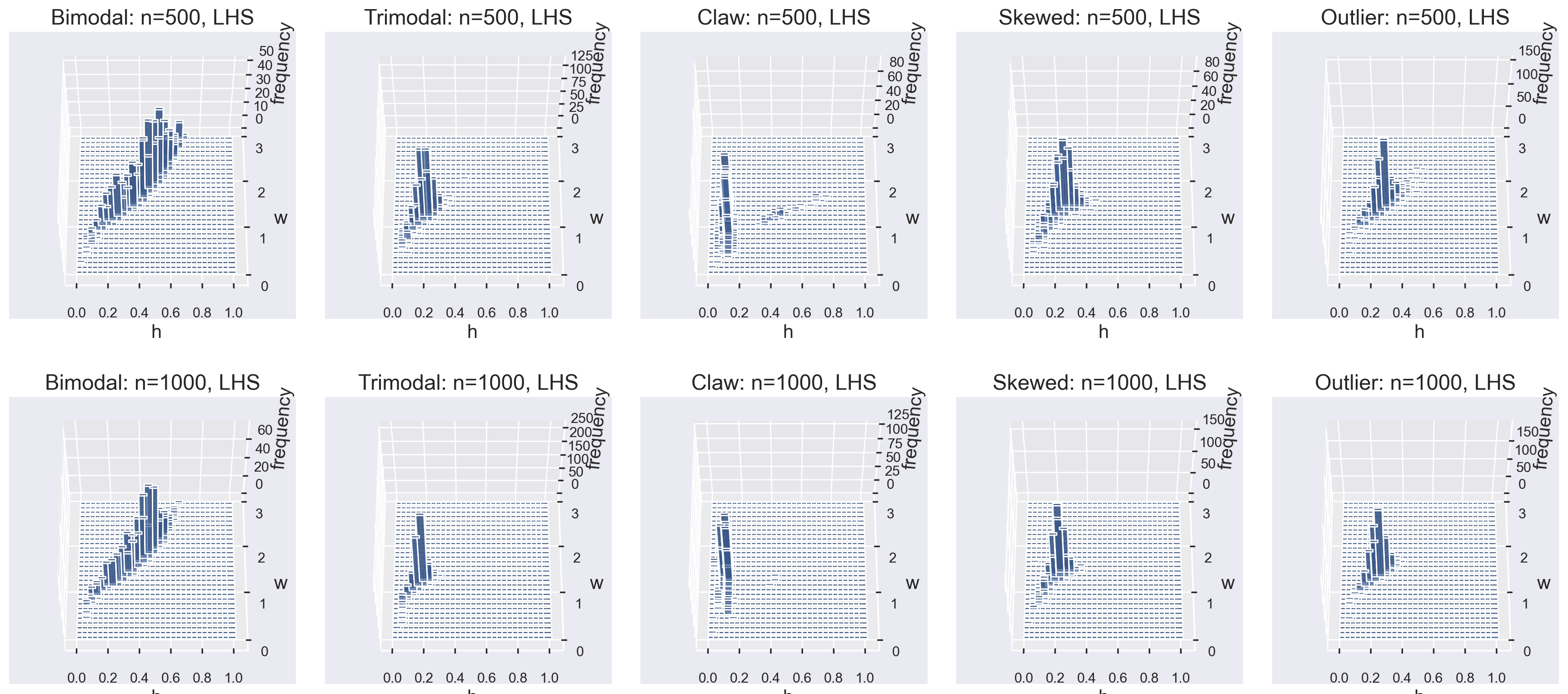}
\caption{Histograms of estimated parameters (LHS) based on 1000 Monte Carlo sample.}
\label{fig:para}
\end{figure}

To investigate the roles of the two hyperparameters $(w_n, h_n)$ in the exponentiated KDE (\ref{KDE2}), we present the two-dimensional histograms of the Monte Carlo replications of the estimated $(w_n, h_n)$ by LHS in Figure~\ref{fig:para}. 
First, the variability between Monte Carlo replications under $n=1000$ is found to be smaller than that under $n=500$, which is consistent with the convergence of the H-score. 
In addition, the optimal $(w_n, h_n)$ is significantly different for the five scenarios, indicating the flexibility of the proposed exponentiated KDE. 
In scenarios (iii) and (iv), the variability of the estimated $\hat{h}_n$ is considerably small, and the form of the final density estimate is dictated only by the additional parameter $w_n$. 
Furthermore, when a sample contains outliers in scenario (v), the optimal bandwidth in standard KDE tends to be unnecessarily small for capturing local fluctuations near outliers; this difficulty can  be addressed by incorporating $w_n$. 
Moreover, the histogram of $w_n$ under scenario (v) shows that the optimal value of $w_n$ in scenario (v) is smaller than those in other scenarios, leading to a smoother density estimate.   
A detailed comparison of optimal values is presented in Section~\ref{sec:supp-add-fig} and explained in Section~\ref{sec:MISE_relative} of Supplementary Material.
According to Figures~\ref{fig:comparison_h}  and \ref{fig:comparison_w}, LHS and IHS tend to provide similar optimal values for $(w_n, h_n)$.
In addition, Figure \ref{fig:comparison_four} shows that the uncertainties of parameters estimated by IHS and LHS are significantly larger than those estimated by PI and CV, whereas the numerical ISE of  IHS and LHS is smaller than those of PI and CV in some cases (Table \ref{tab:sim-MISE}) . 
This suggests the significant benefit of introducing $w_n$.

\subsection{Real data application}\label{sec:application}
The estimation of income distributions is particularly popular in labor market analysis (see \citealp{dinardo1995labor,hajargasht2012inference,kobayashi2022bayesian} for instance).
We apply the proposed methodology to the Current Population Survey (CPS) to estimate income distributions.
The CPS is a monthly survey of approximately $57,000$ households in the United States, carried out by the Bureau of Labor Statistics.
We used a dataset created by B. Hansen, named \verb+cps09mar+, which used certain variables from the CPS for March 2009.
The dataset and its detailed description can be downloaded from \url{https://www.ssc.wisc.edu/~bhansen/econometrics/}.
We applied the proposed method to sub-samples of females and males aged in their 20s or 50s.
The sample sizes of the four sub-samples are 3,426 for females (20s), 4,764 for females (50s), 4,540 for males (20s), and 6,079 for males (50s). 
To avoid the same observed values, we added a small noise following $N(0, (0.05)^2)$, which will be used in the subsequent analysis.

\begin{figure}[htb!]
\centering
\includegraphics[width=13cm,clip]{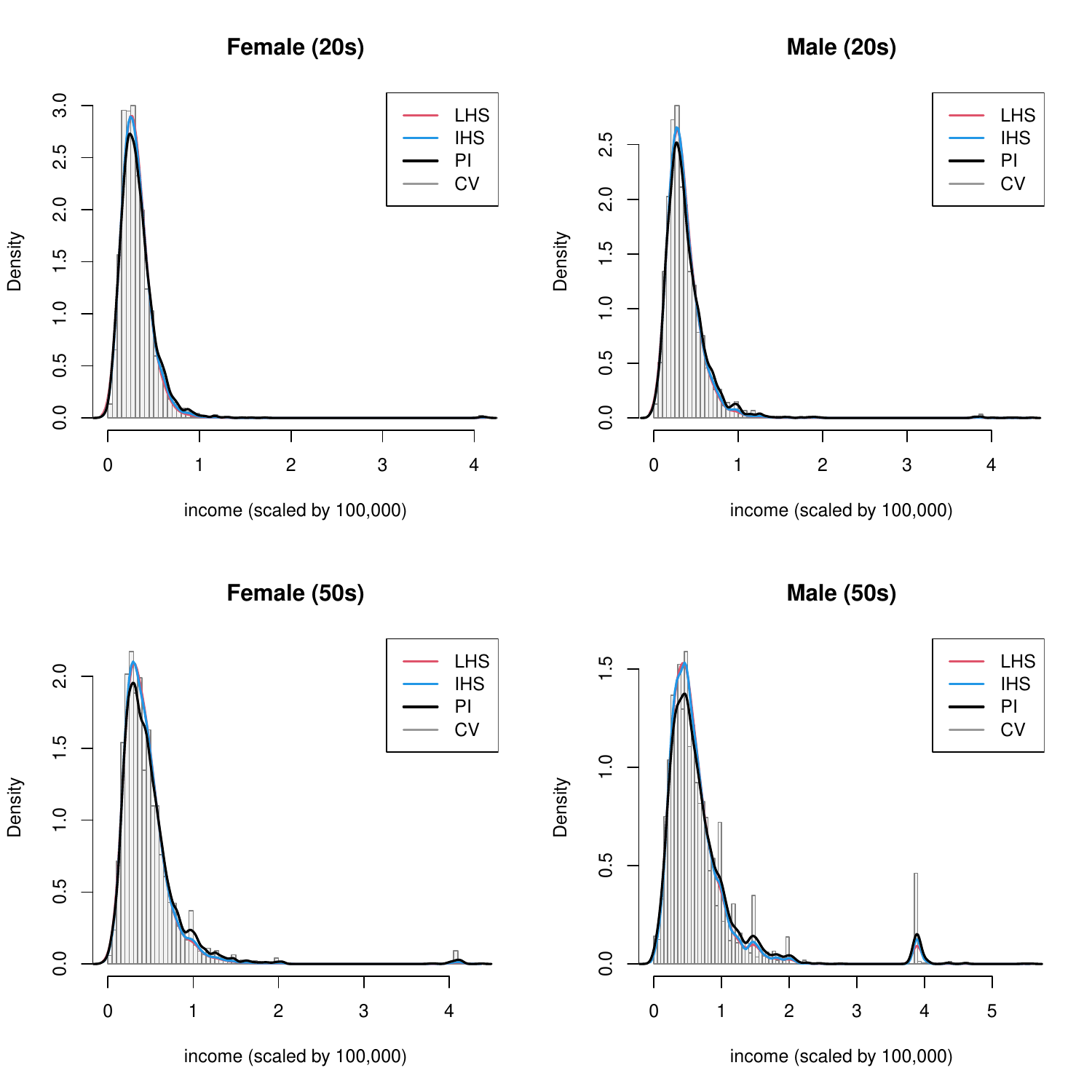}
\caption{Histograms of income data and estimated kernel density.}
\label{fig:income}
\end{figure}

For each sub-sample, we applied the proposed H-score-based KDE along with the standard KDE method used in Section~\ref{sec:sim}, where the minimum bandwidth of LHS and IHS is set in the same way as in Section~\ref{sec:sim}.
The estimated tuning parameters are listed in Table~\ref{tab:income}, and the estimated densities are presented in Figure~\ref{fig:income}. 
The bandwidth selected by PI and CV are similar, leading to almost the same density estimates.
On the other hand, the estimated bandwidth in LHS and IHS are larger than those of PI and CV, and the power parameter $w_n$ is larger than $1$.
Comparing the resulting density estimates, HS methods can capture the main body of the histogram while PI and CV methods underestimate the density around the mode, which can be attributed to outlying households with significantly high income. 
Hence, we can conclude that the proposed HS methods can flexibly capture the underlying density without being affected by outlying households owing to the additional flexibility of $w_n$ in the exponentiated KDE (\ref{KDE2}).

\begin{table}[htb!]
\begin{centering}
\begin{tabular}{cccccccccccc}
\hline 
& & \multicolumn{2}{c}{LHS} & & \multicolumn{2}{c}{IHS} && PI & & CV \\
 & & $h_n$ & $w_n$ && $h_n$ & $w_n$ &&$h_n$ && $h_n$ \\
\hline 
Female (20s) && $6.21\times 10^{-2}$ & $1.27$ &&$4.24\times 10^{-2}$ & $1.15$  && $2.83\times 10^{-2}$ && $3.12\times 10^{-2}$\\ 
Female (50s) && $6.15\times 10^{-2}$ & $1.20$ &&  $4.96\times 10^{-2}$ & $1.17$ && $2.80\times 10^{-2}$ && $2.76\times 10^{-2}$\\ 
Male (20s) && $6.14\times 10^{-2}$ & $1.23$ && $4.74\times 10^{-2}$  & $1.17$ &&$2.80\times 10^{-2}$ && $2.76\times 10^{-2}$\\ 
Male (50s) && $5.86\times 10^{-2}$ & $1.19$ &&$3.77\times 10^{-2}$ & $1.16$ && $4.33\times 10^{-2}$ && $4.20\times 10^{-2}$\\ 
\hline
\end{tabular}
\end{centering}
\centering{}
\caption{
The estimated tuning parameters in the application to the income data.  
}
\label{tab:income}
\end{table}

\section{Concluding Remarks} \label{sec:remark}
In this study, we investigated a new framework for data-driven density estimation with an intractable normalizing constant.
We estimated the tuning parameters by minimizing the Hyv\"arinen score, an empirical version of the Fisher divergence. 
We then scrutinized the theoretical properties of the proposed estimator, including the asymptotic role of tuning parameters and consistency of the proposed estimator. 
The main contribution of this study to the literature on kernel-smoothing-based density estimators is the characterization of the additional tuning parameter $w_n$.
Moreover, we interestingly discovered that the theoretically optimal $w_n$ is not $1$, implying that the accuracy of KDE can be improved by allowing $w_n \neq 1$. 

%
Nevertheless, some remaining theoretical issues must be addressed. 
First, although we demonstrated the consistency of selected parameters with the asymptotic Fisher divergence, the uncertainty of $(\hat{w}_n,\hat{h}_n)$ has not been fully discussed. 
Likewise \cite{hall1987cv}, it would be meaningful to evaluate the closeness between minimizers of the Hyv\"arinen score and exact Fisher divergence. 
Second, the extension to multivariate cases, and the investigation of the properties of $\hat{f}_{w_n,h_n}$ at the boundary point are valuable. 
In addition, consideration of conditional densities may lead to the discovery of properties that differ from the mere analogy of those of density estimation. 
Specifically, \cite{hall2004cross,hall2007nadarayacv} reported that the unbiased least-squares CV smoothes out irrelevant regressors. We suspect that the procedure of \cite{jewson2021general} does too.

%
Numerous studies have been conducted on the effect of bandwidth on the performance of kernel-smoothing-based estimators. 
However, several issues cannot be solved by tuning only a single bandwidth parameter.
A notable example is the estimation of contaminated density, as illustrated by the income distribution in Section~\ref{sec:application}, where the additional tuning parameter $w_n$ makes KDE robust against contamination. 
Hence, the proposed approach could also be useful for other kernel-smoothing methods.

Finally, while two methods (IHS and LHS) are asymptotically equivalent, the finite sample performance is not necessarily similar, as confirmed in our numerical studies. 
Some theoretical investigation regarding this issue is left to a future study.

\section*{Funding}
This study was supported by JSPS KAKENHI (grant no. 21H00699 and 23H00805).

\bibliographystyle{apalike}
\bibliography{references}

\newpage 
\appendix
\setcounter{page}{1}
\setcounter{equation}{0}
\renewcommand{\theequation}{S\arabic{equation}}
\setcounter{section}{0}
\renewcommand{\thelem}{S\arabic{lem}}
\setcounter{thm}{0}
\renewcommand{\thethm}{S\arabic{thm}}
\setcounter{prop}{0}
\renewcommand{\theprop}{S\arabic{prop}}
\setcounter{table}{0}
\renewcommand{\thesection}{S\arabic{section}}
\setcounter{table}{0}
\renewcommand{\thetable}{S\arabic{table}}
\setcounter{figure}{0}
\renewcommand{\thefigure}{S\arabic{figure}}
\setcounter{rem}{0}
\renewcommand{\therem}{S\arabic{rem}}
\fontsize{11pt}{11pt}\selectfont

\begin{center}
{\Large\bf Supplemental Materials for ``Fully Data-driven Normalized and Exponentiated Kernel Density Estimator with Hyv\"arinen Score"}
\end{center}


{\allowdisplaybreaks

\section{Notations}
In this section, we provide the list and notes of the notations used in the proof.

\begin{itemize}
    \item Expectation operator
    \begin{itemize}
        \item $\mathbb{E}_{X_{1:n}}$ : the expectation operator over the whole observation.
        \item $\mathbb{E}_{X_i}$ : the expectation operator over only $i$-th observation.
        \item $\mathbb{E}_{x}[g(x)] = \int_{\mathcal{X}} g(x)f(x)dx$
    \end{itemize}
    \item To simplify notation, we write
    \begin{align*}
        K_i(x) \equiv K\left( \frac{X_i - x}{h_n}\right), \quad K_{ij} \equiv K\left( \frac{X_i - X_j}{h_n}\right).
    \end{align*}
    \item  Constants associated with the kernel function
    \begin{align*}
        \kappa_l \equiv \frac{1}{l!} \int u^lK(u)du , \quad R(K') \equiv \int K'(u)^2du
    \end{align*}
    \item  Constants in the leading terms of Expected Fisher-divergence
    \begin{align*}
        & C_{(B,2L)} \equiv \kappa_L^2  \mathbb{E}_x\Bigl[\{f^{(L+1)}(x)f(x)^{-1}-f'(x)f^{(L)}(x)f(x)^{-2}\}^2\Bigl] \\
        & C_{(B,1)} \equiv \mathbb{E}_x[f'(x)^2f(x)^{-2}] \\
        & C_{(B,L)} \equiv 2\kappa_L \mathbb{E}_x\Bigl[f'(x)f^{(L+1)}(x)f(x)^{-2}-f'(x)^2f^{(L)}(x)f(x)^{-3}\Bigl] \\
        & C_V \equiv R(K')\mathbb{E}_x[f(x)^{-1}] \\
        & C'_{(B,l)} \equiv 2\kappa_l\mathbb{E}_x[f'(x)f^{(l+1)}(x)f(x)^{-2}-f'(x)^2f^{(l)}(x)f(x)^{-3}], \quad \text{for } L\le l \le 2L  \\
        & C'_{(B,2L)} \equiv \kappa_L^2\mathbb{E}_x[f'(x)^2f^{(L)}(x)^2f(x)^{-4}-f'(x)f^{(L)}(x)f^{(L+1)}(x)f(x)^{-3}]
    \end{align*}
    \item  $\mathcal{H}_i(w,h)= \mathcal{H}_{1i}(w,h) + \mathcal{H}_{2i}(w,h)$ : individual Hyv\"arinen Score for $i$-th observation, which is explicitly defined in Section \ref{section:derivation_of_hyvarinen_score}.
\end{itemize}

{\textbf{About the dependence of parameters on $\bf{n}$}:} If $w$ and $h$ depend on $n$, we write them as $w_n$ and $h_n$ respectively to clarify the dependence. When we suppose both the  cases where the parameters depend on $n$ and the cases where they do not in general, we do not give the subscript $n$.

\section{Additional Theoretical Results (MISE)}
In this section, we investigate the asymptotic MISE (Mean Integrated Squared Error) of $\hat{f}_{w,h}(x)$.

\subsection{MISE} \label{sec:MISE}
\begin{thm} \label{thm:MISE}
Under Assumption \ref{ass:DGP}, \ref{ass:interiorpoint}, \ref{ass:fdiffrentiability1}, \ref{ass:Kernel}, \ref{ass:model},
\begin{align*}
    & AMISE \left[ \hat{f}_{w_n, h_n}(x) \right] = M_{(B,2L)}h_n^{2L} +  (w_n-1)M_{(B,L)}h_n^L + (w_n-1)^2M_{(B,1)}  + \frac{M_V}{nh_n},
\end{align*}
with
\begin{align*}
    & M_{(B,2L)} \equiv \kappa_L^2 \int f^{(L)}(x)^2 dx,  \\
    & M_{(B,L)} \equiv 2\kappa_L \int f^{(L)}(x)\left( f(x)\log f(x) - f(x) \int f(x)\log f(x)dx  \right) dx,\\
    & M_{(B,1)} \equiv \int \left( f(x)\log f(x) - f(x) \int f(x)\log f(x)dx  \right)^2 dx,\\
    & M_V \equiv R(K),
\end{align*}
where $R(K) \equiv \int K(u)^2du$. See Section \ref{sec:proof_MISE} for the proof.
\end{thm}

\begin{table}[htb]
    \centering
    \begin{spacing}{1.3}
    \begin{tabular}{c|ccc|c}
    \hline\hline
        DGP & $\int f''(x)f(x)dx$ & $\int f(x)\log f(x)dx$ &  $\int f''(x)f(x)\log f(x)dx$ & $M_{(B,L)}$\\
        \hline\hline
        Bimodal & -0.563 & -1.415 & 0.373 & $<0$\\
        \hline
        Trimodal & -0.931 & -1.321 & 0.481 & $<0$ \\
        \hline 
        Claw & -6.929 & -1.193 & 0.107 & $<0$ \\
        \hline 
        Skewed & -1.038 & -1.260 & 0.120& $<0$ \\
        \hline
        Outline & -3.677 & -0.487 & -1.641 & $<0$ \\
        \hline
    \end{tabular}
    \end{spacing}
    \caption{Constants for $M_{(B,L)}$ and the sign of $M_{(B,L)}$}
    \label{tab:my_label}
\end{table}

\subsection{MISE Optimal Parameters} \label{sec:MISE_opt}
We denote the MISE optimal parameters as $w_n^0$ and $h_n^0$. Minimizing the leading terms in Theorem \ref{thm:MISE} with respect to $w_n$ gives the MISE optimal $w_n$.
\begin{align*}
        & w_n^0  -  1 = \frac{-M_{(B,L)}}{2 M_{(B,1)}}({h_n^0})^L
    \end{align*}
Using this results, we have the MISE optimal parameters as $n$-dependent sequences.
    \begin{align*}
        & h_n^0 = \left(  \frac{4M_{(B,1)}M_V}{2L[4M_{(B,1)}M_{(B,2L)}  - M_{(B,L)}^2]} \right)^{1/(2L+1)} n^{-1/(2L+1)}, \\
        & w_n^0  -  1 = \frac{-M_{(B,L)}}{2 M_{(B,1)}}\left(  \frac{4M_{(B,1)}M_V}{2L[4M_{(B,1)}M_{(B,2L)}  - M_{(B,L)}^2]} \right)^{L/(2L+1)} n^{-L/(2L+1)}.
    \end{align*}

\begin{rem}
    Since $4M_{(B,1)}M_{(B,2L)} - M_{(B,L)}^2 \ge 0$ from the Cauchy-Schwarz inequality, $M_{(B,1)} \ge 0$ by definition, and $M_{(B,L)}<0$ for the DGPs in our numerical studies. These imply that MISE optimal $w_n^0$ is larger than $1$ for the five DGPs.
\end{rem}

\begin{rem}
    The MISE optimal bandwidth converges to $0$  at the rate of  $h_n^{-1/5}$, so the estimated bandwidth via H-score leads to oversmoothing for MISE. 
    However, MISE optimal $w_n$ approaches to $1$ at the rate of $n^{-2/5}$ and is faster than that of Fisher-divergence optimal $w_n^*$ ($w_n^* -1 \propto n^{-2/7}$), so Fisher-divergence optimal $w_n^*$ in turn leads to undersmoothing for MISE.
    Combining these results on $h_n$ and $w_n$, although $h_n^*$ certainly leads to oversmoothing, $w_n^*$ compensate the lack of smoothness.
\end{rem}

\subsection{Relative asymptotic MISE between with and without $w_n$} \label{sec:MISE_relative}

In this section, we consider the benefit from introducing $w_n$ based on the asymptotic MISE. Table \ref{tab:AMISE_500} and \ref{tab:AMISE_1000} describe the asymptotic bias, asymptotic variance, asymptotic MISE and asymptotic relative MISE of $\hat{f}_{h_n}(x)$ and $\hat{f}_{w_n,h_n}(x)$ for each DGP in Section \ref{sec:numerical}. These quantities of $\hat{f}_{h_n}(x)$ are calculated with the MISE optimal bandwidth for the standard KDE (See e.g. \citealp[pp.22]{wand1996smoothing}, \citealp[pp.134]{wasserman2006all}, \citealp[pp.8]{tsybakov2009nonpara}) and those of $\hat{f}_{w_n,h_n}(x)$ are with $(w_0,h_0)$.

In view of Table \ref{tab:AMISE_500} and \ref{tab:AMISE_1000}, consistently with the intuition, the asymptotic bias is reduced by introducing $w_n>1$. 
Because of this asymptotic bias reduction and the fact that $w_n$ has no influence on the asymptotic variance as shown in Theorem \ref{thm:MISE}, the bandwidth $h_0$ is larger than that without $w_n$, so the asymptotic variance is also smaller. 
As a result, the allowing $w_n \neq 1$ improves the asymptotic MISE for all DGPs. 

\subsubsection{Additional Discussion on the Simulation Results}
Based on Table \ref{tab:AMISE_500} and \ref{tab:AMISE_1000}, we provide additional explanation on the numerical studies.
We also refer Figure \ref{fig:comparison_h} and \ref{fig:comparison_w} to compare the parameters selected by IHS and those by LHS and Figure \ref{fig:comparison_four} to compare the estimated bandwidth by IHS, LHS, CV and PI.

Numerical performance of $\hat{f}_{w_n,h_n}$ would depend on (1) how much benefit there is in introducing $w_n$, (2) whether the Fisher-divergence optimal parameters $(w_n^*,h_n^*)$ are close to the MISE optimal parameters $(w_n^0,h_n^0)$, and (3) how much do the estimated parameters vary.
Relative MISE in Table \ref{tab:AMISE_500} and \ref{tab:AMISE_1000} reveals the first element. 
The second and third factors are not theoretically clarified in this paper but Figure \ref{fig:comparison_four} numerically compares the third factor with that of PI and CV for the standard KDE.
With this in mind, we offer some discussion on the results for each DGP.

In scenario (i) Bimodal, Figure \ref{fig:comparison_h} and \ref{fig:comparison_w} show that there is no significant difference in the distributions of both estimated bandwidth $\hat{h}_n$ and learning rate $\hat{w}_n$ between IHS and LHS. 
Accordingly, the $\mathrm{ISE}_f$ are almost identical  between IHS and LHS.
On the other hand,  according  to Table \ref{tab:sim-MISE}, IHS and LHS  performs better than the ISE-oriented fully data-driven method (CV) in the ISE.  
Not only that, H-score based $\hat{f}_{w,h}$ also outperforms PI.
More notably, Figure \ref{fig:comparison_four} shows that the parameters selected by IHS and LHS are considerably more varied than those by PI and CV. 
Nevertheless, the ISE of $\hat{f}_{w_n,h_n}$ by HS-based method listed in Table \ref{tab:sim-MISE} is smaller than those of $\hat{f}_{h_n}$ by PI and CV.
These would be because, in view of Table \ref{tab:AMISE_500} and \ref{tab:AMISE_1000},  the benefit of introducing $w_n$ is large.
For the same reason, IHS and LHS perform better than CV and PI in scenario (v).

In other scenario, $\hat{f}_{w_n,h_n}$ tuned by IHS and LHS are inferior to $\hat{f}_{h_n}$ tuned by PI or to by both PI and CV.
The reason would be that (1) the advantage of introducing $w_n$ is not that great for these DGPs in view of Table \ref{tab:AMISE_500} and \ref{tab:AMISE_1000}, and that (2) CV and PI are designed to minimize MISE, while LHS and IHS are not as well as (3) the variability of estimated  parameters (Figure \ref{fig:comparison_four}).
Among them, scenario (iii) Claw is the remarkable case where IHS and LHS differ in MISE
The LHS estimates $w_n$ smaller than the IHS. 
To compensate for the lack of smoothness, $h_n$ is also estimated to be smaller in LHS, but the frequency of $\hat{w}_n < 1$ is quite high especially when $n=500$.  
Considering that the asymptotic MISE optimal $w_n^0$ is greater than 1, it is consistent with the theory that LHS perform worse than IHS.

\begin{table}[htb!] 
\begin{spacing}{1.5}
    \centering
    \begin{tabular}{c|c|ccc|c} 
    \hline\hline
        DGP & Estimator & Bias & Variance & MISE & Relative MISE\\
        \hline\hline
        \multirow{2}{*}{(i) Bimodal} & $\hat{f}_{h_n}$ & $8.05 \times 10^{-4}$ & $3.22 \times 10^{-3}$  & $4.02 \times 10^{-3}$ & \multirow{2}{*}{$0.197$}\\ \cline{2-5}
        & $\hat{f}_{w_n,h_n}$ & $1.59\times 10^{-4}$ & $6.34\times 10^{-4}$ & $7.93 \times 10^{-4}$ \\
        \hline
        \multirow{2}{*}{(ii) Trimodal} & $\hat{f}_{h_n}$ & $1.10 \times 10^{-3}$ & $4.40 \times 10^{-3}$  & $5.50 \times 10^{-3}$ & \multirow{2}{*}{$0.842$}\\ \cline{2-5}
        & $\hat{f}_{w_n,h_n}$ & $9.27\times 10^{-4}$ & $3.71 \times 10^{-3}$ & $4.64 \times 10^{-3}$ \\
        \hline
        \multirow{2}{*}{(iii) Claw} & $\hat{f}_{h_n}$ & $2.58 \times 10^{-3}$ & $1.03 \times 10^{-2}$  & $1.29 \times 10^{-2}$ & \multirow{2}{*}{$0.827$}\\ \cline{2-5}
        & $\hat{f}_{w_n,h_n}$ & $2.13\times 10^{-3}$ & $8.53\times 10^{-3}$ & $1.07 \times 10^{-2}$ \\
        \hline
        \multirow{2}{*}{(iv) Skewed} & $\hat{f}_{h_n}$ & $1.22 \times 10^{-3}$ & $4.86 \times 10^{-3}$  & $6.08 \times 10^{-3}$ & \multirow{2}{*}{$0.910$}\\ \cline{2-5}
        & $\hat{f}_{w_n,h_n}$ & $1.11\times 10^{-3}$ & $4.42\times 10^{-3}$ & $5.53 \times 10^{-3}$ \\
        \hline
        \multirow{2}{*}{(v) Outlier} & $\hat{f}_{h_n}$ & $1.40 \times 10^{-3}$ & $5.60 \times 10^{-3}$  & $7.00 \times 10^{-3}$ & \multirow{2}{*}{$0.598$}\\ \cline{2-5}
        & $\hat{f}_{w_n,h_n}$ & $8.37\times 10^{-4}$ & $3.35\times 10^{-3}$ & $4.18 \times 10^{-3}$ \\
        \hline
    \end{tabular}
\end{spacing}
    \caption{Asymptotic-Bias, Variance, MISE and Relative MISE ($n=500$). Asymptotic relative MISE is defined as $AMISE(\hat{f}_{w_n,h_n}) / AMISE(\hat{f}_{h_n})$. These values of $\hat{f}_{h_n}$ are calculated with usual MISE optimal bandwidth and those of $\hat{f}_{w_n,h_n}$ are with $w_n^0$ and $h_n^0$.}
    \label{tab:AMISE_500}
\end{table}

\begin{table}[htb!] 
\begin{spacing}{1.5}
    \centering
    \begin{tabular}{c|c|ccc|c}
    \hline\hline
        DGP & Estimator & Bias & Variance & MISE & Relative MISE\\
        \hline\hline
        \multirow{2}{*}{(i) Bimodal} & $\hat{f}_{h_n}$ & $4.62 \times 10^{-4}$ & $1.85 \times 10^{-3}$  & $2.31 \times 10^{-3}$ & \multirow{2}{*}{$0.197$}\\ \cline{2-5}
        & $\hat{f}_{w_n,h_n}$ & $9.11\times 10^{-5}$ & $3.64\times 10^{-4}$ & $4.55 \times 10^{-4}$ \\
        \hline
        \multirow{2}{*}{(ii) Trimodal} & $\hat{f}_{h_n}$ & $6.32 \times 10^{-4}$ & $2.53 \times 10^{-3}$  & $3.16 \times 10^{-3}$ & \multirow{2}{*}{$0.842$}\\ \cline{2-5}
        & $\hat{f}_{w_n,h_n}$ & $5.32\times 10^{-5}$ & $2.13 \times 10^{-3}$ & $2.66 \times 10^{-3}$ \\
        \hline
        \multirow{2}{*}{(iii) Claw} & $\hat{f}_{h_n}$ & $1.48 \times 10^{-3}$ & $5.92 \times 10^{-3}$  & $7.40 \times 10^{-3}$ & \multirow{2}{*}{$0.827$}\\ \cline{2-5}
        & $\hat{f}_{w_n,h_n}$ & $1.22\times 10^{-3}$ & $4.90\times 10^{-3}$ & $6.12 \times 10^{-3}$ \\
        \hline
        \multirow{2}{*}{(iv) Skewed} & $\hat{f}_{h_n}$ & $6.98 \times 10^{-4}$ & $2.79 \times 10^{-3}$  & $3.49 \times 10^{-3}$ & \multirow{2}{*}{$0.910$}\\ \cline{2-5}
        & $\hat{f}_{w_n,h_n}$ & $6.35\times 10^{-4}$ & $2.54\times 10^{-3}$ & $3.18 \times 10^{-3}$ \\
        \hline
        \multirow{2}{*}{(v) Outlier} & $\hat{f}_{h_n}$ & $8.04 \times 10^{-4}$ & $3.22 \times 10^{-3}$  & $4.02 \times 10^{-3}$ & \multirow{2}{*}{$0.598$}\\ \cline{2-5}
        & $\hat{f}_{w_n,h_n}$ & $4.81\times 10^{-4}$ & $1.92\times 10^{-3}$ & $2.40 \times 10^{-3}$ \\
        \hline
    \end{tabular}
\end{spacing}
    \caption{Asymptotic-Bias, Variance, MISE and Relative MISE ($n=1000$). These values of $\hat{f}_{h_n}$ are calculated with usual MISE optimal bandwidth and those of $\hat{f}_{w_n,h_n}$ are with $w_n^0$ and $h_n^0$.}
    \label{tab:AMISE_1000}
\end{table}

\section{Additional Figures of Simulation Results}\label{sec:supp-add-fig}
\begin{figure}[H]
    \centering
    \includegraphics[scale=0.35]{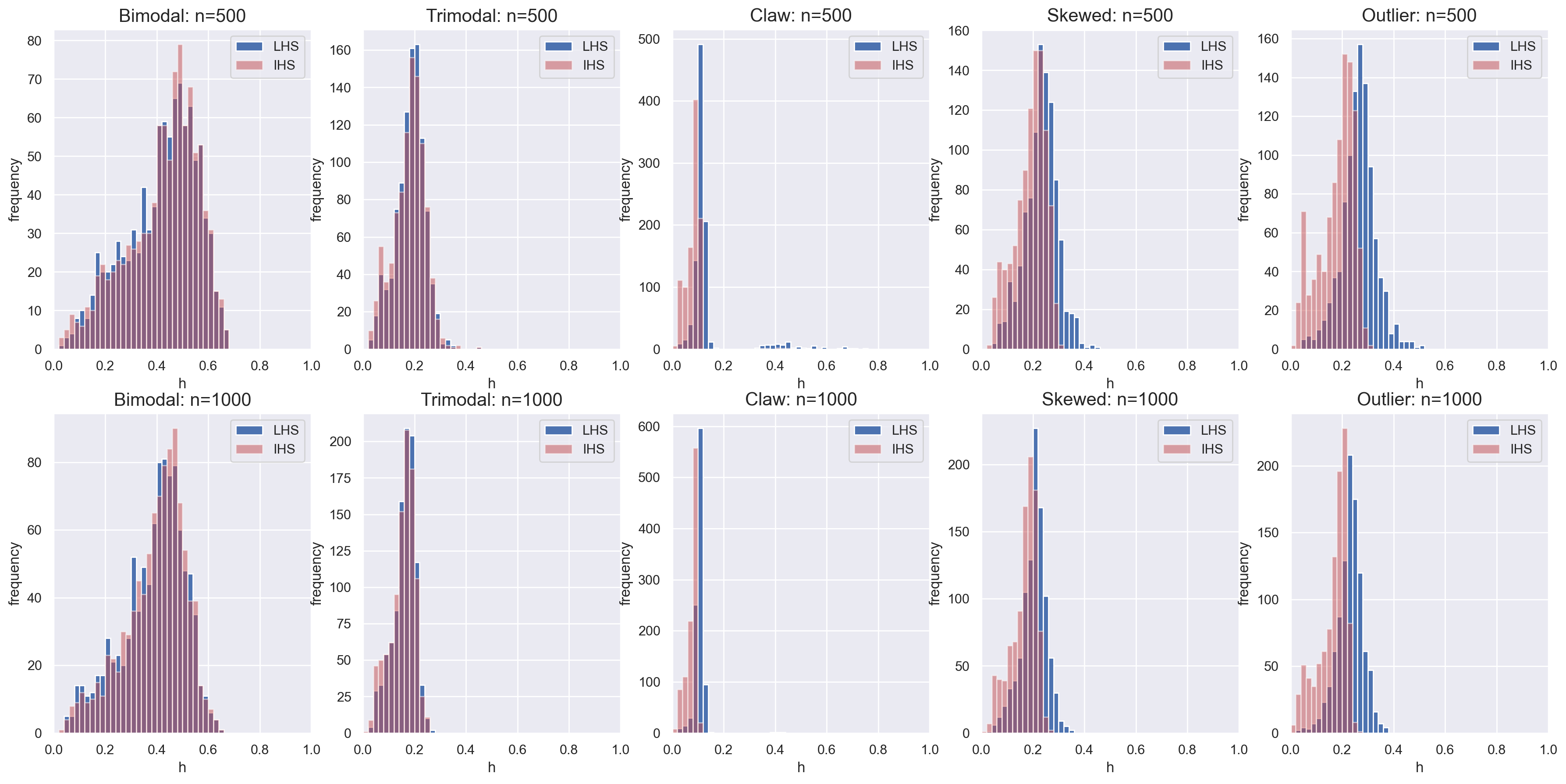}
    \caption{Histograms of estimated bandwidths $\hat{h}_n$ by IHS and LHS based on 1000 Monte Carlo sample under $n=500$ and $n=1000$.}
    \label{fig:comparison_h}
\end{figure}

\begin{figure}[H]
    \centering
    \includegraphics[scale=0.35]{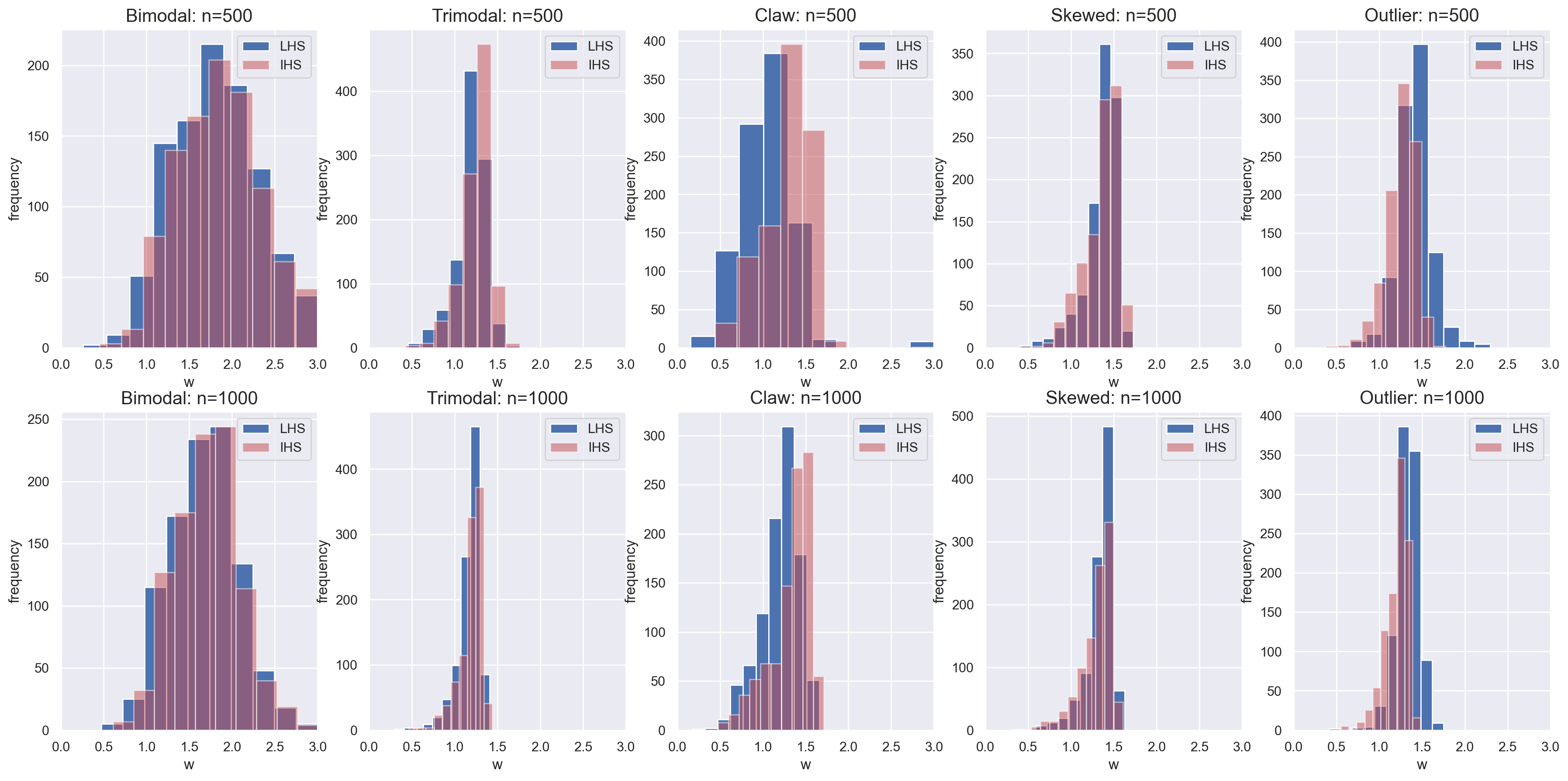}
    \caption{Histograms of estimated learning rates $\hat{w}_n$ by IHS and LHS based on 1000 Monte Carlo sample under $n=500$ and $n=1000$.}
    \label{fig:comparison_w}
\end{figure}

\begin{figure}[H]
    \centering
    \includegraphics[scale=0.35]{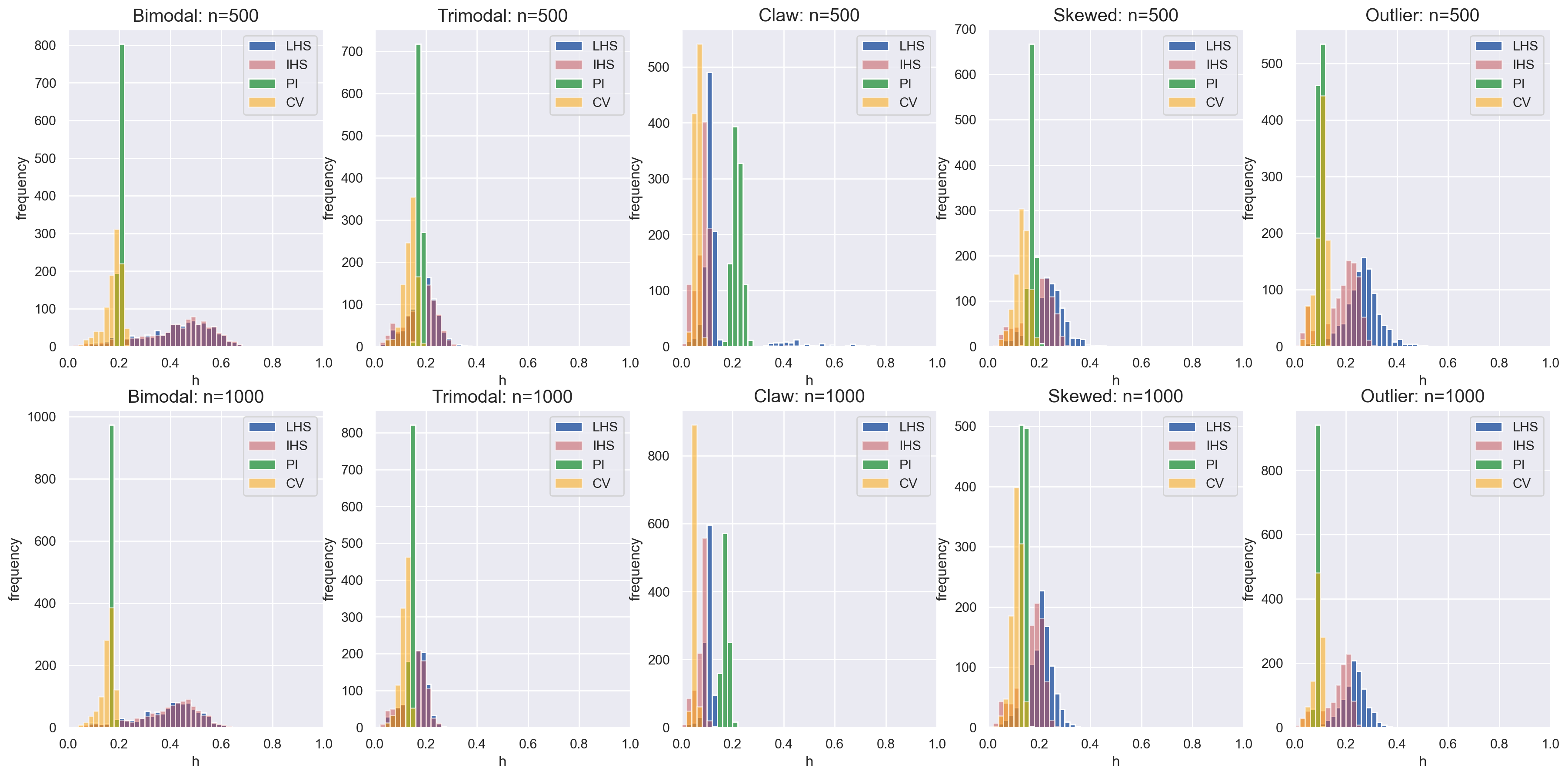}
    \caption{Histograms of estimated bandwidths $\hat{h}_n$ by IHS, LHS, PI and CV based on 1000 Monte Carlo sample under $n=500$ and $n=1000$.}
    \label{fig:comparison_four}
\end{figure}

\section{Outline of the Proofs}
We provide the outline of the proofs in this section and technical details in separated sections, because the proofs are long and tedious rather than difficult.

\subsection{Outline of the Proofs of Theorem \ref{thm:ej} and \ref{thm:ej_fix}} \label{section:outline_ej}
In this subsection, we provide the outline of the proofs for Theorem \ref{thm:ej} and \ref{thm:ej_fix}. We expand the expected Fisher divergence up to higher-order than necessary to prove Theorem \ref{thm:ej}, because Theorem \ref{thm:ej_fix} needs the expansion.

First, we derive the bias like terms of the expected Fisher divergence. From the result of computation in \ref{section:bias}, under Assumption \ref{ass:nhinfty} (Note that $w$ is fixed.), the bias like term is given by
\begin{align}
    & \mathbb{E}_{X_{1:n}}\left[\frac{\partial}{\partial x}\log\hat{f}_{w,h_n}(x)\right] - \frac{\partial}{\partial x}\log f(x) \nonumber\\
    & \quad=  (w-1)f'(x)f(x)^{-1} + w\sum_{l=L}^{2L}\kappa_l[f^{(l+1)}(x)f(x)^{-1}-f'(x)f^{(l)}(x)f(x)^{-2}]h_n^l, \nonumber\\
    &\qquad + w\kappa_L^2[f'(x)f^{(L)}(x)^2f(x)^{-3}-f^{(L)}(x)f^{(L+1)}(x)f(x)^{-2}]h_n^{2L}  + O(n^{-1}h_n^{-2}) + o(h_n^{2L}), \nonumber 
\end{align}
which implies 
\begin{align}
    & \left(\mathbb{E}_{X_{1:n}}\left[\frac{\partial}{\partial x}\log\hat{f}_{w,h_n}(x)\right] - \frac{\partial}{\partial x}\log f(x)\right)^2  \nonumber\\
    & \quad = (w-1)^2f'(x)^2f(x)^{-2} \nonumber\\
    & \quad\quad + 2w(w-1)\sum_{l=L}^{2L}\kappa_l [f'(x)f^{(l+1)}(x)f(x)^{-2}-f'(x)^2f^{(l)}(x)f(x)^{-3}]h_n^l \nonumber\\
    & \quad\quad + 2w(w-1)\kappa_L^2 [f'(x)^2f^{(L)}(x)^2f(x)^{-4}-f'(x)f^{(L)}(x)f^{(L+1)}(x)f(x)^{-3}]h_n^{2L}  \nonumber\\
    & \quad\quad + w^2\kappa_L^2[f^{(L+1)}(x)f(x)^{-1}-f'(x)f^{(L)}(x)f(x)^{-2}]^2h_n^{2L} + o\left\{\frac{1}{nh_n^3}+h_n^{2L}\right\} \label{eq:bias_fix}.
\end{align}
In addition, under Assumption \ref{ass:nhinfty} and \ref{ass:wlocalise}, since $(w_n - 1) \to 0$ as $n \to \infty$, 
\begin{align}
    & \left(\mathbb{E}_{X_{1:n}}\left[\frac{\partial}{\partial x}\log\hat{f}_{w_n,h_n}(x)\right] - \frac{\partial}{\partial x}\log f(x)\right)^2  \nonumber\\
    & \quad= (w_n-1)^2f'(x)^2f(x)^{-2} \nonumber\\
    & \quad\quad + 2w_n(w_n-1)\kappa_L [f'(x)f^{(L+1)}(x)f(x)^{-2}-f'(x)^2f^{(L)}(x)f(x)^{-3}]h_n^L \nonumber\\
    & \quad\quad + \{(w_n-1)^2+2(w-1)+1\}\kappa_L^2[f^{(L+1)}(x)f(x)^{-1}-f'(x)f^{(L)}(x)f(x)^{-2}]^2h_n^{2L} \nonumber\\
    & \quad\quad + o\left\{\frac{1}{nh_n^3}+(w_n-1)^2+(w_n-1)h_n^L+h_n^{2L}\right\},\nonumber \\
    & \quad= (w_n-1)^2f'(x)^2f(x)^{-2} \nonumber\\
    & \quad\quad+ 2w_n(w_n-1) \kappa_L [f'(x)f^{(L+1)}(x)f(x)^{-2} - f'(x)^2f^{(L)}(x)f(x)^{-3}] h_n^L \nonumber\\
    & \quad\quad + \kappa_L^2 [f^{(L+1)}(x)f(x)^{-1} - f'(x)f^{(L)}(x)f(x)^{-2}]^2 h_n^{2L} \nonumber\\
    & \quad\quad + o\left\{\frac{1}{nh_n^3}+(w_n-1)^2+(w_n-1)h_n^L+h_n^{2L}\right\}. \label{equation:squared_bias}
\end{align}

Next, we derive the variance like term. From the result of computation in \ref{section:variance}, under Assumption \ref{ass:nhinfty}, the variance like term is 
\begin{align}
     & \mathbb{V}_{X_{1:n}}\left[\frac{\partial}{\partial x}\log\hat{f}_{w,h_n}(x)\right] = w^2\frac{R(K')}{nh_n^3}f(x)^{-1} + o\left\{\frac{1}{nh_n^3}+h_n^{2L}\right\}. \label{eq:variance_fix}
\end{align}
Under Assumption \ref{ass:wlocalise}, since $(w_n - 1) \to 0$ as $n \to \infty$, 
\begin{align}
    \mathbb{V}_{X_{1:n}}\left[\frac{\partial}{\partial x}\log\hat{f}_{w_n,h_n}(x)\right]
     & \quad =  \{(w_n-1)^2+2(w_n-1)+1\}\frac{R(K')}{nh_n^3}f(x)^{-1} + o\left\{\frac{1}{nh^3}+h_n^{2L}\right\} \nonumber\\
     & \quad = \frac{R(K')}{nh_n^3}f(x)^{-1} + o\left\{\frac{1}{nh_n^3} +( w_n-1)^2 + h_n^{2L}\right\}. \label{equation:variance}
\end{align}
From (\ref{equation:squared_bias}) and (\ref{equation:variance}), Theorem \ref{thm:ej} holds. Similarly, from (\ref{eq:bias_fix}) and (\ref{eq:variance_fix}), Theorem \ref{thm:ej_fix} holds.

\subsection{Proof of Corollary \ref{cor:opt_h}} \label{sec:proof_opt_h}
\begin{proof}
   Since optimal $w_n^*$ is $w_n^* = 1 - \frac{C_{(B,L)}}{2C_{(B,1)}}{h_n^*}^L$, the leading terms of the expectation of $\mathbb{E}_{X_{1:n}}[J(f||\hat{f}_{w_n^*,h_n^*})]$ turns out to be
    \begin{align}
        \mathbb{E}_{X_{1:n}}[J_x(f||\hat{f}_{w_n^*,h_n^*})]
        & \approx \left\{C_{(B,2L)}-\frac{C_{(B,L)}^2}{4C_{(B,1)}}\right\}{h_n^*}^{2L} + \frac{C_{V}}{n{h_n^*}^3} . \nonumber
    \end{align}
    Since $C_V$ and (as shown later) the quantity in the bracket is positive, we can immediately deduce the explicit form of optimal bandwidth.
    
    Next, we show that $C_{(B,2L)}-\frac{C_{(B,L)}^2}{4C_{(B,1)}}$ is non-negative. Recalling the definition of $C_{(B,2L)}, C_{(B,L)}$ and $C_{(B,1)}$, and letting
    \begin{align}
        & A_i = f'(X_i)f(X_i)^{-1} , \quad B_i = f^{(L+1)}(X_i)f(X_i)^{-1} - f'(X_i)f^{(L)}(X_i)f(X_i)^{-2} \nonumber
    \end{align}
    then we can see that
    \begin{align}
        & C_{(B,2L)} = \kappa_L^2\mathbb{E}_{X_i}[B_i^2], \quad C_{(B,L)} = 2\kappa_L\mathbb{E}_{X_i}[A_iB_i], \quad C_{(B,1)} = \mathbb{E}_{X_i}[A_i^2], \label{eq:rep_cs}
    \end{align}
    so
    \begin{align*}
        4C_{(B,1)}C_{(B,2L)}-C_{(B,L)}^2 &= 4\kappa_L^2\mathbb{E}_{X_i}[A_i^2]\mathbb{E}_{X_i}[B_i^2] - 4\kappa_L^2\mathbb{E}_{X_i}[A_iB_i]^2 \\
        & = 4\kappa_L^2\left(\mathbb{E}_{X_i}[A_i^2]\mathbb{E}_{X_i}[B_i^2] - \mathbb{E}_{X_i}[A_iB_i]^2\right) \ge 0,
    \end{align*}
    from Cauchy-Schwarz inequality. Assumption \ref{ass:csequality} exclude the cases where the equality holds.
\end{proof}

\subsection{Proof of Theorem \ref{thm:convex}} \label{sec:proof_hessian}
\begin{proof}
The Hesse matrix of expected Fisher divergence is given by
\begin{align*}
    & \frac{\partial^2 \mathbb{E}_{X_{1:n}}[J_x(f||\hat{f}_{w_n,h_n})]}{\partial(w_n,h_n)^\top\partial(w_n,h_n)} \\
    & \quad \approx \begin{bmatrix} 2C_{(B,1)} & LC_{(B,L)}{h_n}^{L-1} \\ LC_{(B,L)}{h_n}^{L-1} & 2L(2L-1)C_{B,2L}{h_n}^{2L-2}+L(L-1)({w_n}-1)C_{(B,L)}{h_n}^{L-2} + \frac{12 C_V}{n{h_n}^5} \end{bmatrix}
\end{align*}
Then, the Hessian is
\begin{align}
    &  \det\left|\frac{\partial^2 \mathbb{E}_{X_{1:n}}[J_x(f||\hat{f}_{w_n,h_n})]}{\partial(w_n,h_n)^\top\partial(w_n,h_n)}\right| \nonumber\\
    &= 2C_{(B,1)}\left(2L(2L-1)C_{(B,2L)}{h_n}^{2L-2}+L(L-1)({w_n}-1)C_{(B,L)}{h_n}^{L-2} + \frac{12C_V}{n{h_n}^5}\right) \nonumber\\
    & \quad - \left(LC_{(B,L)}{h_n}^{L-1} \right)^2 \nonumber\\
    & = L^2C_{(B,1)}\left[C_{(B,2L)}-\frac{C_{(B,L)}^2}{C_{(B,1)}}\right]h_n^{2L-2} + L(7L-4)C_{(B,1)}C_{(B,2L)}h_n^{2L-2} \nonumber\\
    & \quad + 2L(L-1)(w-1)C_{(B,1)}C_{(B,2L)}h_n^{2L-2} + \frac{24C_{(B,1)}C_V}{nh_n^5} \nonumber\\
    & = L^2C_{(B,1)}\left[C_{(B,2L)}-\frac{C_{(B,L)}^2}{C_{(B,1)}}\right]h_n^{2L-2} \nonumber\\
    & \quad + L\left[(7L-4)+2(L-1)(w-1)\right]C_{(B,1)}C_{(B,2L)}h_n^{2L-2}  +   \frac{24C_{(B,1)}C_V}{nh_n^5} \label{eq:hessian1}
\end{align}
Since $C_{(B,1)}$ is positive from its definition and  $C_{(B,2L)}-\frac{C_{(B,L)}^2}{C_{(B,1)}}$ is non-negative as shown in Section \ref{sec:proof_opt_h}, the first term in (\ref{eq:hessian1}) is non-negative. Also,  $L\ge 2$ from Assumption \ref{ass:Kernel} and the domain of $w_n$ is $w_n>0$, which imply that $\left[(7L-4)+2(L-1)(w_n-1)\right]>0$. Therefore, combined with the fact that $C_{(B,1)}$ is positive and $C_{(B,2L)}$ is non-negative, the second term in (\ref{eq:hessian1}) is non-negative. Finally, since $C_{(B,1)}$ and $C_V$ are strictly positive, the last term in (\ref{eq:hessian1}) is strictly positive. Written in inequalities,
\begin{align*}
    & L^2C_{(B,1)}\left[C_{(B,2L)}-\frac{C_{(B,L)}^2}{C_{(B,1)}}\right]h_n^{2L-2}\ge 0,\\
    & L\left[(7L-4)+2(L-1)(w_n-1)\right]C_{(B,1)}C_{(B,2L)}h_n^{2L-2} \ge 0, \\
    & \frac{24C_{(B,1)}C_V}{nh_n^5} >0.
\end{align*}
Then, in view of (\ref{eq:hessian1}), the Hessian is always strictly positive.

Since, moreover, $|2C_{(B,1)}|$ is always strictly positive, determinants of all leading principle sub-matrices are strictly positive. This implies that the Hesse matrix is strictly positive definite.
\end{proof}

\subsection{Outline of the proof of Theorem \ref{thm:eh} and \ref{thm:eh_fix}} \label{sec:proof_eh}
In this subsection, we provide the outline of the proofs for Theorem \ref{thm:eh} and \ref{thm:eh_fix}. Similarly to the proof for the expansion of expected Fisher divergence, we expand the expected Hyv\"arinen score up to higher-order than necessary to prove Theorem \ref{thm:eh}, becasuse \ref{thm:ej_fix} needs the expansion.

In view of Section \ref{section:derivation_of_hyvarinen_score}, empirical Hyv\"arinen score,  under fixed $w$, has two terms $\mathcal{H}_1(w,h_n)$ and $\mathcal{H}_2(w,h_n)$, then, from the tower propety, the expected Hyv\"arinen score is given by as follows. 
\begin{align*}
    \mathbb{E}_{X_{1:n}}[\mathcal{H}(w,h_n)] 
    &= \mathbb{E}_{X_{1:n}}\left[\mathbb{E}_{X_{1:n}}\left[\frac{1}{n}\sum_{i=1}^n \mathcal{H}_i(w,h_n) \mid X_i\right]\right]\\ 
    &= \mathbb{E}_{X_{1:n}}\left[\frac{1}{n}\sum_{i=1}^n \mathbb{E}_{X_{1:n}}[\mathcal{H}_i(w,h_n) \mid X_i]\right]\\ 
    &= \mathbb{E}_{X_{1:n}}\left[\frac{1}{n}\sum_{i=1}^n\mathbb{E}_{X_{1:n}}[\mathcal{H}_{1i}(w,h_n)+\mathcal{H}_{2i}(w,h_n) \mid X_i]\right].
\end{align*}
From the result of computation in Section \ref{section:eh_1} and \ref{section:eh_2},
\begin{align*}
    & \mathbb{E}_{X_{1:n}} \left[\frac{1}{n}\sum_{i=1}^n\mathbb{E}_{X_{1:n}}[ \mathcal{H}_{1i}(w,h_n) \mid X_i] \right] \\
    &= \frac{2wR(K')}{nh_n^3}\mathbb{E}_{X_i} [f(X_i)^{-1}] + 2w\mathbb{E}_{X_i}[f''(X_i)f(X_i)^{-1}] \\
    & \quad + 2w\sum_{l=L}^{2L}\kappa_l\Bigl(\mathbb{E}_{X_i}[f^{(l+2)}(X_i)f(X_i)^{-1}]-\mathbb{E}_{X_i}[f''(X_i)f^{(l)}(X_i)f(X_i)^{-2}]\Bigl)h_n^l \\
    & \quad -2 w\kappa_L^2\Biggl(\mathbb{E}_{X_i}[f^{(L)}(X_i)f^{(L+2)}(X_i)f(X_i)^{-2}]-\mathbb{E}_{X_i}[f''(X_i)f^{(L)}(X_i)^2f(X_i)^{-3}]\Biggl)h_n^{2L} \\
    & \quad + o\left\{\frac{1}{nh_n^3}+h_n^{2L}\right\}
\end{align*}
and
\begin{align*}
    & \mathbb{E}_{X_{1:n}}\left[\frac{1}{n}\sum_{i=1}^n\mathbb{E}_{X_{1:n}}[\mathcal{H}_{2i}(w,h_n) \mid X_i]\right] \\
    &=
    \frac{(w^2-2w)R(K')}{nh_n^3}\mathbb{E}_{X_i}[f(X_i)^{-1}] + (w^2-2w)\mathbb{E}_{X_i}[f'(X_i)^2f(X_i)^{-2}] \\
    & \quad + 2(w^2-2w)\sum_{l=L}^{2L}\kappa_l\Bigl(\mathbb{E}_{X_i}[f'(X_i)f^{(l+1)}(X_i)f(X_i)^{-2}]-\mathbb{E}_{X_i}[f'(X_i)^2f^{(l)}(X_i)f(X_i)^{-3}]\Bigl)h_n^l \\
    & \quad + (w^2-2w)\kappa_L^2\Bigl(\mathbb{E}_{X_i}[f^{(L+1)}(X_i)^2f(X_i)^{-2}]\\
    & \quad \quad +3\mathbb{E}_{X_i}[f'(X_i)^2f^{(L)}(X_i)^2f(X_i)^{-4}] -4\mathbb{E}_{X_i}[f'(X_i)f^{(L)}(X_i)f^{(L+1)}(X_i)f(X_i)^{-3}]\Bigl) h_n^{2L} \\
    & \quad + o\left\{\frac{1}{nh_n^3}+h_n^{2L}\right\}
\end{align*}
then we have the asymptotic representation of the expected Hyv\"arinen score.
\begin{align*}
    \mathbb{E}_{X_{1:n}}[\mathcal{H}(w,h_n)] 
    &= w^2\frac{R(K')}{nh_n^3}\mathbb{E}_{X_i}[f(X_i)^{-1}] + 2w\mathbb{E}_{X_i}[f''(X_i)f(X_i)^{-1}] + (w^2-2w)\mathbb{E}_{X_i}[f'(X_i)^2f(X_i)^{-2}] \\
    & \quad + 2w\sum_{l=L}^{2L}\kappa_l\Bigl(\mathbb{E}_{X_i}[f^{(l+2)}(X_i)f(X_i)^{-1}]-\mathbb{E}_{X_i}[f''(X_i)f^{(l)}(X_i)f(X_i)^{-2}]\Bigl)h_n^l \\
    & \quad +  2(w^2-2w)\sum_{l=L}^{2L}\kappa_l\Bigl(\mathbb{E}_{X_i}[f'(X_i)f^{(l+1)}(X_i)f(X_i)^{-2}]-\mathbb{E}_{X_i}[f'(X_i)^2f^{(l)}(X_i)f(X_i)^{-3}]\Bigl)h_n^l \\
    & \quad -2 w\kappa_L^2\Biggl(\mathbb{E}_{X_i}[f^{(L)}(X_i)f^{(L+2)}(X_i)f(X_i)^{-2}]-\mathbb{E}_{X_i}[f''(X_i)f^{(L)}(X_i)^2f(X_i)^{-3}]\Biggl)h_n^{2L} \\
    & \quad + (w^2-2w)\kappa_L^2\Bigl(\mathbb{E}_{X_i}[f^{(L+1)}(X_i)^2f(X_i)^{-2}]\\
    & \quad \quad +3\mathbb{E}_{X_i}[f'(X_i)^2f^{(L)}(X_i)^2f(X_i)^{-4}] -4\mathbb{E}_{X_i}[f'(X_i)f^{(L)}(X_i)f^{(L+1)}(X_i)f(X_i)^{-3}]\Bigl) h_n^{2L} \\
    & \quad + o\left\{\frac{1}{nh_n^3}+h_n^{2L}\right\}.
\end{align*}
Next, we transform coefficients in order to prove the asymptotic equivalency of the expected Hyv\"arinen score to the expected Fisher divergence. One can show that, under Assumption \ref{ass:hyvarinen2}, integration by parts yields
\begin{align}
    & \mathbb{E}_{X_i}[f''(X_i)f(X_i)^{-1}] = 0 \label{eq:inv_hyvarinen_1}\\
    & \mathbb{E}_{X_i}[f^{(l+2)}(X_i)f(X_i)^{-1}] = 0 \label{eq:inv_hyvarinen_2}\\
    & \mathbb{E}_{X_i}[f''(X_i)f^{(l)}(X_i)f(X_i)^{-2}] \nonumber\\
    & \quad = -\mathbb{E}_{X_i}[f'(X_i)f^{(l+1)}(X_i)f(X_i)^{-2}] + \mathbb{E}_{X_i}[f'(X_i)^2f^{(l)}(X_i)f(X_i)^{-3}] \label{eq:inv_hyvarinen_3}\\
    & \mathbb{E}_{X_i}[f^{(L)}(X_i)f^{(L+2)}(X_i)f(X_i)^{-2}]  \nonumber\\
    & \quad = -\mathbb{E}_{X_i}[f^{(L+1)}(X_i)^2f(X_i)^{-2}] + \mathbb{E}_{X_i}[f'(X_i)f^{(L)}(X_i)f^{(L+1)}(X_i)f(X_i)^{-3}] \label{eq:inv_hyvarinen_4}\\
    & \mathbb{E}_{X_i}[f''(X_i)f^{(L)}(X_i)^2f(X_i)^{-3}] \nonumber\\
    & \quad = -2\mathbb{E}_{X_i}[f'(X_i)f^{(L)}(X_i)f^{(L+1)}(X_i)f(X_i)^{-3}] + 2\mathbb{E}_{X_i}[f'(X_i)f^{(L)}(X_i)^2f(X_i)^{-4}] \label{eq:inv_hyvarinen_5}
\end{align}
These imply that
\begin{align*}
    \mathbb{E}_{X_{1:n}}[\mathcal{H}(w,h_n)] &= w^2\frac{R(K')}{nh_n^3}\mathbb{E}_{X_i}[f(X_i)^{-1}] + (w^2-2w)\mathbb{E}_{X_i}[f'(X_i)^2f(X_i)^{-2}] \\
    & + 2(w^2-w)\sum_{l=L}^{2L}\kappa_l\Bigl(\mathbb{E}_{X_{i}}[f'(X_i)f^{(L+1)}(X_i)f(X_i)^{-2}]-\mathbb{E}_{X_{i}}[f'(X_i)^2f^{(L)}(X_i)f(X_i)^{-3}]\Bigl)h_n^l \\
    & + 2w\kappa_L^2\Bigl(\mathbb{E}_{X_i}[f^{(L+1)}(X_i)^2f(X_i)^{-1}] \\
    & \quad\quad -3\mathbb{E}_{X_i}[f'(X_i)f^{(L)}(X_i)f^{(L+1)}(X_i)f(X_i)^{-3}] + 2\mathbb{E}_{X_i}[f'(X_i)f^{(L)}(X_i)^2f(X_i)^{-4}]\Bigl)h_n^{2L}\\
    & \quad + (w^2-2w)\kappa_L^2\Bigl(\mathbb{E}_{X_i}[f^{(L+1)}(X_i)^2f(X_i)^{-2}]\\
    & \quad \quad +3\mathbb{E}_{X_i}[f'(X_i)^2f^{(L)}(X_i)^2f(X_i)^{-4}] -4\mathbb{E}_{X_i}[f'(X_i)f^{(L)}(X_i)f^{(L+1)}(X_i)f(X_i)^{-3}]\Bigl) h_n^{2L} \\
    & \quad + o\left\{\frac{1}{nh_n^3}+h_n^{2L}\right\}\\
    & = w^2\frac{R(K')}{nh_n^3}\mathbb{E}_{X_i}[f(X_i)^{-1}] + (w^2-2w)\mathbb{E}_{X_i}[f'(X_i)^2f(X_i)^{-2}] \\
    & \quad + 2(w^2-w)\sum_{l=L}^{2L}\kappa_l\Bigl(\mathbb{E}_{X_i}[f'(X_i)f^{(L+1)}(X_i)f(X_i)^{-2}]-\mathbb{E}_{X_i}[f'(X_i)^2f^{(L)}(X_i)f(X_i)^{-3}]\Bigl)h_n^l \\
    & \quad + w^2\kappa_L^2\mathbb{E}_{X_i}[f^{(L+1)}(X_i)^2f(X_i)^{-1}]h_n^{2L} \\
    & \quad + (3w^2-2w)\kappa_L^2\mathbb{E}_{X_i}[f'(X_i)^2f^{(L)}(X_i)^2f(X_i)^{-4}]h_n^{2L} \\
    & \quad + (-4w^2+2w)\kappa_L^2\mathbb{E}_{X_i}[f'(X_i)f^{(L)}(X_i)f^{(L+1)}(X_i)f(X_i)^{-3}]h_n^{2L} + o\left\{\frac{1}{nh_n^3}+h_n^{2L}\right\}  \\
    & = w^2\frac{R(K')}{nh_n^3}\mathbb{E}_{X_i}[f(X_i)^{-1}] + (w^2-2w)\mathbb{E}_{X_i}[f'(X_i)^2f(X_i)^{-2}] \\
    & \quad + 2(w^2-w)\sum_{l=L}^{2L}\kappa_l\Bigl(\mathbb{E}_{X_i}[f'(X_i)f^{(L+1)}(X_i)f(X_i)^{-2}]-\mathbb{E}_{X_i}[f'(X_i)^2f^{(L)}(X_i)f(X_i)^{-3}]\Bigl)h_n^l \\
    & \quad + 2w(w-1)\kappa_L^2\Bigl(\mathbb{E}_{X_i}[f'(X_i)^2f^{(L)}(X_i)^2f(X_i)^{-2}] - \mathbb{E}_{X_i}[f'(X_i)f^{(L)}(X_i)f^{(L+1)}(X_i)f(X_i)^{-3}]\Bigl)h_n^{2L} \\
    & \quad + w^2\kappa_L^2\mathbb{E}_{X_i}[\{f^{(L+1)}(X_i)f(X_i)^{-1}-f'(X_i)f^{(L)}(X_i)f(X_i)^{-2}\}^2]h_n^{2L} + o\left\{\frac{1}{nh_n^3}+h_n^{2L}\right\} 
\end{align*}
Then, from Assumption \ref{ass:wlocalise}, 
\begin{align*}
    \mathbb{E}_{X_{1:n}}[\mathcal{H}(w_n,h_n)]
    & = w_n^2\frac{R(K')}{nh_n^3}\mathbb{E}_{X_i}[f(X_i)^{-1}] + \{(w_n-1)^2-1\}\mathbb{E}_{X_i}[f'(X_i)^2f(X_i)^{-2}] \\
    & \quad + 2w_n(w_n-1)\kappa_L\Bigl(\mathbb{E}_{X_i}[f'(X_i)f^{(L+1)}(X_i)f(X_i)^{-2}]-\mathbb{E}_{X_i}[f'(X_i)^2f^{(L)}(X_i)f(X_i)^{-3}]\Bigl)h_n^L \\
    & \quad + w_n^2\kappa_L^2\mathbb{E}_{X_i}[\{f^{(L+1)}(X_i)f(X_i)^{-1}-f'(X_i)f^{(L)}(X_i)f(X_i)^{-2}\}^2]h_n^{2L} \\
    & \quad + o\left\{\frac{1}{nh_n^3}+(w_n-1)^2+(w_n-1)h_n^L+h_n^{2L}\right\} 
\end{align*}
Now, we have the asymptotic representation of expected Hyv\"arinen score with the same shape as the expected Fisher divergence.

\subsection{Outline of the Proof of Theorem \ref{thm:uniform}} \label{sec:proof_uniform}
\begin{proof}
See Section~\ref{sec:empirical_hyvarinen_score}, for the derivation of $Q_n$. 

Next, we will show that $\sup_{(w_n,h_n)\in\mathscr{W}_n\times\mathscr{H}_n} \left\{h_n^{2L}+(w_n-1)h_n^L+(w_n-1)^2+\frac{1}{nh_n^3}\right\}^{-1} R_n(w_n,h_n) \xrightarrow{p} 0$, with the discretization technique of parameter space likewise \cite[pp.1509-1510]{hall1987kl}. Details of descretization are too lengthy in our case and the process is almost identical with \cite{hall1987kl}, so we will omit it.
Under \ref{ass:kernelholder}, there exist lattice points $\{h_{n,k}\}_{k=1}^{m_h}$ and $\{w_{n,l}\}_{l=1}^{m_w}$  close enough to each other so that, for some constant $C$
\begin{align}
    |R_n(w_{n,l},h_{n,k})-R_n(w_{n,l+1},h_{n,k+1})| \le Cn^{-1} \nonumber
\end{align}
and, for some $\lambda\in\mathbb{R}_{++}$,
\begin{align}
    C\left\{{h_{n,k}}^{2L}+(w_{n,l}-1)h_{n,k}^L+(w_{n,l}-1)^2+\frac{1}{nh_{n,k}^3}\right\}^{-1}n^{-1} = Cn^{-\lambda}\rightarrow 0 \nonumber
\end{align}
holds. Then 
\begin{align*}
    & \sup_{(w_n,h_n)\in\mathscr{W}_n\times\mathscr{H}_n} \left\{h_n^{2L}+(w_n-1)h_n^L+(w_n-1)^2+\frac{1}{nh_n^3}\right\}^{-1} R_n(w_n,h_n) \\
    & \le \frac{m_hm_w}{\delta^2}\max_{\substack{1\le k \le m_h\\1\le l \le m_w }}\left\{{h_{n,k}}^{2L}+(w_{n,l}-1)h_{n,k}^L+(w_{n,l}-1)^2+\frac{1}{nh_{n,k}^3}\right\}^{-1}R_n(w_{n,k},h_{n,l})  + Cn^{-\lambda} 
\end{align*}
so, from Chebyshev's inequality, we have
\begin{align*}
    & \mathbb{P}\left( \sup_{(w_n,h_n)\in\mathscr{W}_n\times\mathscr{H}_n} \left\{h_n^{2L}+(w_n-1)h_n^L+(w_n-1)^2+\frac{1}{nh_n^3}\right\}^{-1} R_n(w_n,h_n) > \delta + Cn^{-\lambda}\right) \\
    & \quad \le \frac{m_hm_w}{\delta^2}\max_{\substack{1\le k \le m_h\\1\le l \le m_w }}\left\{{h_{n,k}}^{2L}+(w_{n,l}-1)h_{n,k}^L+(w_{n,l}-1)^2+\frac{1}{nh_{n,k}^3}\right\}^{-2}\mathbb{E}[R_n(w_{n,k},h_{n,l})^2] .
\end{align*}
In Section~\ref{sec:empirical_hyvarinen_score}, we have derived the bound on $\mathbb{E}[R_n(w_n,h_n)^2]$ as 
\begin{align}
    & \mathbb{E}[R_n(w_n,h_n)^2] = O\{ n^{-1}h_n^{2L} + n^{-1}(w_n-1)h_n^L + n^{-1}(w_n-1)^2 + n^{-2}h_n^{-5}\} . \nonumber
\end{align}
Recalling the parameter space is limited so that $h_n \rightarrow 0, nh_n^3\rightarrow\infty$ and $w_n\rightarrow 1$ as $n\rightarrow\infty$, since,
\begin{align*}
    & n^{-1}h_n^{2L} \ll \left(h_n^{2L}\times \frac{1}{nh_n^3}\right), \quad  n^{-1}(w_n-1)h_n^L \ll \left((w_n-1)h_n^L\times\frac{1}{nh_n^3}\right) \\&  n^{-1}(w_n-1)^2 \ll  \left((w_n-1)^2\times\frac{1}{nh_n^3}\right), \quad n^{-2}h_n^{-5} \ll \left( \frac{1}{nh_n^3}\right)^{2}, 
\end{align*}
we have a required bound to prove the theorem.
\begin{rem}
    Although it is natural doubt that, for example, $n^{-2}h_n^{-5}h_n^{-4L}$ does not converge even if  $n^{-2}h_n^{-5} (n^2h_n^6)\rightarrow0$, one can show that it is sufficient by deviding the cases, in order to prove the Theorem, to  guarantee that the product of one of the terms in $\left\{h_n^{2L}+(w_n-1)h_n^L+(w_n-1)^2+\frac{1}{nh_n^3}\right\}^{-2}$ and $\mathbb{E}[R_n(w_n,h_n)^2]$ converges. In the following, we provide one of the cases for illustration. We do not provide the proof of all cases to avoid tediousness.
\end{rem}
Let take the case where $n^{-1}h_n^{2L}$ dominants the other terms as example. For arbitrarily small $\epsilon>0$, it holds that
\begin{align}
    n^{-2}h_n^{-5} \ll n^{-1}h_n^{2L} \iff n^{-1/(2L+5)} \ll h_n \lesssim n^{-\epsilon} .\nonumber
\end{align}
In this case, since $h_n^{-4L}$ is dominated by $n^2h_n^6$ as
\begin{align}
    h_n^{-4L} \ll n^{4L/(2L+5)}, \quad n^{(4L+4)/(2L+5)} \ll n^2h_n^6 \implies h_n^{-4L} \ll n^2h_n^6, \nonumber
\end{align}
so, $n^{-2}h_n^{-5} (n^2h_n^6)\rightarrow0$ implies $n^{-2}h_n^{-5}h_n^{-4L}\rightarrow 0$. 
\end{proof}

\subsection{Proof of Corollary \ref{cor:consistency}} \label{sec:proof_consistency}
\begin{proof}
First we define $\tilde{\mathcal{H}}(w_n, h_n)$ as
\begin{align*}
    \widetilde{\mathcal{H}}(w_n, h_n) \equiv \mathcal{H}(w_n, h_n) - \frac{1}{n}\sum_{i=1}^n \{2f''(X_i)f(X_i)^{-1}-f'(X_i)^2f(X_i)^2\}. 
\end{align*}
Using this notation, from Theorem \ref{thm:uniform}, we have
\begin{align*}
    \widetilde{\mathcal{H}}(w_n^*, h_n^*) =  \mathbb{E}_{X_{1:n}}[\widetilde{\mathcal{H}}(w_n^*,h_n^*)] + o_p\left\{{h_n^*}^{2L}+(w_n^*-1){h_n^*}^L+(w_n^*-1)^2+\frac{1}{n{h_n^*}^3} \right\}.
\end{align*}
Also, since $\hat{w}_n$ and $\hat{h}_n$ minimize the empirical Hyv\"arinen Score, it holds that
\begin{align*}
\widetilde{\mathcal{H}}(\hat{w}_n, \hat{h}_n) \le \widetilde{\mathcal{H}}(w_n^*, h_n^*) + o_p\left\{{h_n^*}^{2L}+(w_n^*-1){h_n^*}^L+(w_n^*-1)^2+\frac{1}{n{h_n^*}^3} \right\}.
\end{align*}
These imply that
\begin{align}
    & -\mathbb{E}_{X_{1:n}}[\widetilde{\mathcal{H}}(w_n^*,h_n^*)] + \mathbb{E}_{X_{1:n}}[\widetilde{\mathcal{H}}(\hat{w}_n, \hat{h}_n)]  \nonumber \\
    & \quad \le -\widetilde{\mathcal{H}}(\hat{w}_n, \hat{h}_n) + \mathbb{E}_{X_{1:n}}[\widetilde{\mathcal{H}}(\hat{w}_n, \hat{h}_n)] + o_p\left\{{h_n^*}^{2L}+(w_n^*-1){h_n^*}^L+(w_n^*-1)^2+\frac{1}{n{h_n^*}^3} \right\} \nonumber\\
    \implies
    & \frac{-\mathbb{E}_{X_{1:n}}[\widetilde{\mathcal{H}}(w_n^*,h_n^*)] + \mathbb{E}_{X_{1:n}}[\widetilde{\mathcal{H}}(\hat{w}_n, \hat{h}_n)] }{\mathbb{E}_{X_{1:n}}[\widetilde{\mathcal{H}}(w_n^*,h_n^*)] } \le \frac{-\widetilde{\mathcal{H}}(\hat{w}_n, \hat{h}_n) + \mathbb{E}_{X_{1:n}}[\widetilde{\mathcal{H}}(\hat{w}_n, \hat{h}_n)]}{\mathbb{E}_{X_{1:n}}[\widetilde{\mathcal{H}}(w_n^*,h_n^*)]} + o_p(1)  \nonumber\\
    & \qquad\qquad\qquad\qquad\qquad~~  \le \sup_{(w_n,h_n) \in \mathscr{W}_n \times \mathscr{H}_n} \left| \frac{\widetilde{\mathcal{H}}(w_n, h_n) - \mathbb{E}_{X_{1:n}}[\widetilde{\mathcal{H}}(w_n, h_n)]}{\mathbb{E}_{X_{1:n}}[\widetilde{\mathcal{H}}(w_n,h_n)]}\right| + o_p(1) \nonumber\\
    & \qquad\qquad\qquad\qquad\qquad~~ \xrightarrow{p} 0 \label{eq:uniform_htilde}
\end{align}
Then, from the uniqueness of the optimal  parameters (Theorem \ref{thm:convex}), we can confirm that for any $\delta_h, \delta_w \in \mathbb{R}_{++}$, there exists $\eta$ such that $\{\mathbb{E}_{X_{1:n}}[\widetilde{\mathcal{H}}(w_n, h_n)] - \mathbb{E}_{X_{1:n}}[\widetilde{\mathcal{H}}(w_n^*, h_n^*)]\} / \mathbb{E}_{X_{1:n}}[\widetilde{\mathcal{H}}(w_n^*, h_n^*)] > \eta$ is satisfied for all $w_n$ and $h_n$ such that $|(\hat{h}_n - h_n^*)/h_n^*| > \delta_h$ and $|\{(\hat{w}_n - 1) - (w_n^* - 1)\}/(w_n^* -1) |>\delta_w$. Thus 
\begin{align*}
    & \mathbb{P}\left( \left|\frac{\hat{h}_n - h_n^*}{h_n^*}\right| >  \delta_h \bigcup \left| \frac{(\hat{w}_n - 1) - (w_n^* - 1)}{ (w_n^* -1)} \right| > \delta_w \right) \\
    & \le \mathbb{P}\left( \frac{-\mathbb{E}_{X_{1:n}}\left[\widetilde{\mathcal{H}}(\hat{w}_n, \hat{h}_n)\right] + \mathbb{E}_{X_{1:n}}[\widetilde{\mathcal{H}}(w_n^*, h_n^*)]}{\mathbb{E}_{X_{1:n}}[\widetilde{\mathcal{H}}(w_n^*, h_n^*)]} > \eta \right) \\
    & \to 0
\end{align*}
where the convergence is shown as (\ref{eq:uniform_htilde}).
Now,  we confirm that the Theorem \ref{thm:uniform} implies 
\begin{align*}
    (\hat{h}_n - h_n^*) / h_n^* \to 0, \quad \{(\hat{w}_n - 1) - (w_n^* - 1)\} / (w_n^* -1) \to 0.
\end{align*}
\end{proof}

\section{Properties of Expected Fisher divergence}
In this section, we provide the detail computation for expected Fisher divergence. We need the result in this section to prove Theorem \ref{thm:ej} and \ref{thm:ej_fix}.  
Section~\ref{sec:bv_deco} is bias and variance like decomposition of the expected Fisher divergence.  Sections~\ref{section:bias} and \ref{section:variance} are detail computation of bias and variance like terms of expected Fisher divergence. In this section, all expansion is conducted under Assumption \ref{ass:nhinfty} and with fixed $w$.

\subsection{Bias and Variance Decomposition}\label{sec:bv_deco}
Since we can consider Fisher divergence of $\hat{f}(x)$ as Weighted MISE of $\frac{\partial}{\partial x}\log \hat{f}(x)$, we can decompose $\mathbb{E}_{X_{1:n}}\left[J_x(f||\hat{f})\right]$ into the squared bias and variance terms as follows.
\begin{align*}
    & \mathbb{E}_{X_{1:n}}\left[J_x(f||\hat{f})\right] \\
    \equiv & \mathbb{E}_{X_{1:n}}\left[\int \left\{\frac{\partial}{\partial x}\log \hat{f}(x) - \frac{\partial}{\partial x}\log f(x) \right\}^2f(x)dx\right] \\
    = & \mathbb{E}_{X_{1:n}}\left[\int \left\{\frac{\partial}{\partial x}\log \hat{f}(x) - \mathbb{E}_{X_{1:n}}\left[\frac{\partial}{\partial x}\log \hat{f}(x)\right] + \mathbb{E}_{X_{1:n}}\left[\frac{\partial}{\partial x}\log \hat{f}(x)\right] - \frac{\partial}{\partial x}\log f(x)\right\}^2f(x)dx\right] \\
    = & \int \mathbb{E}_{X_{1:n}}\left[\left\{\frac{\partial}{\partial x}\log \hat{f}(x) - \mathbb{E}_{X_{1:n}}\left[\frac{\partial}{\partial x}\log \hat{f}(x)\right]\right\}^2\right]  f(x)dx + \int \left\{\mathbb{E}_{X_{1:n}}\left[\frac{\partial}{\partial x}\log \hat{f}(x)\right] - \frac{\partial}{\partial x}\log f(x)\right\}^2f(x)dx
\end{align*}

\subsection{Bias}\label{section:bias}
In order to get the asymptotic representation of the bias term, we compute $\mathbb{E}_{X_{1:n}}\left[\frac{\partial}{\partial x}\log\hat{f}_{w,h_n}(x)\right]$ at first. 
\begin{align*}
    & \mathbb{E}_{X_{1:n}}\left[\frac{\partial}{\partial x}\log\hat{f}_{w,h_n}(x)\right] \\
    = & \mathbb{E}_{X_{1:n}}\left[-w \frac{\frac{1}{nh_n^2}\sum_{i=1}^nK'_{i,h_n}(x)}{\frac{1}{nh_n}\sum_{i=1}^nK_{i,h_n}(x)}\right] \\
    = & -w \mathbb{E}_{X_{1:n}}\left[\frac{\frac{1}{nh_n^2}\sum_{i=1}^nK'_{i,h_n}(x)}{\mathbb{E}_{X_{1:n}}\left[\frac{1}{nh_n}\sum_{i=1}^nK_{i,h_n}(x)\right]+\frac{1}{nh_n}\sum_{i=1}^nK_{i,h_n}(x)-\mathbb{E}_{X_{1:n}}\left[\frac{1}{nh_n}\sum_{i=1}^nK_{i,h_n}(x)\right]}\right] \\
    = & -w\mathbb{E}_{X_{1:n}}\left[\frac{\frac{1}{nh_n^2}\sum_{i=1}^nK'_{i,h_n}(x)}{\mathbb{E}_{X_{1:n}}\left[\frac{1}{nh_n}\sum_{i=1}^nK_{i,h_n}(x)\right]}\right] \\
    & \quad +w\mathbb{E}_{X_{1:n}}\left[ \frac{\frac{1}{nh_n^2}\sum_{i=1}^nK'_{i,h_n}(x)\left(\frac{1}{nh_n}\sum_{i=1}^nK_{i,h_n}(x)-\mathbb{E}_{X_{1:n}}\left[\frac{1}{nh_n}\sum_{i=1}^nK_{i,h_n}(x)\right]\right)}{\mathbb{E}_{X_{1:n}}\left[\frac{1}{nh_n}\sum_{i=1}^nK_{i,h_n}(x)\right]^2} \right]  + O\{(nh_n)^{-1}\} \\
    = & -2w\frac{\mathbb{E}_{X_{1:n}}\left[\frac{1}{nh_n^2}\sum_{i=1}^nK'_{i,h_n}(x)\right]}{\mathbb{E}_{X_{1:n}}\left[\frac{1}{nh_n}\sum_{i=1}^nK_{i,h_n}(x)\right]} \\
    & \quad + w \frac{\mathbb{E}_{X_{1:n}}\left[\frac{1}{n^2h_n^3}\sum_{i=1}^n\sum_{j=1}^nK'_{i,h_n}(x)K_{j,h_n}(x)\right]}{\mathbb{E}_{X_{1:n}}\left[\frac{1}{nh_n}\sum_{i=1}^nK_{i,h_n}(x)\right]^2} + o\{(nh_n^3)^{-1}\}
\end{align*}
where the third equality expands the denominator of the right-hand side of the second equality around $\frac{1}{nh_n}\sum_{i=1}^nK_{i,h_n}(x)=\mathbb{E}_{X_{1:n}}\left[\frac{1}{nh_n}\sum_{i=1}^nK_{i,h_n}(x)\right]$.
Since from a straightforward computation,
\begin{align*}
    & \mathbb{E}_{X_{1:n}}\left[\frac{1}{nh_n^2}\sum_{i=1}^nK'_{i,h_n}(x)\right] = -f'(x) - \sum_{l=L}^{2L}\kappa_lf^{(l)}(x)h_n^l + o(h_n^{2L}), \\
    & \mathbb{E}_{X_{1:n}}\left[\frac{1}{nh_n}\sum_{i=1}^n K_{i,h_n}(x)\right] = f(x) + \sum_{l=L}^{2L}\kappa_lf^{(l)}(x)h_n^l + o(h_n^{2L}), 
\end{align*}
combined with Lemma \ref{lem:k'k_x}, which states that 
\begin{align*}
    & \mathbb{E}_{X_{1:n}}\left[\frac{1}{n^2h_n^3}\sum_{i=1}^n\sum_{j=1}^nK'_{i,h_n}(x)K_{j,h_n}(x)\right] \\
    & = -f(x)f'(x) - \sum_{l=L}^{2L}\kappa_l[f'(x)f^{l}(x)+f(x)f^{(l+1)}(x)]h_n^l - \kappa_L^2f^{(L)}(x)f^{(L+1)}(x)h_n^{2L} \\
    & \quad + O\{n^{-1}h_n^{-2}\} + o(h_n^{2L}),
\end{align*}
These imply that $\mathbb{E}_{X_{1:n}}\left[\frac{\partial}{\partial x}\log\hat{f}_{w,h_n}(x)\right]$ is expanded as follows.
\begin{align*}
    & \mathbb{E}_{X_{1:n}}\left[\frac{\partial}{\partial x}\log\hat{f}_{w,h_n}(x)\right] \\
    & = 2w\frac{f'(x) + \sum_{l=L}^{2L}\kappa_lf^{(l+1)}(x)h_n^l + o(h_n^{2L})}{f(x) + \sum_{l=L}^{2L}\kappa_lf^{(l)}(x)h_n^l + o(h_n^{2L})} \\
    & \quad - w\frac{f(x)f'(x) + \sum_{l=L}^{2L}\kappa_l[f'(x)f^{l}(x)+f(x)f^{(l+1)}(x)]h_n^l }{\Bigl(f(x) + \sum_{l=L}^{2L}\kappa_lf^{(l)}(x)h_n^l + o(h_n^{2L})\Bigl)^2} \\
     & \quad - w\frac{ \kappa_L^2f^{(L)}(x)f^{(L+1)}(x)h_n^{2L} + O\{n^{-1}h_n^{-2}\} + o(h_n^{2L})}{\Bigl(f(x) + \sum_{l=L}^{2L}\kappa_lf^{(l)}(x)h_n^l + o(h_n^{2L})\Bigl)^2} \\
    & = 2w\frac{f'(x) + \sum_{l=L}^{2L}\kappa_lf^{(l+1)}(x)h_n^l}{f(x)} \\
    & \qquad - 2w\frac{\sum_{l=L}^{2L}\kappa_lf'(x)f^{(l)}(x)h_n^l + \kappa_L^2f^{(L)}(x)f^{(L+1)}(x)h_n^{2L}}{f(x)^2} \\
    & \qquad +2w\frac{\kappa_L^2f'(x)f^{(L)}(x)^2h_n^{2L}}{f(x)^3}\\
    & \quad -w \frac{f(x)f'(x) + \sum_{l=L}^{2L}\kappa_l[f'(x)f^{(l)}(x) + f(x)f^{(l+1)}(x)]h_n^l+\kappa_L^2f^{(L)}(x)f^{(L+1)}(x)h_n^{2L}}{f(x)^2} \\
    & \qquad +2w\frac{\sum_{l=L}^{2L}\kappa_lf(x)f'(x)f^{(l)}(x)h_n^l + \kappa_L^2f^{(L)}(x)[f'(x)f^{(L)}(x)+f(x)f^{(L+1)}(x)]h_n^{2L}}{f(x)^3} \\
    & \qquad - 3w\frac{\kappa_L^2f(x)f'(x)f^{(L)}(x)^2h_n^{2L}}{f(x)^4} + O(n^{-1}h_n^{-2}) + o(h_n^{2L}) \\
    & = wf'(x)f(x)^{-1} + w\sum_{l=L}^{2L}\kappa_l[f^{(l+1)}(x)f(x)^{-1}-f'(x)f^{(l)}(x)f(x)^{-2}]h_n^l \\
    &\quad + w\kappa_L^2[f'(x)f^{(L)}(x)^2f(x)^{-3}-f^{(L)}(x)f^{(L+1)}(x)f(x)^{-2}]h_n^{2L}  + O(n^{-1}h_n^{-2}) + o(h_n^{2L}) 
\end{align*}
then we have the asymptotic representation of the bias term.
\begin{align*}
    & \mathbb{E}_{X_{1:n}}\left[\frac{\partial}{\partial x}\log\hat{f}_{w,h}(x)\right] - \frac{\partial}{\partial x}\log f(x) \\
    & \quad=  (w-1)f'(x)f(x)^{-1} + w\sum_{l=L}^{2L}\kappa_l[f^{(l+1)}(x)f(x)^{-1}-f'(x)f^{(l)}(x)f(x)^{-2}]h_n^l \\
    &\qquad + w\kappa_L^2[f'(x)f^{(L)}(x)^2f(x)^{-3}-f^{(L)}(x)f^{(L+1)}(x)f(x)^{-2}]h_n^{2L}  + o\left\{\frac{1}{nh_n^3}+h_n^{2L}\right\}
\end{align*}

\subsection{Variance} \label{section:variance}
In order to get the asymptotic representation of the variance term, we expand \\ $\mathbb{E}_{X_{1:n}}\left[\left(\frac{\partial}{\partial x}\log \hat{f}_{w,h_n}(x)\right)^2\right]$ at first.
\begin{align*}
    & \mathbb{E}_{X_{1:n}}\left[\left(\frac{\partial}{\partial x}\log \hat{f}_{w,h_n}(x)\right)^2\right] \\
    & = \mathbb{E}_{X_{1:n}}\left[\left(-w \frac{\frac{1}{nh_n^2}\sum_{i=1}^nK'_{i,h_n}(x)}{\frac{1}{nh_n}\sum_{i=1}^nK_{i,h_n}(x)}\right)^2\right] \\
    & = w^2\mathbb{E}_{X_{1:n}}\left[\frac{\left(\frac{1}{nh_n^2}\sum_{i=1}^nK'_{i,h_n}(x)\right)^2}{\left(\frac{1}{nh_n}\sum_{i=1}^nK_{i,h_n}(x)\right)^2}\right] \\
    & = w^2\mathbb{E}_{X_{1:n}}\left[\frac{\left(\frac{1}{nh_n^2}\sum_{i=1}^nK'_{i,h_n}(x)\right)^2}{\mathbb{E}_{X_{1:n}}\left[\left(\frac{1}{nh_n}\sum_{i=1}^nK_{i,h_n}(x)\right)^2\right]+\left(\frac{1}{nh_n}\sum_{i=1}^nK_{i,h_n}(x)\right)^2-\mathbb{E}_{X_{1:n}}\left[\left(\frac{1}{nh_n}\sum_{i=1}^nK_{i,h_n}(x)\right)^2\right]}\right] \\
    & = 2w^2 \frac{\mathbb{E}_{X_{1:n}}\left[\left(\frac{1}{nh_n^2}\sum_{i=1}^nK'_{i,h_n}(x)\right)^2\right]}{\mathbb{E}_{X_{1:n}}\left[\left(\frac{1}{nh_n}\sum_{i=1}^nK_{i,h_n}(x)\right)^2\right]} \\
    & \quad - w^2 \frac{\mathbb{E}_{X_{1:n}}\left[\left(\frac{1}{nh_n^2}\sum_{i=1}^nK'_{i,h_n}(x)\right)^2\left(\frac{1}{nh_n}\sum_{i=1}^nK_{i,h_n}(x)\right)^2\right]}{\mathbb{E}_{X_{1:n}}\left[\left(\frac{1}{nh_n}\sum_{i=1}^nK_{i,h_n}(x)\right)^2\right]^2} + O\{(nh_n)^{-1}\}.
\end{align*}
where the bound on the remainder term follows from Lemma \ref{lem:bound_remainder} and Cauchy-Schwarz inequality.
Since Lemma \ref{lem:k'_x_square}, \ref{lem:k_x_square} and \ref{lem:k'k_x_square} state that
\begin{align*}
    & \mathbb{E}_{X_{1:n}}\left[\left(\frac{1}{nh_n^2}\sum_{i=1}^n K'_{i,h_n}(x)\right)^2\right] = \frac{R(K')}{nh_n^3}f(x) + f'(x)^2 \\
    & \quad + 2\sum_{l=L}^{2L}\kappa_{l}f'(x)f^{(l+1)}(x)h_n^l + \kappa_L^2f^{(L+1)}(x)^2h_n^{2L} + o\left\{\frac{1}{nh_n^3}+h_n^{2L}\right\}\\
    & \mathbb{E}_{X_{1:n}}\left[\left(\frac{1}{nh_n}\sum_{i=1}^n K_{i,h_n}(x)\right)^2\right] = f(x)^2 + 2\sum_{l=L}^{2L}\kappa_l f(x)f^{(l)}(x)h_n^{l} + \kappa_L^2f^{(L)}(x)^2h_n^{2L} + o\left\{\frac{1}{nh_n^3}+h_n^{2L}\right\}\\
    & \mathbb{E}_{X_{1:n}}\left[\left(\frac{1}{nh_n^2}\sum_{i=1}^n K'_{i,h_n}(x)\right)^2\left(\frac{1}{nh_n}\sum_{i=1}^n K_{i,h_n}(x)\right)^2\right] \\
    & \quad = \frac{R(K')}{nh_n^3}f(x)^3 + f'(x)^2f(x)^2 + 2\sum_{l=L}^{2L}\kappa_l\Bigl(f(x)f'(x)^2f^{(l)}(x) + f(x)^2f'(x)f^{(l+1)}(x)\Bigl)h_n^l \\
    & \qquad + \kappa_L^2\Bigl(4f(x)f'(x)f^{(L)}(x)f^{(L+1)}(x) + f'(x)^2f^{(L)}(x)^2 + f(x)^2f^{(L+1)}(x)^2\Bigl)h_n^{2L} \\
    & \quad + o\left\{\frac{1}{nh_n^3}+h_n^{2L}\right\} 
\end{align*}
These imply that $\mathbb{E}_{X_{1:n}}\left[\left(\frac{\partial}{\partial x}\log \hat{f}_{w,h_n}(x)\right)^2\right]$  is expanded as follows.
\begin{align*}
    & \mathbb{E}_{X_{1:n}}\left[\left(\frac{\partial}{\partial x}\log \hat{f}_{w,h_n}(x)\right)^2\right] \\
    & = 2w^2 \frac{\frac{R(K')}{nh_n^3}f(x) + f'(x)^2 + 2\sum_{l=L}^{2L}\kappa_{l}f'(x)f^{(l+1)}(x)h_n^l + \kappa_L^2f^{(L+1)}(x)^2h_n^{2L}}{f(x)^2 + 2\sum_{l=L}^{2L}\kappa_l f(x)f^{(l)}(x)h_n^{l} + \kappa_L^2f^{(L)}(x)^2h_n^{2L}}\\
    & \quad -w^2 \frac{\frac{R(K')}{nh_n^3}f(x)^3 + f'(x)^2f(x)^2  }{\left(f(x)^2 + 2\sum_{l=L}^{2L}\kappa_l f(x)f^{(l)}(x)h_n^{l} + \kappa_L^2f^{(L)}(x)^2h_n^{2L}\right)^2} \\
    & \quad -w^2 \frac{2\sum_{l=L}^{2L}\kappa_l\Bigl(f(x)f'(x)^2f^{(l)}(x) + f(x)^2f'(x)f^{(l+1)}(x)\Bigl)h_n^l}{\left(f(x)^2 + 2\sum_{l=L}^{2L}\kappa_l f(x)f^{(l)}(x)h_n^{l} + \kappa_L^2f^{(L)}(x)^2h_n^{2L}\right)^2} \\
    & \quad -w^2 \frac{\kappa_L^2\Bigl(4f(x)f'(x)f^{(L)}(x)f^{(L+1)}(x)+f'(x)^2f^{(L)}(x)^2 + f(x)^2f^{(L+1)}(x)^2\Bigl)h_n^{2L}}{\left(f(x)^2 + 2\sum_{l=L}^{2L}\kappa_l f(x)f^{(l)}(x)h_n^{l} + \kappa_L^2f^{(L)}(x)^2h_n^{2L}\right)^2}\\
    &\quad + o\left\{\frac{1}{nh_n^3}+h_n^{2L}\right\} \\
    & = 2w^2\frac{\frac{R(K')}{nh_n^3}f(x) + f'(x)^2 + 2\sum_{l=L}^{2L}\kappa_{l}f'(x)f^{(l+1)}(x)h_n^l + \kappa_L^2f^{(L+1)}(x)^2h_n^{2L}}{f(x)^2 }\\
    & \quad -2w^2\frac{2\sum_{l=L}^{2L}\kappa_lf(x)f'(x)^2f^{(l)}(x)h_n^l+ \kappa_L^2f'(x)^2f^{(L)}(x)^2h_n^{2L}+4\kappa_L^2f(x)f'(x)f^{(L)}(x)f^{(L+1)}(x)h_n^{2L}}{f(x)^4}  \\
    &\quad + 2w^2\frac{4\kappa_L^2f(x)^2f'(x)^2f^{(L)}(x)^2h_n^{2L}}{f(x)^6}\\
    & \quad -w^2 \frac{\frac{R(K')}{nh_n^3}f(x)^3+f'(x)^2f(x)^2}{f(x)^4} + 2w^2\frac{2\sum_{l=L}^{2L}\kappa_lf'(x)^2f(x)^3f^{(l)}(x)h_n^l + \kappa_L^2f'(x)^2f(x)^2f^{(L)}(x)^2h_n^{2L}}{f(x)^6} \\
    & \quad -3w^2 \frac{4\kappa_L^2f'(x)^2f(x)^4f(x)^{(L)}(x)^2h_n^{2L}}{f(x)^8} \\
    & \quad - w^2\frac{2\sum_{l=L}^{2L}\kappa_l\Bigl(f(x)f'(x)^2f^{(l)}(x) + f(x)^2f'(x)f^{(l+1)}(x)\Bigl)h_n^l}{f(x)^4} \\
    & \quad + 2w^2\frac{4\kappa_L^2[f(x)^2f'(x)^2f^{(L)^2}(x)+f(x)^3f'(x)f^{(L)}(x)f^{(L+1)}(x)]h_n^{2L}}{f(x)^6} \\
    & \quad -w^2 \frac{\kappa_L^2\Bigl(4f(x)f'(x)f^{(L)}(x)f^{(L+1)}(x)+f'(x)^2f^{(L)}(x)^2 + f(x)^2f^{(L+1)}(x)^2\Bigl)h_n^{2L}}{f(x)^4} \\
    & \quad +  o\left\{\frac{1}{nh_n^3}+h_n^{2L}\right\}\\
    & = w^2\frac{R(K')}{nh_n^3}f(x)^{-1} + w^2[f'(x)^2f(x)^2] \\
    & \quad + 2w^2\sum_{l=L}^{2L}\kappa_l[(f'(x)f^{(l+1)}(x)f(x)^{-2}-f'(x)^2f^{(l)}(x)f(x)^{-3}]h_n^l \\
    &\quad + w^2\kappa_L^2[f^{(L+1)}(x)^2f(x)^{-2}-4f(x)f'(x)f^{(L)}(x)f^{(L+1)}(x)+3f'(x)^2f^{(L)}(x)^2f(x)^{-4}]h_n^{2L} \\
    & \quad +  o\left\{\frac{1}{nh_n^3}+h_n^{2L}\right\}
\end{align*}
Since from the result of the computation in \ref{section:bias}, the asymptotic representation of $\mathbb{E}_{X_{1:n}}\left[\frac{\partial}{\partial x}\log\hat{f}_{w,h_n}(x)\right]^2 $ is given by
\begin{align*}
    & \mathbb{E}_{X_{1:n}}\left[\frac{\partial}{\partial x}\log\hat{f}_{w,h_n}(x)\right]^2 \\
    & = \Biggl(wf'(x)f(x)^{-1} + w\sum_{l=L}^{2L}\kappa_l[f^{(l+1)}(x)f(x)^{-1}-f'(x)f^{(l)}(x)f(x)^{-2}]h_n^l \\
    &\quad + w\kappa_L^2[f'(x)f^{(L)}(x)^2f(x)^{-3}-f^{(L)}(x)f^{(L+1)}(x)f(x)^{-2}]h_n^{2L}  + O(n^{-1}h_n^{-2}) + o(h_n^{2L}) \Biggl)^2 \\
    & = w^2[f'(x)^2f(x)^2] +2w^2\sum_{l=L}^{2L}\kappa_l[f'(x)f^{(l+1)}(x)f(x)^{-2}-f'(x)^2f^{(l)}(x)f(x)^{-3}]h_n^l \\
    & \quad + w^2 \kappa_L^2[f^{(L+1)}(x)^2f(x)^{-2}-2f'(x)f^{(L)}(x)f^{(L+1)}(x)f(x)^{-3}+f'(x)^2f^{(L)}(x)^2f(x)^{-4}]h_n^{2L} \\
    & \quad + 2w^2 \kappa_L^2 [f'(x)^2f^{(L)}(x)^2f(x)^{-4}-f'(x)f^{(L)}(x)f^{(L+1)}(x)f(x)^{-3}]h_n^{2L} + O(n^{-1}h_n^{-2}) + o(h_n^{2L}) \\
    & = w^2[f'(x)^2f(x)^2] +2w^2\sum_{l=L}^{2L}\kappa_l[f'(x)f^{(l+1)}(x)f(x)^{-2}-f'(x)^2f^{(l)}(x)f(x)^{-3}]h_n^l \\
    & \quad + w^2\kappa_L^2[f^{(L+1)}(x)^2f(x)^{-2}-4f(x)f'(x)f^{(L)}(x)f^{(L+1)}(x)+3f'(x)^2f^{(L)}(x)^2f(x)^{-4}]h_n^{2L} \\
    & \quad + O(n^{-1}h_n^{-2}) + o(h_n^{2L}).
\end{align*}
then we have the asymptotic representation of $\mathbb{V}_{X_{1:n}}\left[\frac{\partial}{\partial x}\log \hat{f}_{w,h_n}(x)\right]$.
\begin{align*}
    & \mathbb{V}_{X_{1:n}}\left[\frac{\partial}{\partial x}\log \hat{f}_{w,h_n}(x)\right] =  w^2\frac{R(K')}{nh_n^3}f(x)^{-1} + o\left\{\frac{1}{nh_n^3}+h_n^{2L}\right\}.
\end{align*}

\section{Properties of Hyv\"arinen score}
In this section, we provide the detail computation to investigate the properties of empirical Hyv\"arinen score. We need the result in this section to prove Theorem \ref{thm:eh}, \ref{thm:uniform} and \ref{thm:eh_fix}.  \ref{section:derivation_of_hyvarinen_score} is the derivation process of empirical Hyv\"arinen score and the generalisation of Section A.2.2 of \cite{jewson2021general}. Section~\ref{sec:the_expectation_of_Hyvarinen_score} is detail computation of the expected Hyv{\"a}rinen Score. Section~\ref{sec:empirical_hyvarinen_score} scrutinise the empirical Hyv\"arinen score in order to prove Theorem \ref{thm:uniform}.

\subsection{Derivation of empirical Hyv\"arinen score} \label{section:derivation_of_hyvarinen_score}
As Section A.2.2 of \cite{jewson2021general}, kernel density loss function to be the log-density
\begin{align}
    & \log\hat{f}_{w,h_n}(X_i) = w \log \left\{\frac{K(0)}{nh_n}+\frac{1}{nh_n}\sum_{j\neq i}^nK\left(\frac{X_i-X_j}{h_n}\right)\right\}. \label{equation:log_density}
\end{align}

\noindent The Hyv\"arinen score is then composed of the first and second derivatives of the log-density, given by
\begin{align}
    & \frac{\partial}{\partial X_i}\log\hat{f}_{w,h_n}(X_i) = w \frac{\frac{1}{nh_n^2}\sum_{j\neq i}^n K'\left(\frac{X_i-X_j}{h_n}\right)}{\frac{K(0)}{nh_n}+\frac{1}{nh_n}\sum_{j\neq i}^nK\left(\frac{X_i-X_j}{h_n}\right)}, \nonumber\\
    & \frac{\partial^2}{\partial X_i^2}\log \hat{f}_{w,h_n}(X_i) = w \frac{\frac{1}{nh_n^3}\sum_{j\neq i}^nK''\left(\frac{X_i-X_j}{h_n}\right)}{\frac{K(0)}{nh_n}+\frac{1}{nh_n}\sum_{j\neq i}^nK\left(\frac{X_i-X_j}{h_n}\right)}-w\frac{\left\{\frac{1}{nh_n^2}\sum_{j\neq i}^nK'\left(\frac{X_i-X_j}{h_n}\right)\right\}^2}{\left\{\frac{K(0)}{nh_n}+\frac{1}{nh_n}\sum_{j\neq i}^nK\left(\frac{X_i-X_j}{h_n}\right)\right\}^2}, \nonumber
\end{align}

\noindent then, the Hyv{\"a}rinen score of the kernel density estimate is
\begin{align*}
    \mathcal{H}_i(w,h_n)
    & = 2\frac{\partial^2}{\partial X_i^2}\log \hat{f}_{w,h_n}(X_i) + \left\{\frac{\partial}{\partial X_i}\log \hat{f}_{w,h_n}(X_i)\right\}^2 \nonumber\\
    & = 2w\frac{\frac{1}{nh_n^3}\sum_{j\neq i}^nK''\left(\frac{X_i-X_j}{h_n}\right)}{\frac{K(0)}{nh_n} + \frac{1}{nh_n}\sum_{j\neq i}^nK\left(\frac{X_i-X_j}{h_n}\right)} + (w^2-2w)\frac{\left\{\frac{1}{nh_n^2}\sum_{j\neq i}^nK'\left(\frac{X_i-X_j}{h_n}\right)\right\}^2}{\left\{\frac{K(0)}{nh_n} + \frac{1}{nh_n}\sum_{j\neq i}^nK\left(\frac{X_i-X_j}{h_n}\right)\right\}^2} \nonumber\\
    & = 2w\left\{\frac{1}{h_n^2}\sum_{j\neq i}^nK''\left(\frac{X_i-X_j}{h_n}\right)\right\}\left\{K(0)+\sum_{j\neq i}^nK\left(\frac{X_i-X_j}{h_n}\right)\right\}^{-1} \\
    & \quad + (w^2-2w)\left\{\frac{1}{h_n}\sum_{j\neq i}^nK'\left(\frac{X_i-X_j}{h_n}\right)\right\}^2\left\{K(0)+\sum_{j\neq i}^nK\left(\frac{X_i-X_j}{h_n}\right)\right\}^{-2}  \nonumber\\
    & = 2w\left\{\frac{1}{h_n^2}\sum_{j\neq i}^nK''_{ij}\right\}\left\{K(0)+\sum_{j\neq i}^nK_{ij}\right\}^{-1} \\
    & \quad + (w^2-2w)\left\{\frac{1}{h_n}\sum_{j\neq i}^nK'_{ij}\right\}^2\left\{K(0)+\sum_{j\neq i}^nK_{ij}\right\}^{-2}.
\end{align*}
We introduce abbreviations for these two terms.
\begin{align*}
    & \mathcal{H}_{1i}(w,h_n) \equiv 2w\left\{\frac{1}{h_n^2}\sum_{j\neq i}^nK''_{ij}\right\}\left\{K(0)+\sum_{j\neq i}^nK_{ij}\right\}^{-1}\\
    & \mathcal{H}_{2i}(w,h_n) \equiv (w^2-2w)\left\{\frac{1}{h_n}\sum_{j\neq i}^nK'_{ij}\right\}^2\left\{K(0)+\sum_{j\neq i}^nK_{ij}\right\}^{-2}
\end{align*}
Then, we call the following statistics $\mathcal{H}$ as 'empirical Hyv\"arinen score'.
\begin{align}
    \mathcal{H}(w,h_n) \equiv \frac{1}{n}\sum_{i=1}^n \mathcal{H}_i(w,h_n) = \frac{1}{n}\sum_{i=1}^n \{\mathcal{H}_{1i}(w,h_n)+\mathcal{H}_{2i}(w,h_n)\} \nonumber
\end{align}
\subsection{Expected Hyv\"arinen score} 

\label{sec:the_expectation_of_Hyvarinen_score} 
In this subsection, we derive the asymptotic representation of expected Hyv\"arinen score.
In view of the definition of empirical Hyv\"arinen score derived in Section \ref{section:derivation_of_hyvarinen_score}, 
\begin{align*}
     \mathbb{E}_{X_{1:n}}[\mathcal{H}(w,h_n)]
    & = \mathbb{E}_{X_{1:n}}\left[ \frac{1}{n}\sum_{i=1}^n \mathbb{E}_{X_{1:n}}\big[\mathcal{H}_{1i}(w,h_n) + \mathcal{H}_{2i}(w,h_n)\mid X_i\big]\right].
\end{align*}
We compute these two term separately.

\subsubsection{$\mathbb{E}_{X_{1:n}}[\mathcal{H}_{1i}(w,h_n)]$} \label{section:eh_1}
First, in order to derive $\mathbb{E}_{X_{1:n}}[\mathcal{H}_{1i}(w,h_n)]$, we transform $\mathcal{H}_{1i}(w,h_n)$ into a tractable form.
\begin{align}
    \mathcal{H}_{1i} &= \frac{\frac{2w}{h_n^2}\sum_{j\neq i}^nK''_{ij}}{K(0)+\sum_{j\neq i}^nK_{ij}} \nonumber\\
    & = \frac{\frac{2w}{h_n^2}\left\{\mathbb{E}_{X_{1:n}}\left[\sum_{j\neq i}^nK''_{ij} \mid X_i \right]+\sum_{j\neq i}^nK''_{ij}-\mathbb{E}_{X_{1:n}}\left[\sum_{j\neq i}^nK''_{ij} \mid X_i\right]\right\}}{K(0)+\mathbb{E}_{X_{1:n}}\left[\sum_{j\neq i}^nK_{ij} \mid X_i \right]+\sum_{j\neq i}^nK_{ij}-\mathbb{E}_{X_{1:n}} \left[\sum_{j\neq i}^nK_{ij} \mid X_i \right]} \nonumber\\
    &= \frac{\frac{2w}{h_n^2}\left\{\mathbb{E}_{X_{1:n}}\left[\sum_{j\neq i}^nK''_{ij} \mid X_i \right]+\sum_{j\neq i}^nK''_{ij}-\mathbb{E}_{X_{1:n}} \left[\sum_{j\neq i}^nK''_{ij} \mid X_i\right]\right\}}{K(0)+\mathbb{E}_{X_{1:n}}\left[\sum_{j\neq i}^nK_{ij} \mid X_i \right]} \nonumber\\
    & \quad \times \left(1+\frac{\sum_{j\neq i}^nK_{ij}-\mathbb{E}_{X_{1:n}}\left[\sum_{j\neq i}^nK_{ij} \mid X_i \right]}{K(0)+\mathbb{E}_{X_{1:n}}\left[\sum_{j\neq i}^nK_{ij} \mid X_i \right]}\right)^{-1} \nonumber\\
    & = \frac{\frac{2w}{h_n^2}\mathbb{E}_{X_{1:n}}\left[\sum_{j\neq i}^nK''_{ij} \mid X_i \right]}{K(0)+\mathbb{E}_{X_{1:n}}\left[\sum_{j\neq i}^nK_{ij} \mid X_i \right]}+\frac{\frac{2w}{h_n^2}\left\{\sum_{j\neq i}^nK''_{ij}-\mathbb{E}_{X_{1:n}}\left[\sum_{j\neq i}^nK''_{ij} \mid X_i \right]\right\}}{K(0)+\mathbb{E}_{X_{1:n}}\left[\sum_{j\neq i}^nK_{ij} \mid X_i\right]} \nonumber\\
    & \qquad - \frac{\frac{2w}{h_n^2}\mathbb{E}_{X_{1:n}}\left[\sum_{j\neq i}^nK''_{ij} \mid X_i \right]\left\{\sum_{j\neq i}^nK_{ij}-\mathbb{E}_{X_{1:n}}\left[\sum_{j\neq i}^nK_{ij} \mid X_i\right]\right\}}{\left\{K(0)+\mathbb{E}_{X_{1:n}}\left[\sum_{j\neq i}^nK_{ij} \mid X_i\right]\right\}^2} \nonumber\\
    & \qquad -\frac{\frac{2w}{h_n^2}\left\{\sum_{j\neq i}^nK''_{ij}-\mathbb{E}_{X_{1:n}}\left[\sum_{j\neq i}^nK''_{ij} \mid X_i \right]\right\}\left\{\sum_{j\neq i}^nK_{ij}-\mathbb{E}_{X_{1:n}}\left[\sum_{j\neq i}^nK_{ij} \mid X_i\right]\right\}}{\left\{K(0)+\mathbb{E}_{X_{1:n}}\left[\sum_{j\neq i}^nK
    _{ij} \mid X_i \right]\right\}^2} + O_p\{(nh_n)^{-1}\} \nonumber
\end{align}
where the third equality multiplies the right hand side by \\$\left\{K(0)+\mathbb{E}_{X_{1:n}}\left[\sum_{j\neq i}^nK_{ij} \mid X_i\right]\right\}\left\{K(0)+\mathbb{E}_{X_{1:n}}\left[\sum_{j\neq i}^nK_{ij} \mid X_i\right]\right\}^{-1}=1$ and the final equlity expands\\ $\left(1+\frac{\sum_{j\neq i}^nK_{ij}-\mathbb{E}_{X_{1:n}}\left[\sum_{j\neq i}^nK_{ij} \mid X_i \right]}{K(0)+\mathbb{E}_{X_{1:n}}\left[\sum_{j\neq i}^nK_{ij} \mid X_i \right]}\right)^{-1}$ around $\sum_{j\neq i}^nK_{ij}=\mathbb{E}_{X_{1:n}}\left[\sum_{j\neq i}^nK_{ij} \mid X_i\right]$. Conditional on $X_i$, because $\frac{1}{(n-1)h_n}\sum_{j\neq i}^nK_{ij}$ is a KDE from a sample with size of $n-1$, $\frac{1}{(n-1)h_n}\sum_{j\neq i}^nK_{ij}=\mathbb{E}_{X_{1:n}}\left[\frac{1}{(n-1)h_n}\sum_{j\neq i}^nK_{ij} \mid X_i \right]$ converges. From this transformation, we have the conditional expectation of $\mathcal{H}_{1i}(w,h_n)$ given $X_i$ as follows. 

\begin{align*}
    & \mathbb{E}_{X_{1:n}}[\mathcal{H}_{1i}(w,h_n) \mid X_i] \\
    &=  \frac{\frac{2w}{h_n^2}\mathbb{E}_{X_{1:n}}\left[\sum_{j\neq i}^nK''_{ij} \mid X_i \right]}{K(0)+\mathbb{E}_{X_{1:n}}\left[\sum_{j\neq i}^nK_{ij} \mid X_i \right]} \\
    & \quad -\frac{\frac{2w}{h_n^2}\mathbb{E}_{X_{1:n}}\left[\left\{\sum_{j\neq i}^nK''_{ij}-\mathbb{E}_{X_{1:n}} \left[\sum_{j\neq i}^nK''_{ij} \mid X_i \right]\right\}\left\{\sum_{j\neq i}^nK_{ij}-\mathbb{E}_{X_{1:n}}\left[\sum_{j\neq i}^nK_{ij} \mid X_i \right]\right\} \mid X_i\right]}{\left\{K(0)+\mathbb{E}_{X_{1:n}}\left[\sum_{j\neq i}^nK_{ij} \mid X_i\right]\right\}^2} \\
    & \qquad + o_p\{(nh_n^3)^{-1}\}\\
    & = \frac{\frac{4w}{h_n^2}\mathbb{E}_{X_{1:n}}\left[\frac{1}{(n-1)h_n}\sum_{j\neq i}^nK''_{ij} \mid X_i \right]}{\mathbb{E}_{X_{1:n}}\left[\frac{1}{(n-1)h_n}\sum_{j\neq i}^nK_{ij} \mid X_i\right]} - \frac{\frac{2w}{h_n^2}\left\{\mathbb{E}_{X_{1:n}}\left[\frac{1}{(n-1)^2h_n^2}\sum_{j\neq i}^n\sum_{k\neq i}^nK''_{ij}K_{ik} \mid X_i \right]\right\}}{\left\{\mathbb{E}_{X_{1:n}}\left[\frac{1}{(n-1)h_n}\sum_{j\neq  i}^n K_{ij} \mid X_i \right]\right\}^2} + o_p\{(nh_n^3)^{-1}\} \\
\end{align*}
where the second equality multiplies the right-hand side by $\{(n-1)h_n\}\{(n-1)h_n\}^{-1}=1$, the third equality expands the denominator. Because $K(0)/\{(n-1)h_n\}=o(1)$, this expansion is convergent.

Since Lemma \ref{lem:k}, \ref{lem:k''} and \ref{lem:k''k} state that
\begin{align*}
    & \mathbb{E}_{X_{1:n}}\left[\frac{1}{(n-1)h_n}\sum_{j\neq i}^n K_{ij} \mid X_i \right] = f(X_i) + \sum_{l=L}^{2L}\kappa_l f^{(l)}(X_i)h_n^l + o_p(h_n^{2L})\\
    & \mathbb{E}_{X_{1:n}}\left[\frac{1}{(n-1)h_n}\sum_{j\neq i}^n K''_{ij} \mid X_i \right] = h^2f''(X_i) + \sum_{l=L}^{2L}\kappa_lf^{(l+2)}(X_i)h_n^{l+2} + o_p(h_n^{2L+2})\\
    & \mathbb{E}_{X_{1:n}}\left[\frac{1}{(n-1)h_n^2}\sum_{j\neq i}^n\sum_{k\neq i}^n K''_{ij}K_{ik} \mid X_i \right] =  \frac{-R(K')}{nh_n}f(X_i) + f(X_i)f''(X_i)h_n^2 \\
    & \quad + \sum_{l=L}^{2L}\kappa_{l}\Bigl(f''(X_i)f^{(l)}(X_i) + f(X_i)f^{(l+2)}(X_i)\Bigl)h_n^{l+2} + \kappa_L^2f^{(L+2)}(X_i)f^{(L)}(X_i)h_n^{2L+2} + o_p\left\{ \frac{1}{nh_n}+h_n^{2L+2} \right\}
\end{align*}
The following equality holds.
\begin{align*}
    & \mathbb{E}_{X_{1:n}}\left[\frac{1}{n}\sum_{i=1}^n\mathbb{E}_{X_{1:n}}\left[\mathcal{H}_{1i}(w,h_n) \mid X_i \right]\right] \\ 
    &= \mathbb{E}_{X_{1:n}}\left[\frac{4w}{n}\sum_{i=1}^n \frac{f''(X_i) + \sum_{l=L}^{2L}\kappa_lf^{(l+2)}(X_i)h_n^{l} + o_p(h_n^{2L})}{f(X_i) + \sum_{l=L}^{2L}\kappa_l f^{(l)}(X_i)h_n^l + o_p(h_n^{2L})}\right] \\
    & \quad - \mathbb{E}_{X_{1:n}}\left[\frac{2w}{n}\sum_{i=1}^n \frac{\frac{-R(K')}{nh_n^3}f(X_i) + f(X_i)f''(X_i)}{\Bigl(f(X_i) + \sum_{l=L}^{2L}\kappa_l f^{(l)}(X_i)h_n^l + o_p(h_n^{2L})\Bigl)^2}\right] \\
    & \quad - \mathbb{E}_{X_{1:n}}\left[\frac{2w}{n}\sum_{i=1}^n\frac{\sum_{l=L}^{2L}\kappa_{l}\Bigl(f''(X_i)f^{(l)}(X_i) + f(X_i)f^{(l+2)}(X_i)\Bigl)h_n^{l}}{\Bigl(f(X_i) + \sum_{l=L}^{2L}\kappa_l f^{(l)}(X_i)h_n^l + o_p(h_n^{2L})\Bigl)^2}\right] \\
    & \quad - \mathbb{E}_{X_{1:n}}\left[\frac{2w}{n}\sum_{i=1}^n\frac{ \kappa_L^2f^{(L+2)}(X_i)f^{(L)}(X_i)h_n^{2L} + o_p(h_n^{2L})}{\Bigl(f(X_i) + \sum_{l=L}^{2L}\kappa_l f^{(l)}(X_i)h_n^l + o_p(h_n^{2L})\Bigl)^2}\right] \\
    & = \mathbb{E}_{X_{1:n}}\left[\frac{4w}{n}\sum_{i=1}^n \frac{\Bigl(f''(X_i) + \sum_{l=L}^{2L}\kappa_lf^{(l+2)}(X_i)h_n^{l}\Bigl)}{f(X_i)} \mid X_i\right] \\
    & \quad - \mathbb{E}_{X_{1:n}}\left[\frac{4w}{n}\sum_{i=1}^n \frac{\Bigl(\sum_{l=L}^{2L}\kappa_lf''(X_i)f^{(l)}(X_i)h_n^l+ \kappa_L^2f^{(L)}(X_i)f^{(L+2)}(X_i)h_n^{2L}\Bigl)}{f(X_i)^2}\right] \\
    & \quad + \mathbb{E}_{X_{1:n}}\left[\frac{4w}{n}\sum_{i=1}^n\frac{\kappa_L^2f''(X_i)f^{(L)}(X_i)^2h_n^{2L}}{f(X_i)^3}\right]\\
    & \quad - \mathbb{E}_{X_{1:n}}\left[\frac{2w}{n}\sum_{i=1}^n \frac{\frac{-R(K')}{nh_n^3}f(X_i) + f(X_i)f''(X_i)}{f(X_i)^2}\right] \\
    & \quad +2 \mathbb{E}_{X_{1:n}}\left[\frac{2w}{n}\sum_{i=1}^n\frac{\sum_{l=L}^{2L}f(X_i)f''(X_i)f^{(l)}(X_i)h_n^l}{f(X_i)^3}\right] - 3\mathbb{E}_{X_{1:n}}\left[\frac{2w}{n}\sum_{i=1}^n\frac{f(X_i)f''(X_i)f^{(L)}(X_i)^2h_n^{2L}}{f(X_i)^4}\right]\\
    & \quad - \mathbb{E}_{X_{1:n}}\left[\frac{2w}{n}\sum_{i=1}^n\frac{\sum_{l=L}^{2L}\kappa_{l}\Bigl(f''(X_i)f^{(l)}(X_i) + f(X_i)f^{(l+2)}(X_i)\Bigl)h_n^{l} + \kappa_L^2f^{(L+2)}(X_i)f^{(L)}(X_i)h_n^{2L}}{f(X_i)^2}\right] \\
    & \quad + 2 \mathbb{E}_{X_{1:n}}\left[\frac{2w}{n}\sum_{i=1}^n \frac{\kappa_L^2\Bigl(f''(X_i)f^{(L)}(X_i)^2 + f(X_i)f^{(L)}(X_i)f^{(L+2)}(X_i)\Bigl)h_n^{2L}}{f(X_i)^3}\right] + o\left\{\frac{1}{nh_n^3}+h_n^{2L}\right\}\\
    & = \frac{2wR(K')}{nh_n^3}\mathbb{E}_{X_i}[f(X_i)^{-1}] + 2w\mathbb{E}_{X_i}[f''(X_i)f(X_i)^{-1}] \\
    & \quad + 2w\sum_{l=L}^{2L}\kappa_l\Bigl(\mathbb{E}_{X_i}[f^{(l+2)}(X_i)f(X_i)^{-1}]-\mathbb{E}[f''(X_i)f^{(l)}(X_i)f(X_i)^{-2}]\Bigl)h_n^l \\
    & \quad -2 w\kappa_L^2\Biggl(\mathbb{E}_{X_i}[f^{(L)}(X_i)f^{(L+2)}(X_i)f(X_i)^{-2}]-\mathbb{E}_{X_i}[f''(X_i)f^{(L)}(X_i)^2f(X_i)^{-3}]\Biggl)h_n^{2L} \\
    & \quad + o\left\{\frac{1}{nh_n^3}+h_n^{2L}\right\}
\end{align*}
where the second equality expands the denominators.

\subsubsection{$\mathbb{E}_{X_{1:n}}[\mathcal{H}_{2i}(w,h_n)]$} \label{section:eh_2}
Noting that conditional on $X_i$,  $\left(\frac{1}{(n-1)h_n}\sum_{j\neq i}^nK_{ij}\right)^2-\mathbb{E}_{X_{1:n}}\left[\left(\frac{1}{(n-1)h_n}\sum_{j\neq i}^nK_{ij}\right)^2 \mid X_i \right]$ converges, in the same process as $\mathcal{H}_{1i}(w,h_n)$, we can see that
\begin{align*}
    \mathbb{E}_{X_{1:n}}[\mathcal{H}_{2i}(w,h_n) \mid X_i] &= \frac{\frac{2(w^2-2w)}{h_n^2}\mathbb{E}_{X_{1:n}}\left[\frac{1}{(n-1)^2h_n^2}\sum_{j\neq i}^n\sum_{k\neq i}^n K'_{ij}K'_{ik} \mid X_i\right]}{\mathbb{E}_{X_{1:n}}\left[\left\{\frac{1}{(n-1)h_n}\sum_{j\neq i}^n K_{ij}\right\}^2 \mid X_i \right]}  \\
    & \quad - \frac{\frac{(w^2-2w)}{h_n^2}\mathbb{E}_{X_{1:n}}\left[\frac{1}{(n-1)^4h_n^4}\sum_{j\neq i}^n\sum_{k\neq i}^n\sum_{l\neq i}^n\sum_{m\neq i}^n K'_{ij}K'_{ik}K_{il}K_{im} \mid X_i \right]}{\mathbb{E}_{X_{1:n}}\left[\left\{\frac{1}{(n-1)h_n}\sum_{j\neq i}^n K_{ij}\right\}^2 \mid X_i \right]^2} \\
    & \quad + O_p\left\{ \frac{1}{nh_n} \right\}
\end{align*}
where the bound on the remainder term follows from Lemma \ref{lem:bound_remainder}. Next, 
from Lemma \ref{lem:k'k'}, \ref{lem:k_square} and \ref{lem:k'k'kk},
\begin{align*}
    & \mathbb{E}_{X_{1:n}}\left[\left\{\frac{1}{(n-1)h_n}\sum_{j\neq i}^n K_{ij}\right\}^2 \mid X_i \right] = f(X_i)^2 + 2\sum_{l=L}^{2L}\kappa_lf(X_i)f^{(l)}(X_i)h_n^l + \kappa_L^2f^{(L)}(X_i)^2h_n^{2L} +  o\left\{\frac{1}{nh_n^3}+h_n^{2L}\right\}\\
    & \mathbb{E}_{X_{1:n}}\left[\frac{1}{(n-1)^2h_n^2}\sum_{j\neq i}^n\sum_{k\neq i}^n K'_{ij}K'_{ik} \mid X_i \right] = \frac{R(K')}{nh_n}f(X_i) + f'(X_i)^2h_n^2 \\ & \quad + 2\sum_{l=L}^{2L}\kappa_lf'(X_i)f^{(l+1)}(X_i)h_n^{l+2} + \kappa_L^2f^{(L+1)}(X_i)^2h_n^{2L+2} + o_p\left\{\frac{1}{nh_n}+h_n^{2L+2}\right\}\\
    &\mathbb{E}_{X_{1:n}}\left[\frac{1}{(n-1)^4h_n^4}\sum_{j\neq i}^n\sum_{k\neq i}^n\sum_{l\neq i}^n\sum_{m\neq i}^n K'_{ij}K'_{ik}K_{il}K_{im} \mid X_i \right] = \frac{R(K')}{nh_n}f(X_i)^3 + f'(X_i)^2f(X_i)^2h_n^2 \\
    & \quad + 2\sum_{l=L}^{2L}\kappa_l\Bigl(f(X_i)f'(X_i)^2f^{(l)}(X_i)+f(X_i)^2f'(X_i)f^{(l+1)}(X_i)\Bigl)h_n^{l+2} \\
    & \quad + \kappa_L^2\Bigl(4f(X_i)f'(X_i)f^{(L)}(X_i)f^{(L+1)}(X_i) + f'(X_i)^2f^{(L)}(X_i)^2 + f(X_i)^2f^{(L+1)}(X_i)^2\Bigl)h_n^{2L+2} \\
    & \quad + o_p\left\{\frac{1}{nh_n}+h_n^{2L+2}\right\}
\end{align*}
These imply that
\begin{align*}
    & \mathbb{E}_{X_{1:n}}\left[\frac{1}{n}\sum_{i=1}^n\mathbb{E}_{X_{1:n}}\left[\mathcal{H}_{2i}(w,h_n) \mid X_i \right]\right] \\ 
    & = \mathbb{E}_{X_{1:n}}\left[\frac{2(w^2-2w)}{n}\sum_{i=1}^n \frac{\frac{R(K')}{nh_n^3}f(X_i) + f'(X_i)^2 + 2\sum_{l=L}^{2L}\kappa_lf'(X_i)f^{(l+1)}(X_i)h_n^{l}}{f(X_i)^2 + 2\sum_{l=L}^{2L}\kappa_lf(X_i)f^{(l)}(X_i)h_n^l + \kappa_L^2f^{(L)}(X_i)^2h_n^{2L} +  o\left\{\frac{1}{nh_n^3}+h_n^{2L}\right\}}\right] \\
    & \quad + \mathbb{E}_{X_{1:n}}\left[\frac{2(w^2-2w)}{n}\sum_{i=1}^n \frac{ \kappa_L^2f^{(L+1)}(X_i)^2h_n^{2L} + o_p\left\{\frac{1}{nh_n^3}+h_n^{2L}\right\}}{f(X_i)^2 + 2\sum_{l=L}^{2L}\kappa_lf(X_i)f^{(l)}(X_i)h_n^l + \kappa_L^2f^{(L)}(X_i)^2h_n^{2L} +  o\left\{\frac{1}{nh_n^3}+h_n^{2L}\right\}}\right] \\
    & \quad - \mathbb{E}_{X_{1:n}}\left[\frac{(w^2-2w)}{n}\sum_{i=1}^n \frac{\frac{R(K')}{nh_n^3}f(X_i)^3 + f'(X_i)^2f(X_i)^2}{\Bigl(f(X_i)^2 + 2\sum_{l=L}^{2L}\kappa_lf(X_i)f^{(l)}(X_i)h_n^l + \kappa_L^2f^{(L)}(X_i)^2h_n^{2L} +  o\left\{\frac{1}{nh_n^3}+h_n^{2L}\right\}\Bigl)^2 }\right] \\
    & \quad - \mathbb{E}_{X_{1:n}}\left[\frac{(w^2-2w)}{n}\sum_{i=1}^n \frac{ 2\sum_{l=L}^{2L}\kappa_l\Bigl(f(X_i)f'(X_i)^2f^{(l)}(X_i)+f(X_i)^2f'(X_i)f^{(l+1)}(X_i)\Bigl)h_n^{l}}{\Bigl(f(X_i)^2 + 2\sum_{l=L}^{2L}\kappa_lf(X_i)f^{(l)}(X_i)h_n^l + \kappa_L^2f^{(L)}(X_i)^2h_n^{2L} +  o\left\{\frac{1}{nh_n^3}+h_n^{2L}\right\}\Bigl)^2 }\right] \\
    & \quad - \mathbb{E}_{X_{1:n}}\left[\frac{(w^2-2w)}{n}\sum_{i=1}^n \frac{4\kappa_L^2f(X_i)f'(X_i)f^{(L)}(X_i)f^{(L+1)}(X_i)h_n^{2L}}{\Bigl(f(X_i)^2 + 2\sum_{l=L}^{2L}\kappa_lf(X_i)f^{(l)}(X_i)h_n^l + \kappa_L^2f^{(L)}(X_i)^2h_n^{2L} +  o\left\{\frac{1}{nh_n^3}+h_n^{2L}\right\}\Bigl)^2 }\right] \\
    & \quad - \mathbb{E}_{X_{1:n}}\left[\frac{(w^2-2w)}{n}\sum_{i=1}^n \frac{\kappa_L^2\Bigl(f'(X_i)^2f^{(L)}(X_i)^2 + f(X_i)^2f^{(L+1)}(X_i)^2\Bigl)h_n^{2L}+ o_p\left\{\frac{1}{nh_n^3}+h_n^{2L}\right\}}{\Bigl(f(X_i)^2 + 2\sum_{l=L}^{2L}\kappa_lf(X_i)f^{(l)}(X_i)h_n^l + \kappa_L^2f^{(L)}(X_i)^2h_n^{2L} +  o\left\{\frac{1}{nh_n^3}+h_n^{2L}\right\}\Bigl)^2 }\right] \\
    & = \mathbb{E}_{X_{1:n}}\left[\frac{2(w^2-2w)}{n}\sum_{i=1}^n \frac{\frac{R(K')}{nh_n^3}f(X_i) + f'(X_i)^2 + 2\sum_{l=L}^{2L}\kappa_lf'(X_i)f^{(l+1)}(X_i)h_n^{l} + \kappa_L^2f^{(L+1)}(X_i)^2h^{2L}}{f(X_i)^2 }\right] \\
    & \quad - \mathbb{E}_{X_{1:n}}\left[\frac{2(w^2-2w)}{n}\sum_{i=1}^n \frac{ 2\sum_{l=L}^{2L}\kappa_lf(X_i)f'(X_i)^2f^{(l)}(X_i)h_n^l}{f(X_i)^4}\right] \\
    & \quad - \mathbb{E}_{X_{1:n}}\left[\frac{2(w^2-2w)}{n}\sum_{i=1}^n \frac{ \kappa_L^2\Bigl(f'(X_i)^2f^{(L)}(X_i)^2+4f(X_i)f'(X_i)f^{(L)}(X_i)f^{(L+1)}(X_i)\Bigl)h_n^{2L}}{f(X_i)^4}\right] \\
    & \quad + \mathbb{E}_{X_{1:n}}\left[\frac{2(w^2-2w)}{n}\sum_{i=1}^n\frac{4\kappa_L^2f(X_i)^2f'(X_i)^2f^{(L)}(X_i)^2h_n^{2L}}{f(X_i)^6}\right] \\
    & \quad - \mathbb{E}_{X_{1:n}}\left[\frac{(w^2-2w)}{n}\sum_{i=1}^n \frac{\frac{R(K')}{nh_n^3}f(X_i)^3 + f'(X_i)^2f(X_i)^2}{f(X_i)^4}\right] \\
    & \quad - \mathbb{E}_{X_{1:n}}\left[\frac{(w^2-2w)}{n}\sum_{i=1}^n \frac{ 2\sum_{l=L}^{2L}\kappa_l\Bigl(f(X_i)f'(X_i)^2f^{(l)}(X_i)+f(X_i)^2f'(X_i)f^{(l+1)}(X_i)\Bigl)h_n^{l}}{f(X_i)^4}\right] \\
    & \quad + 2\mathbb{E}_{X_{1:n}}\left[\frac{(w^2-2w)}{n}\sum_{i=1}^n \frac{ 2\sum_{l=L}^{2L}\kappa_lf'(X_i)^2f(X_i)^3f^{(l)}(X_i)h_n^l}{f(X_i)^6}\right] \\
    & \quad + 2\mathbb{E}_{X_{1:n}}\left[\frac{(w^2-2w)}{n}\sum_{i=1}^n \frac{\kappa_L^2\Bigl( 5f(X_i)^2f'(X_i)^2f^{(L)}(X_i)^2+4f(X_i)^3f'(X_i)f^{(L)}(X_i)f^{(L+1)}(X_i)\Bigl)h_n^{2L}}{f(X_i)^6}\right] \\
    & \quad - 3\mathbb{E}_{X_{1:n}}\left[\frac{(w^2-2w)}{n}\sum_{i=1}^n\frac{4\kappa_L^2f(X_i)^4f'(X_i)^2f^{(L)}(X_i)^2h_n^{2L}}{f(X_i)^8}\right]\\
    & \quad - \mathbb{E}_{X_{1:n}}\left[\frac{(w^2-2w)}{n}\sum_{i=1}^n \frac{4\kappa_L^2f(X_i)f'(X_i)f^{(L)}(X_i)f^{(L+1)}(X_i)h_n^{2L}}{f(X_i)^4 }\right] \\
    & \quad - \mathbb{E}_{X_{1:n}}\left[\frac{(w^2-2w)}{n}\sum_{i=1}^n \frac{\kappa_L^2\Bigl( f'(X_i)^2f^{(L)}(X_i)^2 + f(X_i)^2f^{(L+1)}(X_i)^2\Bigl)h_n^{2L}}{f(X_i)^4 }\right] + o\left\{\frac{1}{nh_n^3}+h_n^{2L}\right\} \\
    & = \frac{(w^2-2w)R(K')}{nh^3}\mathbb{E}_{X_i}[f(X_i)^{-1}] + (w^2-2w)\mathbb{E}_{X_i}[f'(X_i)^2f(X_i)^{-2}] \\
    & \quad + 2(w^2-2w)\sum_{l=L}^{2L}\kappa_l\Bigl(\mathbb{E}_{X_i}[f'(X_i)f^{(L+1)}(X_i)f(X_i)^{-2}]-\mathbb{E}_{X_i}[f'(X_i)^2f^{(L)}(X_i)f(X_i)^{-3}]\Bigl)h_n^l \\
    & \quad + (w^2-2w)\kappa_L^2\Bigl(\mathbb{E}_{X_i}[f^{(L+1)}(X_i)^2f(X_i)^{-2}]\\
    & \quad \quad +3\mathbb{E}_{X_i}[f'(X_i)^2f^{(L)}(X_i)^2f(X_i)^{-4}] -4\mathbb{E}_{X_i}[f'(X_i)f^{(L)}(X_i)f^{(L+1)}(X_i)f(X_i)^{-3}]\Bigl) h_n^{2L} \\
    & \quad + o\left\{\frac{1}{nh_n^3}+h_n^{2L}\right\}
\end{align*}
where the third equality expands the denominators.

\subsection{Empirical Hyv\"arinen score} \label{sec:empirical_hyvarinen_score}
In order to get the asymptotic representation of the process $\mathcal{H}(w_n,h_n)-\mathbb{E}_{X_{1:n}}[\mathcal{H}(w_n,h_n)]$, in other words, to derive $Q_n$ and to bound $\mathbb{E}_{X_{1:n}}[R_n(w_n,h_n)^2]$, we expand the empirical Hyv\"arinen score $\mathcal{H}(w_n,h_n)$. Recall that $\mathcal{H}(w_n,h_n)$ is given by as follows.
    \begin{align}
    \mathcal{H}(w_n,h_n) &= \frac{1}{n}\sum_{i=1}^n\left[2w_n\frac{\frac{1}{(n-1)h_n^3}\sum_{j\neq i}^nK''_{ij}}{\frac{K(0)}{(n-1)h_n}+\frac{1}{(n-1)h_n}\sum_{j\neq i}K_{ij}} + (w_n^2-2w_n)\frac{\left\{\frac{1}{(n-1)h_n^2}\sum_{j\neq i}^nK'_{ij}\right\}^2}{\left\{\frac{K(0)}{(n-1)h_n}+\frac{1}{(n-1)h_n}\sum_{j\neq i}K_{ij}\right\}^2}\right] \nonumber
    \end{align}
    Then, expanding $\mathcal{H}(w_n,h_n)$ around $K(0)/[(n-1)h_n]$ yields
    \begin{align*}
    & \mathcal{H}(w_n,h_n) \\
    & = \frac{1}{n}\sum_{i=1}^n\left[\frac{2w_n\left(\frac{1}{(n-1)h_n^3}\sum_{j\neq i}^n K''_{ij}\right)\left(\frac{1}{(n-1)h_n}\sum_{j\neq i}^n K_{ij}\right)+(w_n^2-2w_n)\left(\frac{1}{(n-1)h_n^2}\sum_{j\neq i}^n K'_{ij}\right)^2}{\left(\frac{1}{(n-1)h_n}\sum_{j\neq i}^n K_{ij}\right)^2}\right] \\
    & \quad + O_p\{(nh_n)^{-1}\}, \\
    & = \frac{1}{n}\sum_{i=1}^n\left[\frac{2w_n\left(\frac{1}{(n-1)h_n^3}\sum_{j\neq i}^n K''_{ij}\right)\left(\frac{1}{(n-1)h_n}\sum_{j\neq i}^n K_{ij}\right)+(w_n^2-2w_n)\left(\frac{1}{(n-1)h_n^2}\sum_{j\neq i}^n K'_{ij}\right)^2}{\mathbb{E}_{X_{1:n}}\left[\left(\frac{1}{(n-1)h_n}\sum_{j\neq i}^n K_{ij}\right)^2 \mid X_i \right]+\left(\frac{1}{(n-1)h_n}\sum_{j\neq i}^n K_{ij}\right)^2-\mathbb{E}_{X_{1:n}}\left[\left(\frac{1}{(n-1)h_n}\sum_{j\neq i}^n K_{ij}\right)^2 \mid X_i \right]}\right]  \\
    & \quad + O_p\{(nh_n)^{-1}\}.
\end{align*}
    Next, we expand the denominator in the bracket around \\$\left(\frac{1}{(n-1)h_n}\sum_{j\neq i}^n K_{ij}\right)^2-\mathbb{E}_{X_{1:n}}\left[\left(\frac{1}{(n-1)h_n}\sum_{j\neq i}^n K_{ij}\right)^2 \mid X_i \right]=0$. From Lemma \ref{lem:bound_remainder}, 
    \begin{align*}
    \mathbb{E}_{X_{1:n}}\left[\left\{\left(\frac{1}{(n-1)h_n}\sum_{j\neq i}^n K_{ij}\right)^2-\mathbb{E}_{X_{1:n}}\left[\left(\frac{1}{(n-1)h_n}\sum_{j\neq i}^n K_{ij}\right)^2 \mid X_i \right]\right\}^4 \right]^{1/2} = O\{(nh_n)^{-1}\}, \end{align*}
    so the expansion is valid. In addition, we have straightforwardly the bound of 
    \begin{align*}
        & \mathbb{E}_{X_{1:n}}\left[\mathbb{E}_{X_{1:n}}\left[\left(\frac{1}{(n-1)h_n}\sum_{j\neq i}^n K_{ij}\right)^2 \mid X_i \right]^{-6}\right.\\
        & \quad \times \left. \left\{2w_n\left(\frac{1}{(n-1)h_n^3}\sum_{j\neq i}^n K''_{ij}\right)\left(\frac{1}{(n-1)h_n}\sum_{j\neq i}^n K_{ij}\right)+(w_n^2-2w_n)\left(\frac{1}{(n-1)h_n^2}\sum_{j\neq i}^n K'_{ij}\right)^2\right\}^2\right]^{1/2}\\
        & =O(1),
    \end{align*}
    then Cauchy-Schwarz's inequality guarantees
    \begin{align*}
        &\mathbb{V}_{X_{1:n}}\left[\mathbb{E}_{X_{1:n}}\left[\left(\frac{1}{(n-1)h_n}\sum_{j\neq i}^n K_{ij}\right)^2 \mid X_i \right]^{-3} \right.\\
        & \quad \times \left\{2w_n\left(\frac{1}{(n-1)h_n^3}\sum_{j\neq i}^n K''_{ij}\right)\left(\frac{1}{(n-1)h_n}\sum_{j\neq i}^n K_{ij}\right)+(w_n^2-2w_n)\left(\frac{1}{(n-1)h_n^2}\sum_{j\neq i}^n K'_{ij}\right)^2\right\} \\
        & \quad \times \left.\left\{\left(\frac{1}{(n-1)h_n}\sum_{j\neq i}^n K_{ij}\right)^2-\mathbb{E}_{X_{1:n}}\left[\left(\frac{1}{(n-1)h_n}\sum_{j\neq i}^n K_{ij}\right)^2 \mid X_i \right]\right\}^2\right] = O\{(nh_n)^{-1}\} = o\{(nh_n^3)^{-1}\}
    \end{align*}
    This implies that it is sufficient to consider the expansion up to the second term. Then, we have
\begin{align*}
    \mathcal{H}(w_n,h_n) & = \frac{1}{n}\sum_{i=1}^n\left[\frac{2w_n\left(\frac{1}{(n-1)h_n^3}\sum_{j\neq i}^n K''_{ij}\right)\left(\frac{1}{(n-1)h_n}\sum_{j\neq i}^n K_{ij}\right)+(w_n^2-2w_n)\left(\frac{1}{(n-1)h_n^2}\sum_{j\neq i}^n K'_{ij}\right)^2}{\mathbb{E}_{X_{1:n}}\left[\left(\frac{1}{(n-1)h_n}\sum_{j\neq i}^n K_{ij}\right)^2 \mid X_i \right]}\right] \\
    & \quad - \frac{1}{n}\sum_{i=1}^n\left[\frac{\left[2w_n\left(\frac{1}{(n-1)h_n^3}\sum_{j\neq i}^n K''_{ij}\right)\left(\frac{1}{(n-1)h_n}\sum_{j\neq i}^n K_{ij}\right)+(w_n^2-2w_n)\left(\frac{1}{(n-1)h_n^2}\sum_{j\neq i}^n K'_{ij}\right)^2\right]}{\mathbb{E}_{X_{1:n}}\left[\left(\frac{1}{(n-1)h_n}\sum_{j\neq i}^n K_{ij}\right)^2 \mid X_i \right]^2} \right.\\
    & \quad\quad \times \left. \left(\frac{1}{(n-1)h_n}\sum_{j\neq i}^n K_{ij}-\mathbb{E}_{X_{1:n}}\left[\frac{1}{(n-1)h_n}\sum_{j\neq i}^n K_{ij} \mid X_i \right]\right)\right] +  O_p\{(nh_n)^{-1}\} + o_p\{(nh_n^3)^{-1}\} \\
    & = \frac{2}{n}\sum_{i=1}^n\left[\frac{2w_n\left(\frac{1}{(n-1)h_n^3}\sum_{j\neq i}^n K''_{ij}\right)\left(\frac{1}{(n-1)h_n}\sum_{j\neq i}^n K_{ij}\right)+(w_n^2-2w_n)\left(\frac{1}{(n-1)h_n^2}\sum_{j\neq i}^n K'_{ij}\right)^2}{\mathbb{E}_{X_{1:n}}\left[\left(\frac{1}{(n-1)h_n}\sum_{j\neq i}^n K_{ij}\right)^2 \mid X_i \right]}\right] \\
    & \quad - \frac{1}{n}\sum_{i=1}^n\left[\frac{\left[2w_n\left(\frac{1}{(n-1)h_n^3}\sum_{j\neq i}^n K''_{ij}\right)\left(\frac{1}{(n-1)h_n}\sum_{j\neq i}^n K_{ij}\right)\right]\left[\frac{1}{(n-1)h_n}\sum_{j\neq i}^n K_{ij}\right]}{\mathbb{E}_{X_{1:n}}\left[\left(\frac{1}{(n-1)h_n}\sum_{j\neq i}^n K_{ij}\right)^2 \mid X_i \right]^2}\right] \\
    & \quad - \frac{1}{n}\sum_{i=1}^n\left[\frac{\left[(w_n^2-2w_n)\left(\frac{1}{(n-1)h_n^2}\sum_{j\neq i}^n K'_{ij}\right)^2\right]\left[\frac{1}{(n-1)h_n}\sum_{j\neq i}^n K_{ij}\right]}{\mathbb{E}_{X_{1:n}}\left[\left(\frac{1}{(n-1)h_n}\sum_{j\neq i}^n K_{ij}\right)^2 \mid X_i \right]^2}\right] + o_p\{(nh_n^3)^{-1}\} \\
    & \equiv (D.I) + (D.II)  + o_p\{(nh_n^3)^{-1}\}
\end{align*}
In the following, we transform $(D.I)$ and $(D.II)$ into a tractable form. First, we deal with $(D.I)$.
\begin{align}
    (D.I) &= \frac{2}{n}\sum_{i=1}^n\left[\frac{2w_n\left(\frac{1}{(n-1)h_n^3}\sum_{j\neq i}^n K''_{ij}\right)\left(\frac{1}{(n-1)h_n}\sum_{j\neq i}^n K_{ij}\right)+(w_n^2-2w_n)\left(\frac{1}{(n-1)h_n^2}\sum_{j\neq i}^n K'_{ij}\right)^2}{\mathbb{E}_{X_{1:n}}\left[\left(\frac{1}{(n-1)h_n}\sum_{j\neq i}^n K_{ij}\right)^2 \mid X_i \right]}\right] \nonumber \\
    & = \frac{2}{n}\sum_{i=1}^n \frac{\frac{2w_n}{(n-1)^2h_n^4}\sum_{j\neq i}^n\sum_{k\neq i}^nK''_{ij}K_{ik} + \frac{(w_n^2-2w_n)}{(n-1)^2h_n^4}\sum_{j\neq i}^n\sum_{k\neq i}^n K'_{ij}K'_{ik}}{\left(f(X_i)+\sum_{l=L}^{2L}\kappa_lf^{(l)}(X_i)h_n^l+o_p(h_n^{2L})\right)^2 + O_p\{(nh_n)^{-1}\}} \nonumber\\
    & = \frac{2}{n(n-1)(n-2)h_n^4}\underset{j\neq k\neq i}{\sum\sum\sum} \{2w_nK''_{ij}K_{ik}+(w_n^2-2w_n)K'_{ij}K'_{ik}\}f(X_i)^{-2} + o_p\{(D.I)\} \label{equation:d1}
\end{align}
Letting the summand in the right hand side of (\ref{equation:d1}) be $I_{ijk}\equiv \{2w_nK''_{ij}K_{ik}+(w_n^2-2w_n)K'_{ij}K'_{ik}\}f(X_i)^{-2}$ and its symmetrised version be $\widetilde{I}_{ijk}\equiv\frac{1}{3!}\sum_{\sigma}I_{\sigma(1)\sigma(2)\sigma(3)}$, then we have U-statistics representation of the leading term of $(D.I)$ as
\begin{align*}
    &\frac{2}{n(n-1)(n-2)h_n^4}\underset{j\neq k\neq i}{\sum\sum\sum} \widetilde{I}_{ijk} \\
    = & \frac{3!\times 2}{n(n-1)(n-2)h_n^4}\underset{ i<j<k}{\sum\sum\sum} \widetilde{I}_{ijk} = \frac{2}{h^4}\binom{n}{3}^{-1}\underset{ i<j<k}{\sum\sum\sum} \widetilde{I}_{ijk}.
\end{align*}
Since the leading term of $(D.I)$ has third-order U-statistics form, Hoeffding-decomposition yields
\begin{align}
    (D.I) &= \frac{2}{h^4}\Biggl[\mathbb{E}_{X_{1:n}}[\widetilde{I}_{ijk}]+\frac{3}{n}\sum_{i=1}^n\left\{\mathbb{E}_{X_{1:n}}[\widetilde{I}_{ijk}|X_i]-\mathbb{E}_{X_{1:n}}[\widetilde{I}_{ijk}]\right\} \nonumber\\
    & \quad +\frac{6}{n(n-1)}\sum_{i=1}^{n-1}\sum_{j=i+1}^n \left\{\mathbb{E}_{X_{1:n}}[\widetilde{I}_{ijk}|X_i,X_j]-\mathbb{E}_{X_{1:n}}[\widetilde{I}_{ijk}|X_i]-\mathbb{E}_{X_{1:n}}[\widetilde{I}_{ijk}|X_j]+\mathbb{E}_{X_{1:n}}[\widetilde{I}_{ijk}]\right\} \nonumber\\
    & \quad + O_p(n^{-3/2}h_n)\Biggl]. \nonumber
\end{align}
In the same process as $(D.I)$, we have U-statistic representation of $(D.II)$ and its Hoeffding-decomposition.
\begin{align}
    (D.II) &= - \frac{1}{n}\sum_{i=1}^n\left[\frac{\left[2w_n\left(\frac{1}{(n-1)h_n^3}\sum_{j\neq i}^n K''_{ij}\right)\left(\frac{1}{(n-1)h_n}\sum_{j\neq i}^n K_{ij}\right)\right]\left[\frac{1}{(n-1)h_n}\sum_{j\neq i}^n K_{ij}\right]}{\mathbb{E}_{X_{1:n}}\left[\left(\frac{1}{(n-1)h_n}\sum_{j\neq i}^n K_{ij}\right)^2 \mid X_i \right]^2}\right] \nonumber \\
    & \quad - \frac{1}{n}\sum_{i=1}^n\left[\frac{\left[(w_n^2-2w_n)\left(\frac{1}{(n-1)h_n^2}\sum_{j\neq i}^n K'_{ij}\right)^2\right]\left[\frac{1}{(n-1)h_n}\sum_{j\neq i}^n K_{ij}\right]}{\mathbb{E}_{X_{1:n}}\left[\left(\frac{1}{(n-1)h_n}\sum_{j\neq i}^n K_{ij}\right)^2 \mid X_i \right]^2}\right] \nonumber \\
    & = -\frac{1}{n}\sum_{i=1}^n \frac{\frac{2w_n}{(n-1)^4h_n^6}\sum_{j\neq i}^n\sum_{k\neq i}^n\sum_{l\neq i}^n\sum_{m\neq i}^nK''_{ij}K_{ik}K_{il}K_{im}}{\left[\left(f(X_i)+\sum_{l=L}^{2L}\kappa_lf^{(l)}(X_i)h_n^l+o_p(h_n^{2L})\right)^2 + O_p\{(nh_n)^{-1}\}\right]^2} \nonumber\\
    & \quad -\frac{1}{n}\sum_{i=1}^n \frac{ \frac{(w_n^2-2w_n)}{(n-1)^4h_n^6}\sum_{k\neq i}^n\sum_{l\neq i}^n\sum_{m\neq i}^nK'_{ij}K'_{ik}K_{il}K_{im}}{\left[\left(f(X_i)+\sum_{l=L}^{2L}\kappa_lf^{(l)}(X_i)h_n^l+o_p(h_n^{2L})\right)^2 + O_p\{(nh_n)^{-1}\}\right]^2} \nonumber\\
    & = \frac{-1}{n(n-1)(n-2)(n-3)(n-4)h_n^6} \nonumber\\
    & \quad\quad \times\underset{j\neq k\neq l\neq m\neq i}{\sum\sum\sum\sum\sum}\{2w_nK''_{ij}K_{ik}K_{il}K_{im}+(w_n^2-2w_n)K'_{ij}K'_{ik}K_{il}K_{im}\}f(X_i)^{-4} \nonumber\\
    & \quad + o_p\{(D.II)\} \label{equation:d2}
\end{align}
Letting the summand in the right hand side of (\ref{equation:d2}) be $II_{ijklm}\equiv \{2w_nK''_{ij}K_{ik}K_{il}K_{im}+(w_n^2-2w_n)K'_{ij}K'_{ik}K_{il}K_{im}\}f(X_i)^{-2}$ and its symmetrised version be $\widetilde{II}_{ijk}\equiv\frac{1}{5!}\sum_{\sigma}II_{\sigma(1)\sigma(2)\sigma(3)\sigma(4)\sigma(5)}$, then we have U-statistics representation of the leading term of $(D.II)$ as
\begin{align*}
     & \frac{-1}{n(n-1)(n-2)(n-3)(n-4)h_n^6}\underset{j\neq k\neq l\neq m\neq i}{\sum\sum\sum\sum\sum} \widetilde{II}_{ijklm} = \frac{-1}{h_n^6}\binom{n}{5}^{-1}\underset{i<j<k<l<m}{\sum\sum\sum\sum\sum} \widetilde{II}_{ijklm}
\end{align*}
Hoeffding-decomposition yields
\begin{align}
    (D.II) &= \frac{-1}{h_n^6}\Biggl[\mathbb{E}_{X_{1:n}}[\widetilde{II}_{ijklm}]+\frac{5}{n}\sum_{i=1}^n\left\{\mathbb{E}_{X_{1:n}}[\widetilde{II}_{ijklm}|X_i]-\mathbb{E}_{X_{1:n}}[\widetilde{II}_{ijklm}]\right\} \nonumber\\
    & \quad +\frac{20}{n(n-1)}\sum_{i=1}^{n-1}\sum_{j=i+1}^n \left\{\mathbb{E}_{X_{1:n}}[\widetilde{II}_{ijklm}|X_i,X_j]-\mathbb{E}_{X_{1:n}}[\widetilde{II}_{ijklm}|X_i]-\mathbb{E}_{X_{1:n}}[\widetilde{II}_{ijklm}|X_j]+\mathbb{E}_{X_{1:n}}[\widetilde{II}_{ijklm}]\right\} \nonumber\\
    & \quad + O_p(n^{-3/2}h_n^3)\Biggl]. \nonumber
\end{align}
\begin{rem}
The bound on the cubic term follows from Lemma \ref{Lemma:fifth_order_U_evaluation}. This bound is not sharp but sufficient for our aim, because obviously $n^{-3/2}h_n^{-3}\ll n^{-1}h_n^{-3}$.
\end{rem}

Since $[R_n(w_n,h_n) = \mathcal{H}(w_n,h_n) - \mathbb{E}_{X_{1:n}}[\mathcal{H}(w_n,h_n)] - Q_n]$ and $\mathbb{E}_{X_{1:n}}[R_n(w_n,h_n)]=0$, in order to bound $\mathbb{E}_{X_{1:n}}[R_n(w_n,h_n)^2]$,  we compute the convergence rate of the variance of $[(D.I) - \frac{2}{h_n^4}\mathbb{E}_{X_{1:n}}[\widetilde{I}_{ijk}] - Q_{1,n}]$ and $[(D.II) + \frac{1}{h_n^6}\mathbb{E}_{X_{1:n}}[\widetilde{II}_{ijklm}] - Q_{2,n}]$ and their covariance.

The constant terms $\mathbb{E}_{X_{1:n}}[\widetilde{I}_{ijk}]$, and $\mathbb{E}_{X_{1:n}}[\widetilde{II}_{ijklm}]$ are cancelled out by the corresponding term of $\mathbb{E}_{X_{1:n}}[\mathcal{H}(w_n,h_n)]$, we only consider the other terms. Define $Q_n\equiv n^{-1}\sum_{i=1}^n \{2f''(X_i)f(X_i)^{-1}-f'(X_i)^2f(X_i)^{-2}-\mathbb{E}_{X_i}[\{2f''(X_i)f(X_i)^{-1}-f'(X_i)^2f(X_i)^{-2}]\}$. In addition, letting $\rho_{1}(X_i)$ and $\rho_2(X_i)$ be some functions of $X_i$, then from Lemma \ref{lem:hp_1} and \ref{lem:hp_2},
\begin{align*}
    & \frac{6}{nh^4}\sum_{i=1}^n\left\{\mathbb{E}_{X_{1:n}}[\widetilde{I}_{ijk}|X_i]-\mathbb{E}_{X_{1:n}}[\widetilde{I}_{ijk}]\right\} \\
    &= 6Q_{n} + \frac{6}{n}\sum_{i=1}^n\left\{h^L\rho_{1,1}(X_i)+(w_n-1)\rho_{1,2}(X_i)-\mathbb{E}_{X_i}[h_n^L\rho_{1,1}(X_i)+(w_n-1)\rho_{1,2}(X_i)]\right\}\\
    & \quad + o_p\{n^{-1/2}h_n^L+n^{-1/2}(w_n-1)\} \\
    &  \frac{-5}{nh_n^6}\sum_{i=1}^n\left\{\mathbb{E}_{X_{1:n}}[\widetilde{II}_{ijklm}|X_i]-\mathbb{E}_{X_{1:n}}[\widetilde{II}_{ijklm}]\right\} \\
    & = -5Q_{n} - \frac{5}{n}\sum_{i=1}^n\left\{h^L\rho_{2,1}(X_i)+(w_n-1)\rho_{2,2}(X_i)-\mathbb{E}_{X_i}[h_n^L\rho_{2,1}(X_i)+(w_n-1)\rho_{2,2}(X_i)]\right\} \\
    & \quad + o_p\{n^{-1/2}h_n^L+n^{-1/2}(w_n-1)\}
\end{align*}
these imply that
\begin{align}
    &(D.I) - \frac{2}{h^4}\mathbb{E}_{X_{1:n}}[\widetilde{I}_{ijk}] - 6Q_{n} \nonumber\\
    &= \frac{6}{n}\sum_{i=1}^n\left\{h_n^L\rho_{1,1}(X_i)+(w_n-1)\rho_{1,2}(X_i)-\mathbb{E}_{X_i}[h_n^L\rho_{1,1}(X_i)+(w_n-1)\rho_{1,2}(X_i)]\right\} \nonumber\\
    & \quad + \frac{12}{n(n-1)h_n^4}\sum_{i=1}^{n-1}\sum_{j=i+1}^n \left\{\mathbb{E}_{X_{1:n}}[\widetilde{I}_{ijk}|X_i,X_j]-\mathbb{E}_{X_{1:n}}[\widetilde{I}_{ijk}|X_i]-\mathbb{E}_{X_{1:n}}[\widetilde{I}_{ijk}|X_j]+\mathbb{E}_{X_{1:n}}[\widetilde{I}_{ijk}]\right\} \nonumber\\
    & \quad + O_p(n^{-3/2}h_n^{-3}) + o_p\{n^{-1/2}h_n^L+n^{-1/2}(w_n-1)\} \label{equation:empirical_hyvarinen_1}
\end{align}
\begin{align}
    &(D.II) + \frac{1}{h^6}\mathbb{E}_{X_{1:n}}[\widetilde{II}_{ijklm}] + 5Q_{n} \nonumber\\
    &= \frac{-5}{n}\sum_{i=1}^n\left\{h_n^L\rho_{2,1}(X_i)+(w_n-1)\rho_{2,2}(X_i)-\mathbb{E}_{X_i}[h_n^L\rho_{2,1}(X_i)+(w_n-1)\rho_{2,2}(X_i)]\right\}\nonumber\\
    & \quad - \frac{20}{n(n-1)h_n^4}\sum_{i=1}^{n-1}\sum_{j=i+1}^n \left\{\mathbb{E}_{X_{1:n}}[\widetilde{II}_{ijklm}|X_i,X_j]-\mathbb{E}_{X_{1:n}}[\widetilde{II}_{ijklm}|X_i]-\mathbb{E}_{X_{1:n}}[\widetilde{II}_{ijklm}|X_j]+\mathbb{E}_{X_{1:n}}[\widetilde{II}_{ijklm}]\right\} \nonumber\\
    & \quad + O_p(n^{-3/2}h_n^{-3}) + o_p\{n^{-1/2}h_n^L+n^{-1/2}(w_n-1)\} \label{equation:empirical_hyvarinen_2}
\end{align}

First, we compute the variace of $(D.I) - \frac{2}{h_n^4}\mathbb{E}_{X_{1:n}}[\widetilde{I}_{ijk}] - 6Q_{n}$. As a standard property of Hoefdding-decomposition, the covariance of the terms in the decomposition are all $0$. Obviously, the variance of the first term is $O_p\{n^{-1}h_n^{2L}+n^{-1}(w_n-1)h_n^L+n^{-1}(w_n-1)^2\}$. In addition, Lemma \ref{lem:quad_1} states that $\mathbb{E}_{X_{1:n}}[\widetilde{I}_{ijk}|X_i,X_j]=O_p(h_n^{3/2})$, combined with Lemma \ref{Lemma:third_order_U_evaluation}, the variance of $(D.I) - \frac{2}{h_n^4}\mathbb{E}_{X_{1:n}}[\widetilde{I}_{ijk}] - 6Q_{n}$ is 
\begin{align}
    &\mathbb{V}_{X_{1:n}}\left[(D.I) - \frac{2}{h_n^4}\mathbb{E}_{X_{1:n}}[\widetilde{I}_{ijk}] - 6Q_{n}\right] \nonumber\\
    &\lesssim \mathbb{V}_{X_{1:n}}\left[\frac{1}{n}\sum_{i=1}^n\left\{h_n^L\rho_{1,1}(X_i)+(w_n-1)\rho_{1,2}(X_i)-\mathbb{E}_{X_{1:n}}[h_n^L\rho_{1,1}(X_i)+(w_n-1)\rho_{1,2}(X_i)]\right\} \right] \nonumber\\
    & \quad + \mathbb{V}_{X_{1:n}}\left[\frac{12}{n(n-1)h^4}\sum_{i=1}^{n-1}\sum_{j=i+1}^n \left\{\mathbb{E}_{X_{1:n}}[\widetilde{I}_{ijk}|X_i,X_j]-\mathbb{E}_{X_{1:n}}[\widetilde{I}_{ijk}|X_i]-\mathbb{E}_{X_{1:n}}[\widetilde{I}_{ijk}|X_j]+\mathbb{E}_{X_{1:n}}[\widetilde{I}_{ijk}]\right\}\right] \nonumber\\
    & \quad + O(n^{-3}h_n^{-6}) \nonumber\\
    & = O\{ n^{-1}h_n^{2L} + n^{-1}(w_n-1)h_n^L + n^{-1}(w_n-1)^2 + n^{-2}h_n^{-5}\}
\end{align}
Similarly, since Lemma \ref{lem:quad_2} state that $\mathbb{E}_{X_{1:n}}[\widetilde{II}_{ijklm}|X_i,X_j]=O_p(h_n^{7/2})$, combined with Lemma \ref{Lemma:fifth_order_U_evaluation}, the variance of $(D.II) + \frac{1}{h^6}\mathbb{E}_{X_{1:n}}[\widetilde{II}_{ijklm}] + 5Q_{n}$ is 
\begin{align}
    &\mathbb{V}_{X_{1:n}}\left[(D.II) + \frac{1}{h_n^6}\mathbb{E}_{X_{1:n}}[\widetilde{II}_{ijklm}] + 5Q_{n}\right] \nonumber\\
    & \lesssim \mathbb{V}_{X_{1:n}}\left[\frac{1}{n}\sum_{i=1}^n\left\{h_n^L\rho_{2,1}(X_i)+(w_n-1)\rho_{2,2}(X_i)-\mathbb{E}_{X_{1:n}}[h_n^L\rho_{2,1}(X_i)+(w_n-1)\rho_{2,2}(X_i)]\right\} \right]\nonumber \\
    & ~ + \mathbb{V}_{X_{1:n}}\left[\frac{12}{n(n-1)h_n^4}\sum_{i=1}^{n-1}\sum_{j=i+1}^n \left\{\mathbb{E}_{X_{1:n}}[\widetilde{II}_{ijklm}|X_i,X_j]-\mathbb{E}_{X_{1:n}}[\widetilde{II}_{ijklm}|X_i]-\mathbb{E}_{X_{1:n}} [\widetilde{II}_{ijklm}|X_j]+\mathbb{E}_{X_{1:n}}[\widetilde{II}_{ijklm}]\right\} \right] \nonumber\\
    & ~ + O(n^{-3}h_n^{-6})\nonumber \\
    & = O(n^{-1}h_n^{2L} + n^{-1}(w_n-1)h_n^L + n^{-1}(w_n-1)^2 + n^{-2}h_n^{-5})
\end{align}
From Cauchy-Schwarz inequality,
\begin{align}
    & Cov\left[(D.I) - \frac{2}{h_n^4}\mathbb{E}_{X_{1:n}}[\widetilde{I}_{ijk}] - 6Q_{n}, \quad (D.II) + \frac{1}{h_n^6}\mathbb{E}_{X_{1:n}}[\widetilde{II}_{ijklm}] + 5Q_{n}\right] \nonumber\\
    & \quad \le \mathbb{V}_{X_{1:n}}\left[(D.I) - \frac{2}{h_n^4}\mathbb{E}_{X_{1:n}}[\widetilde{I}_{ijk}] - 6Q_{n}\right]^{1/2}\mathbb{V}_{X_{1:n}}\left[(D.II) + \frac{1}{h_n^6}\mathbb{E}_{X_{1:n}}[\widetilde{II}_{ijklm}] + 5Q_{n}\right]^{1/2} \nonumber\\
    & =  O(n^{-1}h_n^{2L} + n^{-1}(w_n-1)h_n^L + n^{-1}(w_n-1)^2 + n^{-2}h_n^{-5})
\end{align}
Then, $\mathbb{V}_{X_{1:n}}[R_n(w_n,h_n)]$ is bounded as
\begin{align}
    \mathbb{V}_{X_{1:n}}[R_n(w_n,h_n)] &\lesssim \mathbb{V}_{X_{1:n}}\left[(D.I) - \frac{2}{h^4}\mathbb{E}_{X_{1:n}}[\widetilde{I}_{ijk}] - 6Q_{n}\right] + \mathbb{V}_{X_{1:n}}\left[(D.II) + \frac{1}{h^6}\mathbb{E}_{X_{1:n}}[\widetilde{II}_{ijklm}] + 5Q_{n}\right] \nonumber\\
    & \quad + 2 Cov\left[(D.I) - \frac{2}{h_n^4}\mathbb{E}_{X_{1:n}}[\widetilde{I}_{ijk}] - 6Q_{n}, (D.II) + \frac{1}{h_n^6}\mathbb{E}_{X_{1:n}}[\widetilde{II}_{ijklm}] + 5Q_{n}\right] \nonumber\\
    & =  O(n^{-1}h_n^{2L} + n^{-1}(w_n-1)h_n^L + n^{-1}(w_n-1)^2 + n^{-2}h_n^{-5}).
\end{align}
Therefore in view of (\ref{equation:empirical_hyvarinen_1}) and (\ref{equation:empirical_hyvarinen_2}), we have
\begin{align}
    & \mathcal{H}(w_n,h_n) = (D.I) + (D.II) + o_p\{(nh^3)^{-1}\} \nonumber\\
    & \qquad = \underbrace{\frac{2}{h_n^4}\mathbb{E}_{X_{1:n}}[\widetilde{I}_{ijk}] + \frac{1}{h_n^6}\mathbb{E}_{X_{1:n}}[\widetilde{II}_{ijklm}]}_{\approx\mathbb{E}_{X_{1:n}}[\mathcal{H}(w_n,h_n)]} + Q_n + R_n(w_n,h_n) + o_p\{(nh_n^3)^{-1}\} \nonumber\\
    \implies & \mathcal{H}(w_n,h_n)- \mathbb{E}_{X_{1:n}}[\mathcal{H}(w_n,h_n)] \approx Q_n + R_n(w_n,h_n)
\end{align}
and  $\mathbb{E}_{X_{1:n}}[R_n(w_n,h_n)^2]$ is bounded as $\mathbb{E}_{X_{1:n}}[R_n(w_n,h_n)^2]=\mathbb{V}_{X_{1:n}}[R_n(w_n,h_n)]=O(n^{-1}h_n^{2L} + n^{-1}(w_n-1)h_n^L + n^{-1}(w_n-1)^2 + n^{-2}h_n^{-5})$.

\section{Proof of Theorem \ref{thm:MISE} (MISE)} \label{sec:proof_MISE}
In this section, we expand the MISE of $\hat{f}_{w,h}(x)$.

\subsection{Preliminary Expansion}
In this subsection, we provide the preliminary result for Section \ref{sec:MISE_bias} and \ref{sec:MISE_variance}. 
Recall the definition of $\hat{f}_{w_n, h_n}(x)$.
\begin{align*}
    \hat{f}_{w_n, h_n}(x) \equiv \left\{ \int \hat{f}_{h_n}(x)^{w_n} dx\right\}^{-1} \hat{f}_{h_n}(x)^{w_n}
\end{align*}
Expanding $\hat{f}_{h_n}(x)^{w_n} $ around $w_n=1$ yields
\begin{align*}
    \hat{f}_{h_n}(x)^{w_n} 
    &= \hat{f}_{h_n}(x) +  (w_n - 1) \hat{f}_{h_n}(x) \log \hat{f}_{h_n}(x) + O_p\{(w_n  - 1)^2\} \\
    &= \mathbb{E}_{X_{1:n}} \left[  \hat{f}_{h_n}(x) \right] +  \hat{f}_{h_n}(x) - \mathbb{E}_{X_{1:n}} \left[  \hat{f}_{h_n}(x) \right] + (w_n - 1) \mathbb{E}_{X_{1:n}} \left[ \hat{f}_{h_n}(x) \log \hat{f}_{h_n}(x) \right] \\
    & \quad +  o_p\{(w_n  - 1)  + (nh_n)^{-1/2}\} .
\end{align*}
Using this result, we have
\begin{align*}
     \left\{ \int \hat{f}_{h_n}(x)^{w_n} dx\right\}^{-1} 
     &= \left( \int  \mathbb{E}_{X_{1:n}} \left[  \hat{f}_{h_n}(x) \right]  dx \right)^{-1} \\
     & \quad - \left( \int  \mathbb{E}_{X_{1:n}} \left[  \hat{f}_{h_n}(x) \right]  dx \right)^{-2} \\
     & \quad\quad \times \int
     \Biggl\{ \left( \hat{f}_{h_n}(x) - \mathbb{E}_{X_{1:n}} \left[  \hat{f}_{h_n}(x) \right] \right) +  (w_n-1) \mathbb{E}_{X_{1:n}} \left[ \hat{f}_{h_n}(x) \log \hat{f}_{h_n}(x) \right]  \Biggr\} dx  \\
     &  \quad +  o_p\{(w_n  - 1)  + (nh_n)^{-1/2}\} .
\end{align*}
Therefore
\begin{align}
    & \left\{ \int \hat{f}_{h_n}(x)^{w_n} dx\right\}^{-1} \hat{f}_{h_n}(x)^{w_n} \nonumber \\
    & = \frac{\mathbb{E}_{X_{1:n}} \left[  \hat{f}_{h_n}(x) \right]}{\int  \mathbb{E}_{X_{1:n}} \left[  \hat{f}_{h_n}(x) \right]  dx} 
    + \frac{  \hat{f}_{h_n}(x) - \mathbb{E}_{X_{1:n}} \left[  \hat{f}_{h_n}(x) \right] }{ \int  \mathbb{E}_{X_{1:n}} \left[  \hat{f}_{h_n}(x) \right]  dx }  
    + \frac{  (w_n-1) \mathbb{E}_{X_{1:n}} \left[ \hat{f}_{h_n}(x) \log \hat{f}_{h_n}(x) \right] }{ \int  \mathbb{E}_{X_{1:n}} \left[  \hat{f}_{h_n}(x) \right]  dx  } \nonumber \\
    & \quad - \frac{ \mathbb{E}_{X_{1:n}} \left[  \hat{f}_{h_n}(x) \right] \times \int
     \Biggl\{ \left( \hat{f}_{h_n}(x) - \mathbb{E}_{X_{1:n}} \left[  \hat{f}_{h_n}(x) \right] \right) +  (w_n-1) \mathbb{E}_{X_{1:n}} \left[ \hat{f}_{h_n}(x) \log \hat{f}_{h_n}(x) \right]  \Biggr\} dx}{  \left( \int  \mathbb{E}_{X_{1:n}} \left[  \hat{f}_{h_n}(x) \right]  dx \right)^2 } \nonumber \\
    & \quad +  o_p\{(w_n  - 1)  + (nh_n)^{-1/2}\} \label{eq:MISE_expansion}
\end{align}

\subsection{Bias} \label{sec:MISE_bias}
From (\ref{eq:MISE_expansion}), the expectation of $\hat{f}_{w_n,h_n}(x)$ is
\begin{align*}
    \mathbb{E}[\hat{f}_{w_n,h_n}(x)]
    &= \mathbb{E}_{X_{1:n}} \left[  \left\{ \int \hat{f}_{h_n}(x)^{w_n} dx\right\}^{-1} \hat{f}_{h_n}(x)^{w_n} \right] \\
    & = \frac{\mathbb{E}_{X_{1:n}} \left[  \hat{f}_{h_n}(x) \right]}{\int  \mathbb{E}_{X_{1:n}} \left[  \hat{f}_{h_n}(x) \right]  dx} + \frac{  (w_n-1) \mathbb{E}_{X_{1:n}} \left[ \hat{f}_{h_n}(x) \log \hat{f}_{h_n}(x) \right] }{ \int  \mathbb{E}_{X_{1:n}} \left[  \hat{f}_{h_n}(x) \right]  dx  } \\
    & \quad - \frac{ \mathbb{E}_{X_{1:n}} \left[  \hat{f}_{h_n}(x) \right] \times \int
    (w_n-1) \mathbb{E}_{X_{1:n}} \left[ \hat{f}_{h_n}(x) \log \hat{f}_{h_n}(x) \right]  dx}{  \left( \int  \mathbb{E}_{X_{1:n}} \left[  \hat{f}_{h_n}(x) \right]  dx \right)^2 } +  o\{(w_n  - 1)  + (nh_n)^{-1/2}\}
\end{align*}
Since
\begin{align*}
    & \mathbb{E}_{X_{1:n}} \left[  \hat{f}_{h_n}(x) \right] = f(x) + \kappa_L f^{(L)}(x) h_n^L + o(h^L), 
\end{align*}
and Lemma \ref{lem:flogf} states that
\begin{align*}
    & \mathbb{E}_{X_{1:n}} \left[ \hat{f}_{h_n}(x) \log \hat{f}_{h_n}(x) \right] \\
    & \quad = f(x) \log  f(x) + \kappa_L f^{(L)}(x) \left\{ \log f(x) + 1 \right\}h_n^L + o(h^L)  + O\left( \frac{1}{nh_n} \right),
\end{align*}
it follows that
\begin{align*}
    & \mathbb{E}_{X_{1:n}} \left[  \left\{ \int \hat{f}_{h_n}(x)^{w_n} dx\right\}^{-1} \hat{f}_{h_n}(x)^{w_n} \right] \\
    & = \frac{f(x) + \kappa_L f^{(L)}(x) h_n^L + o(h^L)}{\int   \left\{ f(x) + \kappa_L f^{(L)}(x) h_n^L + o(h^L) \right\} dx} \\
    & \quad+  \frac{(w_n-1)  f(x)\log f(x)}{\int   \left\{ f(x) + \kappa_L f^{(L)}(x) h_n^L + o(h^L) \right\} dx} \\
    & \quad - \frac{(w_n-1) f(x) \int f(x)\log  f(x)dx}{\left( \int   \left\{ f(x) + \kappa_L f^{(L)}(x) h_n^L + o(h^L) \right\} dx \right)^2} + s.o. \\
    & = f(x) + \kappa_L \left( f^{(L)}(x) - f(x) \int f^{(L)}(x)dx \right)h_n^L \\
    & \quad + (w_n - 1) \left( f(x)\log f(x) - f(x) \int f(x)\log f(x)dx  \right) + o\{h^L +  (w_n-1)\} \\
    & = f(x) + \kappa_L f^{(L)}(x)h_n^L + (w_n - 1) \left( f(x)\log f(x) - f(x) \int f(x)\log f(x)dx  \right) + o\{h^L +  (w_n-1)\}
\end{align*}
where the final equality follows from the fact that $\int f^{(L)}(x)dx = 0$ since $f$ and its derivatives are bounded (Assumption \ref{ass:fdiffrentiability1}).
This implies that the squared bias is
\begin{align*}
    & Bias[\hat{f}_{w_n,h_n}(x)]^2 \\
    &= \kappa_L^2 f^{(L)}(x)^2h_n^{2L} \\
    & \quad + 2\kappa_L(w_n-1)f^{(L)}(x)\left( f(x)\log f(x) - f(x) \int f(x)\log f(x)dx  \right) h_n^L  \\
    & \quad + (w_n - 1)^2 \left( f(x)\log f(x) - f(x) \int f(x)\log f(x)dx  \right)^2
\end{align*}
Integration of these term with respect to $x$ over $\mathcal{X}$ gives the bias term in Theorem \ref{thm:MISE}.

\subsection{Variance}  \label{sec:MISE_variance}
From (\ref{eq:MISE_expansion}), the variance of $\hat{f}_{w_n,h_n}(x)$ is
\begin{align*}
     & \mathbb{V}_{X_{1:n}} \left[  \left\{ \int \hat{f}_{h_n}(x)^{w_n} dx\right\}^{-1} \hat{f}_{h_n}(x)^{w_n} \right] \\
     & = \frac{  \mathbb{V}_{X_{1:n}}\left[ \hat{f}_{h_n}(x) \right] }{ \left( \int  \mathbb{E}_{X_{1:n}} \left[  \hat{f}_{h_n}(x) \right]  dx\right)^2 } \\
     & \quad + \frac{\mathbb{E}_{X_{1:n}}\left[  \hat{f}_{h_n}(x) \right]^2 \mathbb{V}_{X_{1:n}}\left[ \int  \hat{f}_{h_n}(x) dx\right]}{ \left( \int  \mathbb{E}_{X_{1:n}} \left[  \hat{f}_{h_n}(x) \right]  dx\right)^4  }  - 2 \mathbb{E}_{X_{1:n}} \left[  \hat{f}_{h_n}(x) \right]\frac{Cov\left(\hat{f}_{h_n}(x), \int \hat{f}_{h_n}(x) dx  \right)}{  \left( \int  \mathbb{E}_{X_{1:n}} \left[  \hat{f}_{h_n}(x) \right]  dx\right)^3 } + o\{(nh_n)^{-1}\}.
\end{align*}
Since
\begin{align*}
    & \int  \mathbb{E}_{X_{1:n}} \left[  \hat{f}_{h_n}(x) \right]  dx = 1 + o(1),
\end{align*}
and Lemma \ref{lem:V_intf} and \ref{lem:cov_fintf} state that
\begin{align*}
    & \mathbb{V}_{X_{1:n}}\left[ \int  \hat{f}_{h_n}(x) dx\right] = o\{(nh_n)^{-1}\}, \\
    &  Cov\left(\hat{f}_{h_n}(x), \int \hat{f}_{h_n}(x) dx  \right) = o\{(nh_n)^{-1}\}, 
\end{align*}
we have the asymptotic variance as
\begin{align*}
    \mathbb{V}_{X_{1:n}} \left[  \left\{ \int \hat{f}_{h_n}(x)^{w_n} dx\right\}^{-1} \hat{f}_{h_n}(x)^{w_n} \right]  =  \mathbb{V}_{X_{1:n}}\left[ \hat{f}_{h_n}(x) \right] + o\{(nh_n)^{-1}\} = \frac{R(K)f(x)}{nh_n}  + o\{(nh_n)^{-1}\}
\end{align*}
Integration of this term with respet to $x$ over $\mathcal{X}$ gives the variance term in Theorem \ref{thm:MISE}.

\section{Lemmas}
\subsection{Lemmas for the Computation of Expected Fisher divergence}

\subsubsection{Lemmas for Computation of the Bias Term}
\begin{lem} \label{lem:k'k_x}
Under Assumption \ref{ass:DGP}, \ref{ass:fdiffrentiability1}, \ref{ass:interiorpoint}, \ref{ass:Kernel}, \ref{ass:integrationbyparts} and \ref{ass:nhinfty}, it holds that
\begin{align*}
    & \mathbb{E}_{X_{1:n}}\left[\frac{1}{n^2h_n^3}\sum_{i=1}^n\sum_{j=1}^nK'_{i,h_n}(x)K_{j,h_n}(x)\right] \\
    & = -f(x)f'(x) - \kappa_L[f'(x)f^{L}(x)+f(x)f^{(L+1)}(x)]h_n^L  - \kappa_L^2f^{(L)}(x)f^{(L+1)}(x)h_n^{2L} + O\{n^{-1}h_n^{-2}\} + o(h_n^{2L}),
\end{align*}
and strengthening  Assumption \ref{ass:fdiffrentiability1} to \ref{ass:fdiffrentiability2}, it holds that
\begin{align*}
  & \mathbb{E}_{X_{1:n}}\left[\frac{1}{n^2h_n^3}\sum_{i=1}^n\sum_{j=1}^nK'_{i,h_n}(x)K_{j,h_n}(x)\right] \\
  & = -f(x)f'(x) - \sum_{l=L}^{2L}\kappa_l[f'(x)f^{l}(x)+f(x)f^{(l+1)}(x)]h_n^l - \kappa_L^2f^{(L)}(x)f^{(L+1)}(x)h_n^{2L}\\
  & \quad + O\{n^{-1}h_n^{-2}\} + o(h_n^{2L}). 
\end{align*}
\begin{proof}
We prove only the second claim. One can prove the first claim in the same way and the proof is easier.
\begin{align*}
    & \mathbb{E}_{X_{1:n}}\left[\frac{1}{n^2h_n^3}\sum_{i=1}^n\sum_{j=1}^nK'_{i,h_n}(x)K_{j,h_n}(x)\right] \\
    & = \mathbb{E}_{X_{1:n}}\left[\frac{1}{n^2h_n^3}\sum_{i=1}^nK'_{i,h_n}(x)K_{i,h_n}(x) + \frac{1}{n^2h_n^3}\underset{i\neq j}{\sum^n\sum^n}K'_{i,h_n}(x)K_{j,h_n}(x)\right] \\
    & = \frac{1}{h_n^3}\left[\int K'\left(\frac{z-x}{h_n}\right)f(z)dz\right]\left[\int K\left(\frac{z-x}{h_n}\right)f(z)dz\right] + O\{n^{-1}h_n^{-2}\} \\
    & = \frac{1}{h_n}\left[\int K'(u)f(x+uh)du\right]\left[\int K(u)f(x+uh_n)du\right]+ O\{n^{-1}h_n^{-2}\} \\
    & = \left[-\int K(u)f'(x+uh_n)du\right]\left[\int K(u)f(x+uh_n)du\right]+ O\{n^{-1}h_n^{-2}\} \\
    & = \left[-f'(x)-\sum_{l=L}^{2L}\{\kappa_lf^{(l+1)}(x)h_n^l\} + o(h_n^{2L})\right]\left[f(x)+\sum_{l=L}^{2L}\{\kappa_lf^{(l)}(x)h_n^l\} + o(h_n^{2L})\right] \\
    & \quad + O\{n^{-1}h_n^{-2}\} \\
    & = -f(x)f'(x) - \sum_{l=L}^{2L}\kappa_l[f'(x)f^{(l)}(x)+f(x)f^{(l+1)}(x)]h_n^l - \kappa_L^2f^{(L)}(x)f^{(L+1)}(x)h_n^{2L} \\
    & \quad + O\{n^{-1}h_n^{-2}\} + o(h_n^{2L})
\end{align*}
where the second equality follows from the assumption that $\{X_i\}_{i=1}^n$ is i.i.d. sample (Assumption \ref{ass:DGP}). The third equality follows from the change of variables $(z-x)/h_n=u$, the fourth equality follows from integration by parts combined with Assumption \ref{ass:integrationbyparts}, and the fifth equality expand $f'(x+uh_n)$ and $f(x+uh_n)$ around $x$ combined with the assumption of $L$-th order kernel (Assumption \ref{ass:Kernel}) and the smoothness assumption (Assumption \ref{ass:fdiffrentiability2}).
\end{proof}
\end{lem}

\subsubsection{Lemmas for Computation of the Variance Term}
\begin{lem} \label{lem:k'_x_square}
Under Assumption  \ref{ass:DGP},  \ref{ass:fdiffrentiability1}, \ref{ass:interiorpoint} , \ref{ass:Kernel}, \ref{ass:integrationbyparts} and \ref{ass:nhinfty}, it holds that
\begin{align*}
    & \mathbb{E}_{X_{1:n}}\left[\left(\frac{1}{nh_n^2}\sum_{i=1}^n K'_{i,h_n}(x)\right)^2\right] \\
    & = \frac{R(K')}{nh_n^3}f(x) + f'(x)^2 + 2\kappa_{L}f'(x)f^{(L+1)}(x)h_n^L + \kappa_L^2f^{(L+1)}(x)^2h_n^{2L} + o\left\{\frac{1}{nh_n^3}+h_n^{2L}\right\}
\end{align*}
and strengthening  Assumption \ref{ass:fdiffrentiability1} to \ref{ass:fdiffrentiability2}, it holds that
\begin{align*}
    & \mathbb{E}_{X_{1:n}}\left[\left(\frac{1}{nh_n^2}\sum_{i=1}^n K'_{i,h_n}(x)\right)^2\right] \\
    & = \frac{R(K')}{nh_n^3}f(x) + f'(x)^2 + 2\sum_{l=L}^{2L}\kappa_{l}f'(x)f^{(l+1)}(x)h_n^l + \kappa_L^2f^{(L+1)}(x)^2h_n^{2L} + o\left\{\frac{1}{nh_n^3}+h_n^{2L}\right\}
\end{align*}
\begin{proof}
We prove only the second claim. One can prove the first claim in the same way and the proof is easier.
\begin{align*}
    & \mathbb{E}_{X_{1:n}}\left[\left(\frac{1}{nh_n^2}\sum_{i=1}^n K'_{i,h_n}(x)\right)^2\right] \\
    & = \mathbb{E}_{X_{1:n}}\left[\frac{1}{n^2h_n^4}\sum_{i=1}^nK'_{i,h_n}(x)^2 + \frac{1}{n^2h_n^4}\underset{i\neq j}{\sum^n\sum^n}K'_{i,h_n}(x)K'_{j,h_n}(x)\right] \\
    & = \frac{1}{nh_n^3} \int K'(u)^2f(x+uh_n)du + \frac{1}{h_n^2}\left(\int K'(u)f(x+uh_n)du\right)^2 + o\left\{\frac{1}{nh_n^3}\right\} \\
    & = \frac{1}{nh_n^3} \int K'(u)^2f(x+uh_n)du +  \left(-\int K(u)f'(x+uh_n)du\right)^2 + o\left\{\frac{1}{nh_n^3}\right\} \\
    & = \frac{1}{nh_n^3} \left(\int K'(u)^2du\right)f(x) +  \left(-f'(x) - \sum_{l=L}^{2L}\kappa_lf^{(l+1)}(x)h_n^l + o(h_n^{2L})\right)^2 + o\left\{\frac{1}{nh_n^3}\right\} \\
    & = \frac{R(K')}{nh_n^3}f(x) + f'(x)^2 + 2\sum_{l=L}^{2L}\kappa_{l}f'(x)f^{(l+1)}(x)h_n^l + \kappa_L^2f^{(L+1)}(x)^2h_n^{2L} + o\left\{\frac{1}{nh_n^3}+h_n^{2L}\right\}
\end{align*}
Since the process of the expansion is similar to of Lemma \ref{lem:k'k_x}, we omit the detail explanation.
\end{proof}
\end{lem}

\begin{lem} \label{lem:k_x_square}
Under Assumption  \ref{ass:DGP},  \ref{ass:fdiffrentiability1}, \ref{ass:interiorpoint}, \ref{ass:Kernel}, \ref{ass:integrationbyparts} and \ref{ass:nhinfty}, it holds that
\begin{align}
    & \mathbb{E}_{X_{1:n}}\left[\left(\frac{1}{nh_n}\sum_{i=1}^n K_{i,h_n}(x)\right)^2\right] \nonumber\\
    & = f(x)^2 + 2\kappa_L f(x)f^{(L)}(x)h_n^{L} + \kappa_L^2f^{(L)}(x)^2h_n^{2L} + o\left\{\frac{1}{nh_n^3}+h_n^{2L}\right\} \nonumber
\end{align}
and strengthening  Assumption \ref{ass:fdiffrentiability1} to \ref{ass:fdiffrentiability2}, it holds that
\begin{align}
    & \mathbb{E}_{X_{1:n}}\left[\left(\frac{1}{nh_n}\sum_{i=1}^n K_{i,h_n}(x)\right)^2\right] \nonumber\\
    & = f(x)^2 + 2\sum_{l=L}^{2L}\kappa_l f(x)f^{(l)}(x)h_n^{l} + \kappa_L^2f^{(L)}(x)^2h_n^{2L} + o\left\{\frac{1}{nh_n^3}+h_n^{2L}\right\} \nonumber
\end{align}
\begin{proof}
We prove only the second claim. One can prove the first claim in the same way and the proof is easier.
\begin{align*}
    & \mathbb{E}_{X_{1:n}}\left[\left(\frac{1}{nh_n}\sum_{i=1}^n K_{i,h_n}(x)\right)^2\right] \\
    & = \mathbb{E}_{X_{1:n}}\left[\frac{1}{n^2h_n^2}\sum_{i=1}^n K_{i,h_n}(x)^2 + \frac{1}{n^2h_n^2}\underset{i\neq j}{\sum^n\sum^n}K_{i,h_n}(x)K_{j,h_n}(x)\right] \\
    & = \left(\int K(u)f(x+uh_n)du\right)^2 + o\left\{\frac{1}{nh_n^3}\right\}\\
    & =  \left(f(x) + \sum_{l=L}^{2L}\kappa_lf^{(l)}(x)h_n^l + o(h_n^{2L})\right)^2 + o\left\{\frac{1}{nh_n^3}\right\}\\
    & = f(x)^2 + 2\sum_{l=L}^{2L}\kappa_l f(x)f^{(l)}(x)h_n^{l} + \kappa_L^2f^{(L)}(x)^2h_n^{2L} + o\left\{\frac{1}{nh_n^3}+h_n^{2L}\right\}
\end{align*}
Since the process of the expansion is similar to of Lemma \ref{lem:k'k_x}, we omit the detail explanation.
\end{proof}
\end{lem}

\begin{lem} \label{lem:k'k_x_square}
Under Assumption  \ref{ass:DGP},  \ref{ass:fdiffrentiability1}, \ref{ass:interiorpoint}, \ref{ass:Kernel}, \ref{ass:integrationbyparts} and \ref{ass:nhinfty},
\begin{align*}
    & \mathbb{E}_{X_{1:n}}\left[\left(\frac{1}{nh_n^2}\sum_{i=1}^n K'_{i,h_n}(x)\right)^2\left(\frac{1}{nh_n}\sum_{i=1}^n K_{i,h_n}(x)\right)^2\right] \\
    & = \frac{R(K')}{nh_n^3}f(x)^3 + f'(x)^2f(x)^2 + 2\kappa_L\Bigl(f(x)f'(x)^2f^{(L)}(x) + f(x)^2f'(x)f^{(L+1)}(x)\Bigl)h_n^L \\
    & \quad + \kappa_L^2\Bigl(4f(x)f'(x)f^{(L)}(x)f^{(L+1)}(x) + f'(x)^2f^{(L)}(x)^2 + f(x)^2f^{(L+1)}(x)^2\Bigl)h_n^{2L} + o\left\{\frac{1}{nh_n^3}+h_n^{2L}\right\} 
\end{align*}
and strengthening  Assumption \ref{ass:fdiffrentiability1} to \ref{ass:fdiffrentiability2}, it holds that
\begin{align*}
    & \mathbb{E}_{X_{1:n}}\left[\left(\frac{1}{nh_n^2}\sum_{i=1}^n K'_{i,h_n}(x)\right)^2\left(\frac{1}{nh_n}\sum_{i=1}^n K_{i,h_n}(x)\right)^2\right] \\
    & = \frac{R(K')}{nh_n^3}f(x)^3 + f'(x)^2f(x)^2 + 2\sum_{l=L}^{2L}\kappa_l\Bigl(f(x)f'(x)^2f^{(l)}(x) + f(x)^2f'(x)f^{(l+1)}(x)\Bigl)h_n^l \\
    & \quad + \kappa_L^2\Bigl(4f(x)f'(x)f^{(L)}(x)f^{(L+1)}(x) + f'(x)^2f^{(L)}(x)^2 + f(x)^2f^{(L+1)}(x)^2\Bigl)h_n^{2L} + o\left\{\frac{1}{nh_n^3}+h_n^{2L}\right\} 
\end{align*}
\begin{proof}
We prove only the second claim. One can prove the first claim in the same way and the proof is easier.
\begin{align*}
    & \mathbb{E}_{X_{1:n}}\left[\left(\frac{1}{nh_n^2}\sum_{i=1}^n K'_{i,h_n}(x)\right)^2\left(\frac{1}{nh_n}\sum_{i=1}^n K_{i,h_n}(x)\right)^2\right] \\
    & = \mathbb{E}_{X_{1:n}}\left[\frac{1}{n^4h_n^6}\sum_{i=1}^n\sum_{j=1}^n\sum_{k=1}^n\sum_{l=1}^n K'_{i,h_n}(x)K'_{j,h_n}(x)K_{k,h_n}(x)K_{l,h_n}(x)\right] \\
    & = \mathbb{E}_{X_{1:n}}\left[\frac{1}{n^4h_n^6}\underset{i\neq j \neq k \neq l}{\sum^n\sum^n\sum^n\sum^n}K'_{i,h_n}(x)K'_{j,h_n}(x)K_{k,h_n}(x)K_{l,h_n}(x)\right] \\
    & \quad  + \mathbb{E}_{X_{1:n}}\Biggl[\frac{1}{n^4h_n^6}\underset{i\neq j\neq k}{\sum^n\sum^n\sum^n}\{K'_{i,h_n}(x)^2K_{j,h_n}(x)K_{k,h_n}(x) \\
    & \quad\quad  +4K'_{i,h_n}(x)K_{i,h_n}(x)K'_{j,h_n}(x)K_{k,h_n}(x)+K_{i,h_n}(x)^2K'_{j,h_n}(x)K'_{k,h_n}(x)\}\Biggl] + o\left\{\frac{1}{nh_n^3}\right\} \\
    & = \frac{1}{h_n^2}\left(\int K'(u)f(x+uh_n)du\right)^2\left(\int K(u)f(x+uh_n)du\right)^2 \\
    & \quad + \frac{1}{nh_n^3}\left(\int K'(u)^2f(x+uh_n)du\right)\left(\int K(u)f(x+uh_n)du\right)^2 \\
    & \quad + \frac{4}{nh_n^3}\left(\int K'(u)K(u)f(x+uh_n)du\right)\left(\int K'(u)f(x+uh_n)du\right)\left(\int K(u)f(x+uh_n)du\right)  \\
    & \quad + \frac{1}{nh_n^3}\left(\int K(u)^2f(x+uh_n)du\right)\left(\int K'(u)f(x+uh_n)du\right)^2 + o\left\{\frac{1}{nh_n^3}\right\} \\
    & = \left(\int K(u)f'(x+uh_n)du\right)^2\left(\int K(u)f(x+uh_n)du\right)^2 \\
    & \quad + \frac{1}{nh_n^3}\left(\int K'(u)^2f(x+uh_n)du\right)\left(\int K(u)f(x+uh_n)du\right)^2 + o\left\{\frac{1}{nh_n^3}\right\} \\
    & = \left(f'(x) + \sum_{l=L}^{2L}\kappa_lf^{(l+1)}(x)h_n^l + o(h_n^l)\right)^2\left(f(x) + \sum_{l=L}^{2L}\kappa_lf^{(l)}(x)h_n^l + o(h_n^l)\right)^2 \\
    & \quad + \frac{R(K')}{nh_n^3}f(x)^3 + o\left\{\frac{1}{nh_n^3}\right\} \\
    & = \left(f'(x)^2 + 2\sum_{l=L}^{2L}\kappa_lf'(x)f^{(l+1)}(x)h_n^l + \kappa_L^2f^{(L+1)}(x)^2h_n^{2L} + o(h_n^{2L})\right) \\
    & \quad \times \left(f(x)^2 + 2\sum_{l=L}^{2L}\kappa_lf(x)f^{(l)}(x)h_n^l +\kappa_L^2f^{(L)}(x)h_n^{2L}+ o(h_n^{2L})\right)  + \frac{R(K')}{nh_n^3}f(x)^3 + o\left\{\frac{1}{nh_n^3}\right\} \\
    & = \frac{R(K')}{nh_n^3}f(x)^3 + f'(x)^2f(x)^2 + 2\sum_{l=L}^{2L}\kappa_l\Bigl(f(x)f'(x)^2f^{(l)}(x) + f(x)^2f'(x)f^{(l+1)}(x)\Bigl)h_n^l \\
    & \quad + 4\kappa_L^2f(x)f'(x)f^{(L)}(x)f^{(L+1)}(x)h_n^{2L} + \kappa_L^2\Bigl(f'(x)^2f^{(L)}(x)^2 + f(x)^2f^{(L+1)}(x)^2\Bigl)h_n^{2L} \\
    & \quad + o\left\{\frac{1}{nh_n^3}+h_n^{2L}\right\} 
\end{align*}
Since the process of the expansion is similar to of Lemma \ref{lem:k'k_x}, we omit the detail explanation.
\end{proof}
\end{lem}

\subsection{Lemmas for the Computation of Expected Hyv\"arinen score}
\begin{lem} \label{lem:k}
Under Assumption  \ref{ass:DGP},  \ref{ass:fdiffrentiability1}, \ref{ass:interiorpoint}, \ref{ass:Kernel}, \ref{ass:integrationbyparts} and \ref{ass:nhinfty}, it holds that
\begin{align*}
    \mathbb{E}_{X_{1:n}}\left[\frac{1}{(n-1)h_n}\sum_{j\neq i}^n K_{ij} \mid X_i \right] = f(X_i) + \kappa_L f^{(L)}(X_i)h_n^L + o_p(h_n^{L})
\end{align*}
and strengthening  Assumption \ref{ass:fdiffrentiability1} to \ref{ass:fdiffrentiability2}, it holds that
\begin{align*}
    \mathbb{E}_{X_{1:n}}\left[\frac{1}{(n-1)h_n}\sum_{j\neq i}^n K_{ij} \mid X_i \right] = f(X_i) + \sum_{l=L}^{2L}\kappa_l f^{(l)}(X_i)h_n^l + o_p(h_n^{2L})
\end{align*}
\begin{proof}
We prove only the second claim. One can prove the first claim in the same way and the proof is easier.
\begin{align*}
    \mathbb{E}_{X_{1:n}}\left[\frac{1}{(n-1)h_n}\sum_{j\neq i}^n K_{ij} \mid X_i \right] &= \frac{1}{h_n}\int K\left(\frac{X_i-z}{h_n}\right)f(z)dz \\
    & = \int K(u)f(X_i+uh_n)du  = f(X_i) + \sum_{l=L}^{2L}\kappa_l f^{(l)}(X_i)h_n^l + o_p(h_n^{2L})
\end{align*}
Since the process of the expansion is similar to of Lemma \ref{lem:k'k_x}, we omit the detail explanation.
\end{proof}
\end{lem}

\begin{lem} \label{lem:k''}
Under Assumption  \ref{ass:DGP},  \ref{ass:fdiffrentiability1}, \ref{ass:interiorpoint}, \ref{ass:Kernel}, \ref{ass:integrationbyparts} and \ref{ass:nhinfty}, it holds that
\begin{align*}
    \mathbb{E}_{X_{1:n}}\left[\frac{1}{(n-1)h_n}\sum_{j\neq i}^n K''_{ij} \mid X_i \right] = h_n^2f''(X_i) + \kappa_Lf^{(L+2)}(X_i)h_n^{L+2} + o_p(h_n^{L+2})
\end{align*}
and strengthening  Assumption \ref{ass:fdiffrentiability1} to \ref{ass:fdiffrentiability2}, it holds that
\begin{align*}
    \mathbb{E}_{X_{1:n}}\left[\frac{1}{(n-1)h_n}\sum_{j\neq i}^n K''_{ij} \mid X_i \right] = h_n^2f''(X_i) + \sum_{l=L}^{2L}\kappa_lf^{(l+2)}(X_i)h_n^{l+2} + o_p(h_n^{2L+2})
\end{align*}
\begin{proof}
We prove only the second claim. One can prove the first claim in the same way and the proof is easier.
\begin{align*}
    \mathbb{E}_{X_{1:n}}\left[\frac{1}{(n-1)h_n}\sum_{j\neq i}^n K''_{ij} \mid X_i \right] &= \frac{1}{h_n}\int K''\left(\frac{X_i-z}{h_n}\right)f(z)dz \\
    &= \int K''(u)f(X_i+uh_n)du \\
    & = h_n^2\int K(u)f''(X_i+uh_n)du \\
    & = h_n^2f''(X_i) + \sum_{l=L}^{2L}\kappa_lf^{(l+2)}(X_i)h_n^{l+2} + o_p(h_n^{2L+2})
\end{align*}
Since the process of the expansion is similar to of Lemma \ref{lem:k'k_x}, we omit the detail explanation.
\end{proof}
\end{lem}

\begin{lem} \label{lem:k''k}
Under Assumption  \ref{ass:DGP},  \ref{ass:fdiffrentiability1}, \ref{ass:interiorpoint}, \ref{ass:Kernel}, \ref{ass:integrationbyparts} and \ref{ass:nhinfty}, it holds that
\begin{align*}
    & \mathbb{E}_{X_{1:n}}\left[\frac{1}{(n-1)h_n^2}\sum_{j\neq i}^n\sum_{k\neq i}^n K''_{ij}K_{ik} \mid X_i \right] = \frac{-R(K')}{nh_n}f(X_i) + f(X_i)f''(X_i)h^2 \\
    & \quad + \kappa_{L}\Bigl(f''(X_i)f^{(L)}(X_i) + f(X_i)f^{(L+2)}(X_i)\Bigl)h_n^{L+2} + \kappa_L^2f^{(L+2)}(X_i)f^{(L)}(X_i)h_n^{2L+2} \\
    & \quad + o_p\left\{\frac{1}{nh_n}+h_n^{2L+2} \right\}
\end{align*}
and strengthening  Assumption \ref{ass:fdiffrentiability1} to \ref{ass:fdiffrentiability2}, it holds that
\begin{align*}
    & \mathbb{E}_{X_{1:n}}\left[\frac{1}{(n-1)h_n^2}\sum_{j\neq i}^n\sum_{k\neq i}^n K''_{ij}K_{ik} \mid X_i \right] = \frac{-R(K')}{nh_n}f(X_i) + f(X_i)f''(X_i)h^2 \\
    & \quad + \sum_{l=L}^{2L}\kappa_{l}\Bigl(f''(X_i)f^{(l)}(X_i) + f(X_i)f^{(l+2)}(X_i)\Bigl)h_n^{l+2} + \kappa_L^2f^{(L+2)}(X_i)f^{(L)}(X_i)h_n^{2L+2} \\
    & \quad + o_p\left\{\frac{1}{nh_n}+h_n^{2L+2} \right\}
\end{align*}
\begin{proof}
We prove only the second claim. One can prove the first claim in the same way and the proof is easier.
\begin{align*}
    & \mathbb{E}_{X_{1:n}}\left[\frac{1}{(n-1)^2h_n^2}\sum_{j\neq i}^n\sum_{k\neq i}^n K''_{ij}K_{ik} \mid X_i \right] \\
    & =  \mathbb{E}_{X_{1:n}}\left[\frac{1}{(n-1)^2h_n^2}\sum_{j\neq i}^n K''_{ij}K_{ij} \mid X_i \right] +  \mathbb{E}_{X_{1:n}}\left[\frac{1}{(n-1)^2h_n^2}\underset{i\neq j\neq k}{\sum^n\sum^n}K''_{ij}K_{ik} \mid X_i \right] \\
    & = \frac{1}{(n-1)h_n}\int K''(u)K(u)f(X_i+uh_n)du \\
    & \quad + \left(\int K''(u)f(X_i+uh_n)du\right)\left(\int K(u)f(X_i+uh_n)du\right) \\
    & = \frac{-1}{nh_n}\left(\int K'(u)K'(u)du\right)f(X_i) + o_p\left\{\frac{1}{nh_n} \right\} \\
    & \quad + \left(h_n^2\int K(u)f''(X_i+uh_n)du\right)\left(\int K(u)f(X_i+uh_n)du\right) \\
    & = \frac{-R(K')}{nh_n}f(X_i) + o_p\left\{\frac{1}{nh_n} \right\} \\
    & \quad + \Bigl(f''(X_i)h_n^2 + \sum_{l=L}^{2L}\kappa_lf^{(l+2)}(X_i)h_n^l + o_p(h_n^{2L+2})\Bigl) \Bigl(f(X_i) + \sum_{l=L}^{2L}\kappa_lf^{(l)}(X_i)h_n^l + o_p(h_n^{2L})\Bigl) \\
    & = \frac{-R(K')}{nh_n}f(X_i) + f(X_i)f''(X_i)h_n^2 \\
    & \quad + \sum_{l=L}^{2L}\kappa_{l}\Bigl(f''(X_i)f^{(l)}(X_i) + f(X_i)f^{(l+2)}(X_i)\Bigl)h_n^{l+2} + \kappa_L^2f^{(L+2)}(X_i)f^{(L)}(X_i)h_n^{2L+2} \\
    & \quad + o_p\left\{\frac{1}{nh_n}+h_n^{2L+2} \right\}
\end{align*}
Since the process of the expansion is similar to of Lemma \ref{lem:k'k_x}, we omit the detail explanation.
\end{proof}
\end{lem}

\begin{lem} \label{lem:k'k'}
Under Assumption  \ref{ass:DGP},  \ref{ass:fdiffrentiability1}, \ref{ass:interiorpoint}, \ref{ass:Kernel}, \ref{ass:integrationbyparts} and \ref{ass:nhinfty}, it holds that
\begin{align*}
    & \mathbb{E}_{X_{1:n}}\left[\frac{1}{(n-1)^2h_n^2}\sum_{j\neq i}^n\sum_{k\neq i}^n K'_{ij}K'_{ik} \mid X_i\right] \\ 
    & = \frac{R(K')}{(n-1)h_n}f(X_i) + f'(X_i)^2h_n^2 + 2\kappa_Lf'(X_i)f^{(L+1)}(X_i)h_n^{L+2} + \kappa_L^2f^{(L+1)}(X_i)^2h_n^{2L+2} \\
    & \quad + o_p\left\{\frac{1}{nh_n}+h_n^{2L+2} \right\}
\end{align*}
and strengthening  Assumption \ref{ass:fdiffrentiability1} to \ref{ass:fdiffrentiability2}, it holds that
\begin{align*}
    & \mathbb{E}_{X_{1:n}}\left[\frac{1}{(n-1)^2h_n^2}\sum_{j\neq i}^n\sum_{k\neq i}^n K'_{ij}K'_{ik} \mid X_i\right] \\ 
    & = \frac{R(K')}{(n-1)h_n}f(X_i) + f'(X_i)^2h_n^2 + 2\sum_{l=L}^{2L}\kappa_lf'(X_i)f^{(l+1)}(X_i)h_n^{l+2} + \kappa_L^2f^{(L+1)}(X_i)^2h_n^{2L+2} \\
    & \quad + o_p\left\{\frac{1}{nh_n}+h_n^{2L+2} \right\}
\end{align*}
\begin{proof}
We prove only the second claim. One can prove the first claim in the same way and the proof is easier.
\begin{align*}
    & \mathbb{E}_{X_{1:n}}\left[\frac{1}{(n-1)^2h_n^2}\sum_{j\neq i}^n\sum_{k\neq i}^n {K'}_{ij}{K'}_{ik} \mid X_i\right] \\
    & = \mathbb{E}_{X_{1:n}}\left[\frac{1}{(n-1)^2h_n^2}\sum_{j\neq i}^n{K'}_{ij}^2 + \frac{1}{(n-1)^2h_n^2}\underset{i\neq j\neq k}{\sum^n\sum^n}{K'}_{ij}{K'}_{ik} \mid X_i \right] \\
    & = \frac{1}{(n-1)h_n}\Bigl(K'(u)^2f(X_i+uh_n)du\Bigl) + \frac{1}{h_n^2}\left(\int K'(u)f(X_i+uh_n)du\right)^2 \\
    & = \frac{1}{(n-1)h_n}\left(\int K'(u)^2du\right)f(X_i) \\
    & \quad + \Bigl(-h_nf'(X_i) - \sum_{l=L}^{2L}\kappa_l f^{(l+1)}(X_i)h_n^{l+1} + o_p(h_n^{L+2})\Bigl)^2 \\
    & = \frac{R(K')}{(n-1)h_n}f(X_i) + f'(X_i)^2h^2 + 2\sum_{l=L}^{2L}\kappa_l f'(X_i)f^{(l+1)}(X_i)h^{l+2} + {\kappa_L}^2f^{(L+1)}(X_i)^2h_n^{2L+2} \\
    & \quad + o_p\left\{\frac{1}{nh_n}+h_n^{2L+2} \right\}
\end{align*}
Since the process of the expansion is similar to of Lemma \ref{lem:k'k_x}, we omit the detail explanation.
\end{proof}
\end{lem}

\begin{lem} \label{lem:k_square}
Under Assumption  \ref{ass:DGP},  \ref{ass:fdiffrentiability1}, \ref{ass:interiorpoint}, \ref{ass:Kernel}, \ref{ass:integrationbyparts} and \ref{ass:nhinfty}, it holds that
\begin{align}
    & \mathbb{E}_{X_{1:n}}\left[\left\{\frac{1}{(n-1)h_n}\sum_{j\neq i}^n K_{ij}\right\}^2 \mid X_i \right] \nonumber\\
    & = f(X_i)^2 + 2\kappa_Lf(X_i)f^{(L)}(X_i)h_n^L + \kappa_L^2f^{(L)}(X_i)^2h_n^{2L} +  o\left\{\frac{1}{nh_n^3}+h_n^{2L}\right\} \nonumber
\end{align}
and strengthening  Assumption \ref{ass:fdiffrentiability1} to \ref{ass:fdiffrentiability2}, it holds that
\begin{align}
    & \mathbb{E}_{X_{1:n}}\left[\left\{\frac{1}{(n-1)h_n}\sum_{j\neq i}^n K_{ij}\right\}^2 \mid X_i \right] \nonumber\\
    & = f(X_i)^2 + 2\sum_{l=L}^{2L}\kappa_lf(X_i)f^{(l)}(X_i)h_n^l + \kappa_L^2f^{(L)}(X_i)^2h_n^{2L} +  o\left\{\frac{1}{nh_n^3}+h_n^{2L}\right\} \nonumber
\end{align}
\begin{proof}
We prove only the second claim. One can prove the first claim in the same way and the proof is easier.
\begin{align*}
    & \mathbb{E}_{X_{1:n}}\left[\left\{\frac{1}{(n-1)h_n}\sum_{j\neq i}^n K_{ij}\right\}^2 \mid X_i \right]\\
    & = \mathbb{E}_{X_j}\left[\frac{1}{(n-1)^2h_n^2}\sum_{j\neq i}^n K_{ij}^2 + \frac{1}{(n-1)^2h_n^2 }\underset{i\neq j\neq k}{\sum^n\sum^n}K_{ij}K_{ik} \mid X_i \right] \\
    & = \frac{1}{h_n^2}\left(h_n \int K(u)f(X_i+uh_n)du\right)^2  + O\left\{\frac{1}{nh_n} \right\}\\
    & = \left(f(X_i) + \sum_{l=L}^{2L}\kappa_lf^{(l)}(X_i)h_n^{l} + o(h_n^{2L})\right)^2 + O\left\{\frac{1}{nh_n} \right\}\\
    & = f(X_i)^2 + 2\sum_{l=L}^{2L}\kappa_lf(X_i)f^{(l)}(X_i)h_n^{l} + \kappa_L^2f^{(L)}(X_i)^2h_n^{2L} +  o\left\{\frac{1}{nh_n^3}+h_n^{2L}\right\}
\end{align*}
Since the process of the expansion is similar to of Lemma \ref{lem:k'k_x}, we omit the detail explanation.
\end{proof}
\end{lem}

\begin{lem} \label{lem:k'k'kk}
Under Assumption  \ref{ass:DGP},  \ref{ass:fdiffrentiability1}, \ref{ass:interiorpoint}, \ref{ass:Kernel}, \ref{ass:integrationbyparts} and \ref{ass:nhinfty},
\begin{align*}
    &\mathbb{E}_{X_{1:n}}\left[\frac{1}{(n-1)^4h_n^4}\sum_{j\neq i}^n\sum_{k\neq i}^n\sum_{l\neq i}^n\sum_{m\neq i}^n K'_{ij}K'_{ik}K_{il}K_{im} \mid X_i \right] = \frac{R(K')}{(n-1)h_n}f(X_i)^3 + f'(X_i)^2f(X_i)^2h_n^2 \\
    & \quad + 2\kappa_l\Bigl(f(X_i)f'(X_i)^2f^{(L)}(X_i)+f(X_i)^2f'(X_i)f^{(L+1)}(X_i)\Bigl)h_n^{L+2} \\
    & \quad + \kappa_L^2\Bigl(4f(X_i)f'(X_i)f^{(L)}(X_i)f^{(L+1)}(X_i) + f'(X_i)^2f^{(L)}(X_i)^2 + f(X_i)^2f^{(L+1)}(X_i)^2\Bigl)h_n^{2L+2} \\
    & \quad + o_p\left\{\frac{1}{nh_n}+h_n^{2L+2}\right\}
\end{align*}
and strengthening  Assumption \ref{ass:fdiffrentiability1} to \ref{ass:fdiffrentiability2}, it holds that
\begin{align*}
    &\mathbb{E}_{X_{1:n}}\left[\frac{1}{(n-1)^4h_n^4}\sum_{j\neq i}^n\sum_{k\neq i}^n\sum_{l\neq i}^n\sum_{m\neq i}^n K'_{ij}K'_{ik}K_{il}K_{im} \mid X_i \right] = \frac{R(K')}{(n-1)h_n}f(X_i)^3 + f'(X_i)^2f(X_i)^2h_n^2 \\
    & \quad + 2\sum_{l=L}^{2L}\kappa_l\Bigl(f(X_i)f'(X_i)^2f^{(l)}(X_i)+f(X_i)^2f'(X_i)f^{(l+1)}(X_i)\Bigl)h_n^{l+2} \\
    & \quad + \kappa_L^2\Bigl(4f(X_i)f'(X_i)f^{(L)}(X_i)f^{(L+1)}(X_i) + f'(X_i)^2f^{(L)}(X_i)^2 + f(X_i)^2f^{(L+1)}(X_i)^2\Bigl)h_n^{2L+2} \\
    & \quad + o_p\left\{\frac{1}{nh_n}+h_n^{2L+2}\right\}
\end{align*}
\begin{proof}
We prove only the second claim. One can prove the first claim in the same way and the proof is easier.
\begin{align*}
    & \mathbb{E}_{X_{1:n}}\left[\frac{1}{(n-1)^4h_n^4}\sum_{j\neq i}^n\sum_{k\neq i}^n\sum_{l\neq i}^n\sum_{m\neq i}^n K'_{ij}K'_{ik}K_{il}K_{im} \mid X_i \right] \\
    & = \frac{1}{(n-1)h_n^4}\mathbb{E}_{X_{1:n}}\left[\{{K'_{ij}}^2K_{ik}K_{il}+4K'_{ij}K_{ij}K'_{ik}K_{il}+{K_{ij}}^2K'_{ik}K'_{il}\} \mid X_i \right] + \frac{1}{h_n^4}\mathbb{E}_{X_{1:n}}[K'_{ij}K'_{ik}K_{il}K_{im} \mid X_i ] \\
    & = \frac{1}{(n-1)h_n^4}\mathbb{E}_{X_{1:n}}\left[{K'_{ij}}^2K_{ik}K_{il} \mid X_i \right] + \frac{1}{h_n^4}\mathbb{E}_{X_{1:n}}[K'_{ij}K'_{ik}K_{il}K_{im} \mid X_i] + o_p\{(nh)^{-1}\}\\
    & = \frac{1}{(n-1)h_n}\left(\int K'(u)^2f(X_i+uh_n)du\right)\left(\int K(u)f(X_i+uh_n)du\right)^2 \\
    & \quad + \left(\int K'(u)f(X_i+uh_n)du\right)^2\left(\int K(u)f(X_i+uh_n)du\right)^2 \\
    & = \frac{1}{(n-1)h_n}\left(\int K'(u)^2f(X_i)du + o_p(1)\right)\Bigl(f(X_i)+O_p(h_n^L)\Bigl)^2 \\
    & \quad + \left(f'(X_i)h_n + \sum_{l=L}^{2L}\kappa_lf^{(l+1)}(X_i)h_n^{l+1} + o_p(h_n^{2L+1})\right)^2\left(f(X_i)+ \sum_{l=L}^{2L}\kappa_lf^{(l)}(X_i)h_n^{l} + o_p(h_n^{2L})\right)^2 \\
    & = \frac{R(K')}{(n-1)h_n}f(X_i)^3 + o_p\left\{ \frac{1}{nh_n} \right\}\\
    & \quad + \left(f'(X_i)^2h_n^2 + 2\sum_{l=L}^{2L}{\kappa_l}f'(X_i)f^{(L+1)}(X_i)h_n^{l+2} + {\kappa_L}^2f^{(L+1)}(X_i)^2h_n^{2L+2}\right) \\
    & \quad\quad \times \left(f(X_i)^2 + 2\sum_{l=L}^{2L}{\kappa_l}f(X_i)f^{(L)}(X_i)h_n^{l} + {\kappa_L}^2f^{(L)}(X_i)^2h_n^{2L}\right) +o_p(h_n^{2L+2})\\
    & = \frac{R(K')}{(n-1)h_n}f(X_i)^3 + f'(X_i)^2f(X_i)^2h_n^2 \\
    & \quad + 2\sum_{l=L}^{2L}\kappa_l\Bigl(f(X_i)f'(X_i)^2f^{(l)}(X_i)+f(X_i)^2f'(X_i)f^{(l+1)}(X_i)\Bigl)h_n^{l+2} \\
    & \quad + {\kappa_L}^2\Bigl(4f(X_i)f'(X_i)f^{(L)}(X_i)f^{(L+1)}(X_i) + f'(X_i)^2f^{(L)}(X_i)^2 + f(X_i)^2f^{(L+1)}(X_i)^2\Bigl)h_n^{2L+2} \\
    & \quad + o_p\left\{\frac{1}{nh_n}+h_n^{2L+2}\right\}
\end{align*}
Since the process of the expansion is similar to of Lemma \ref{lem:k'k_x}, we omit the detail explanation.
\end{proof}
\end{lem}

\subsection{Lemmas for the Uniform Convergence of Empirical Hyv\"arinen score}
\subsubsection{Lemmas for the expansion of Empirical Hyv\"arinen score}
For notational simplicity, we rewrite $\frac{1}{(n-1)h_n}\sum_{j\neq i}^nK_{ij}$ as $\hat{f}_{(-i)}$.
\begin{lem}  \label{lem:bound_remainder}
Under Assumption \ref{ass:DGP},  \ref{ass:Kernel} and  \ref{ass:nhinfty},
\begin{align}
    \mathbb{E}_{X_{1:n}}\left[\left\{\{\hat{f}_{(-i)}\}^2-\mathbb{E}_{X_{1:n}}\left[\{\hat{f}_{(-i)}\}^2 \mid X_i \right]\right\}^4\right] = O\left\{ \frac{1}{n^2h_n^2} \right\} \nonumber
\end{align}
\begin{proof}
\begin{align*}
    & \mathbb{E}_{X_{1:n}}\left[\left\{\{\hat{f}_{(-i)}\}^2-\mathbb{E}_{X_{1:n}}\left[\{\hat{f}_{(-i)}\}^2 \mid X_i \right]\right\}^4 \right]\\
    & = \mathbb{E}_{X_{1:n}}\left[2\mathbb{E}_{X_{1:n}}[\{\hat{f}_{(-i)}\}^8 \mid X_i ]-8\mathbb{E}_{X_{1:n}}[\{\hat{f}_{(-i)}\}^6 \mid X_i ]\mathbb{E}_{X_{1:n}}[\{\hat{f}_{(-i)}\}^2 \mid X_i ]+6\mathbb{E}_{X_{1:n}}[\{\hat{f}_{(-i)}\}^4 \mid X_i ]^2\right]
\end{align*}
From a straightforward computation, 
\begin{align*}
    \mathbb{E}_{X_{1:n}}[\{\hat{f}_{(-i)}\}^2 \mid X_i ]
    & = \frac{1}{(n-1)h_n} \left(\int K(u)^2f(X_i-uh_n)du\right) \\
    & \quad + \left(\int K(u) f(X_i-uh_n)du\right)^2\\
    \mathbb{E}_{X_{1:n}}[\{\hat{f}_{(-i)}\}^4 \mid X_i ] 
    & = \frac{6}{(n-1)h_n}\left(\int K(u)^2f(X_i-uh_n)du\right)\left(\int K(u)f(X_i-uh_n)du\right)^2\\
    & \quad +\left(\int K(u)f(X_i-uh_n)du\right)^4 + O_p\{(nh_n)^{-2}\}\\
    \mathbb{E}_{X_{1:n}}[\{\hat{f}_{(-i)}\}^6 \mid X_i ]
    &= \frac{15}{(n-1)h_n}\left(\int K(u)^2f(X_i-uh_n)du\right)\left(\int K(u)f(X_i-uh_n)du\right)^4 \\
    & \quad +\left(\int K(u)f(X_i-uh_n)du\right)^6 + O_p\{(nh_n)^{-2}\}\\
    \mathbb{E}_{X_{1:n}}[\{\hat{f}_{(-i)}\}^8 \mid X_i ] &= \frac{28}{(n-1)h_n}\left(\int K(u)^2f(X_i-uh_n)du\right)\left(\int K(u)f(X_i-uh_n)du\right)^6 \\
    & \quad +\left(\int K(u)f(X_i-uh_n)du\right)^8 + O_p\{(nh_n)^{-2}\}
\end{align*}
then we have,
\begin{align*}
    \mathbb{E}_{X_{1:n}}[\{\hat{f}_{(-i)}\}^8 \mid X_i] 
    &= \frac{28}{(n-1)h_n}\left(\int K(u)^2f(X_i-uh_n)du\right)\left(\int K(u)f(X_i-uh_n)du\right)^6 \\
    &\quad +\left(\int K(u)f(X_i-uh_n)du\right)^8 + O_p\{(nh_n)^{-2}\} \\
    \mathbb{E}_{X_{1:n}}[\{\hat{f}_{(-i)}\}^6 \mid X_i ]\mathbb{E}_{X_{1:n}}[\{\hat{f}_{(-i)}\}^2 \mid X_i ] 
    &= \frac{16}{(n-1)h_n}\left(\int K(u)^2f(X_i-uh_n)du\right)\left(\int K(u)f(X_i-uh_n)du\right)^6\\
    & \quad +\left(\int K(u)f(X_i-uh_n)du\right)^8 + O_p\{(nh_n)^{-2}\}\\
    \mathbb{E}_{X_{1:n}}[\{\hat{f}_{(-i)}\}^4 \mid X_i ]^2
    & = \frac{12}{(n-1)h_n}\left(\int K(u)^2f(X_i-uh_n)du\right)\left(\int K(u)f(X_i-uh_n)du\right)^6 \\
    & \quad +\left(\int K(u)f(X_i-uh_n)du\right)^8 + O_p\{(nh_n)^{-2}\}.
\end{align*}
Since $(2\times 28 - 8\times 16 + 6\times 12) = 0$ and $(2-8+6)=0$,
\begin{align}
    \mathbb{E}_{X_{1:n}}\left[\left\{\{\hat{f}_{(-i)}\}^2-\mathbb{E}_{X_{1:n}}\left[\{\hat{f}_{(-i)}\}^2 \mid X_i \right]\right\}^4\right] = O\{(nh_n)^{-2}\} \nonumber
\end{align}
\end{proof}
\end{lem}
\subsubsection{Lemmas for the Haj\'ek Projection Terms of Empirical Hyv\"arinen score}
\begin{lem} \label{lem:hp_1}
For some function $\rho_{1,1}$ and $\rho_{1,2}$ of $X_i$, under Assumption \ref{ass:DGP} -  \ref{ass:fdiffrentiability1}, \ref{ass:Kernel}, \ref{ass:integrationbyparts}, \ref{ass:model}, it holds that
\begin{align}
    \mathbb{E}_{X_{1:n}}[\widetilde{I}_{ijk}|X_i] 
    &= \{2f''(X_i)f(X_i)^{-1}-f'(X_i)^2f(X_i)^{-2}\} \nonumber \\ 
    & \quad + h_n^{L+4}\rho_{1,1}(X_i)  + (w_n-1)h^4\rho_{1,2}(X_i) + o_p\{h_n^{L+4}+(w_n-1)h_n^4\} \nonumber
\end{align}
\begin{proof}
From the following computations of $\mathbb{E}[I_{\sigma(1)\sigma(2)\sigma(3)}|X_i]$, for some function $\rho_{1,1}$ of $X_i$, we have
\begin{align}
     \mathbb{E}_{X_{1:n}}[\widetilde{I}_{ijk}|X_i] 
     &= \frac{4w_n}{3}h_n^4f''(X_i)f(X_i)^{-1} + (w_n^2-2w_n)h_n^4\{f'(X_i)^2f(X_i)^{-2}-\frac{2}{3}f''(X_i)f(X_i)^{-1}\} \nonumber \\
     & \quad + h_n^{L+4}\rho_{1,1}(X_i) + o_p(h_n^{L+4}), \nonumber
\end{align}
then, rearranging the right hand side for $(w_n-1)$ yields
\begin{align*}
    \mathbb{E}_{X_{1:n}}[\widetilde{I}_{ijk}|X_i] 
    & = \{(w_n-1)+1\}h^4\{2f''(X_i)f(X_i)^{-1}-f'(X_i)^2f(X_i)^{-2}\} \\
    & \quad + w_n(w_n-1)h_n^4\{f'(X_i)^2f(X_i)^{-2}-\frac{2}{3}f''(X_i)f(X_i)^{-1}\}+ h_n^{L+4}\rho_{1,1}(X_i) + o_p(h_n^{L+4}) \\
    & =  \{2f''(X_i)f(X_i)^{-1}-f'(X_i)^2f(X_i)^{-2}\} \\
    & \quad + h_n^{L+4}\rho_{1,1}(X_i) + (w_n-1)h_n^4\rho_{1,2}(X_i) + o_p\{h_n^{L+4}+(w_n-1)h_n^4\} 
\end{align*}
In the following, we compute $\mathbb{E}_{X_{1:n}}[I_{\sigma(1)\sigma(2)\sigma(3)}|X_i]$ for each permutation separately, for some function $\rho_{1,1,1}, \rho_{1,1,2}$ and $\rho_{1,1,3}$ of $X_i$. 
\begin{itemize}
    \item $\mathbb{E}_{X_{1:n}}[I_{i\sigma(1)\sigma(2)}|X_i]$
    \begin{align*}
        \mathbb{E}_{X_{1:n}}[I_{ijk}|X_i]
        &= \mathbb{E}_{X_{1:n}}[\{2w_nK''_{ij}K_{ik}+(w_n^2-2w_n)K'_{ij}K'_{ik}\}f(X_i)^{-2}|X_i] \\
        &= 2w_nf(X_i)^{-2}\mathbb{E}_{X_{1:n}}[K''_{ij}|X_i]\mathbb{E}_{X_{1:n}}[K_{ij}|X_i] + (w_n^2-2w_n)f(X_i)^{-2}\mathbb{E}_{X_{1:n}}[K'_{ij}|X_i]^2 \\
        & = 2w_nf(X_i)^{-2}\left(h^3\int K(u)f''(X_i-uh_n)du\right)\left(h_n\int K(u)f(X_i-uh_n)du\right)  \\
        & \quad + (w_n^2-2w_n)f(X_i)^{-2}\left(h_n^2\int K(u)f'(X_i-uh_n)du\right)^2 \\
        & = 2w_nh_n^4f''(X_i)f(X_i)^{-1} \\
        & \quad + (w_n^2-2w_n)h_n^4f'(X_i)^2f(X_i)^{-2} + h_n^{L+4}\rho_{1,1,1}(X_i) +o_p(h_n^{L+4})
    \end{align*}
    \item $\mathbb{E}_{X_{1:n}}[I_{\sigma(1)i\sigma(2)}|X_i]$
    \begin{align*}
        \mathbb{E}_{X_{1:n}}[I_{jik}|X_i] 
        &= \mathbb{E}_{X_{1:n}}[\{2w_nK''_{ji}K_{jk}+(w_n^2-2w_n)K'_{ji}K'_{jk}\}f(X_j)^{-2}|X_i] \\
        & = 2w_n\left(\int\int K''\left(\frac{z_1-X_i}{h_n}\right)K\left(\frac{z_1-z_2}{h_n}\right)f(z_1)^{-1}f(z_2)dz_1dz_2\right) \\
        & \quad + (w_n^2-2w_n)\left(\int\int K'\left(\frac{z_1-X_i}{h_n}\right)K'\left(\frac{z_1-z_2}{h_n}\right)f(z_1)^{-1}f(z_2)dz_1dz_2\right) \\
        & = 2w_n\left(h_n^2\int K''(u)K(v)f(X_i+uh_n)^{-1}f\{X_i+(u-v)h_n\}dudv\right) \\
        & \quad + (w_n^2-2w_n)\left(h_n^2\int K'(u)K'(v)f(X_i+uh_n)^{-1}f\{X_i+(u-v)h_n\}dudv\right) \\
        & = 2w\left(h_n^4\int\int K(u)K(v) \right.\\
        & \quad\quad \times\Bigl[2f(X_i+uh_n)^{-3}f'(X_i+uh_n)^2f\{X_i+(u-v)h_n\}\\
        & \quad\quad\quad -f(X_i+uh_n)^{-2}f''(X_i+uh_n)f\{X_i+(u-v)h_n\} \\
        & \quad\quad\quad  + f(X_i+uh_n)^{-1}f''\{X_i+(u-v)h_n\}\\
        & \quad\quad\quad \left. -2f(X_i+uh_n)^{-2}f'(X_i+uh_n)f'\{X_i+(u-v)h_n\}\Bigl]dudv\right) \\
        & \quad + (w_n^2-2w_n)\left(\int h_n^4 K(u)K(v) \right. \\
        & \quad\quad \times \Bigl[f(X_i+uh_n)^{-2}f'(X_i+uh_n)f'\{X_i+(u-v)h_n\}\\
        & \quad\quad\quad \left. -f(X_i+uh_n)^{-1}f''\{X_i+(u-v)h_n\}\Bigl]dudv\right) \\
        & = (w_n^2-2w_n)h_n^4\{f'(X_i)^2f(X_i)^{-2}-f''(X_i)f(X_i)^{-1}\} \\
        & \quad + h_n^{L+4}\rho_{1,1,2}(X_i) +o_p(h_n^{L+4}) 
    \end{align*}
    \item $\mathbb{E}_{X_{1:n}}[I_{\sigma(1)\sigma(2)i}|X_i]$
    \begin{align*}
        \mathbb{E}_{X_{1:n}}[I_{jki}|X_i] 
        &= \mathbb{E}_{X_{1:n}}[\{2w_nK''_{jk}K_{ji}+(w_n^2-2w_n)K'_{jk}K'_{ji}\}f(X_j)^{-2}|X_i] \\
        & = 2w_n\left(\int\int K''\left(\frac{z_1-z_2}{h_n}\right)K\left(\frac{z_1-X_i}{h_n}\right)f(z_1)^{-1}f(z_2)dz_1dz_2\right) \\
        & \quad + (w_n^2-2w_n)\left(\int\int K'\left(\frac{z_1-z_2}{h_n}\right)K'\left(\frac{z_1-X_i}{h_n}\right)f(z_1)^{-1}f(z_2)dz_1dz_2\right) \\
        & = 2w_n\left(h_n^2\int K(u)K''(v)f(X_i+uh_n)^{-1}f\{X_i+(u-v)h_n\}dudv\right) \\
        & \quad + (w_n^2-2w_n)\left(h_n^2\int K'(u)K'(v)f(X_i+uh_n)^{-1}f\{X_i+(u-v)h_n\}dudv\right) \\
        & = 2w_n\left(h_n^4\int K(u)K(v)f(X_i+uh_n)^{-1}f''\{X_i+(u-v)h_n\}dudv\right) \\
        & \quad +(w_n^2-2w_n)\left(\int h_n^4 K(u)K(v)\right.\\
        & \left. \quad\quad \times \Bigl[f(X_i+uh_n)^{-2}f'(X_i+uh_n)f'\{X_i+(u-v)h_n\} \right.\\
        & \quad\quad\quad \left. -f(X_i+uh_n)^{-1}f''\{X_i+(u-v)h_n\}\Bigl]dudv\right) \\
        & = 2w_nh_n^4f''(X_i)f(X_i)^{-1} \\
        & \quad + (w_n^2-2w_n)h_n^4\{f'(X_i)^2f(X_i)^{-2}-f''(X_i)f(X_i)^{-1}\} \\
        & \quad + h_n^{L+4}\rho_{1,1,3}(X_i)+o_p(h_n^{L+4}) 
    \end{align*}
\end{itemize}
\end{proof}
\end{lem}

\begin{lem} \label{lem:hp_2}
For some function $\rho_{2,1}$ and $\rho_{2,2}$ of $X_i$, under Assumption \ref{ass:DGP} -  \ref{ass:fdiffrentiability1}, \ref{ass:Kernel}, \ref{ass:integrationbyparts}, \ref{ass:model}, it holds that
\begin{align}
    \mathbb{E}_{X_{1:n}}[\widetilde{II}_{ijklm}|X_i] &= \{2f''(X_i)f(X_i)^{-1}-f'(X_i)^2f(X_i)^{-2}\}+ h_n^{L+6}\rho_{2,1}(X_i) \nonumber\\
    & \quad + (w_n-1)h_n^6\rho_{2,2}(X_i) + o_p\{h_n^{L+6}+(w_n-1)h_n^6\}. \nonumber
\end{align}
\begin{proof}
From the following computation of $\mathbb{E}_{X_{1:n}}[II_{\sigma(1)\sigma(2)\sigma(3)\sigma(4)\sigma(5)}|X_i]$, we have,
\begin{align}
    \mathbb{E}_{X_{1:n}}[\widetilde{II}_{ijklm}|X_i] 
    &= \frac{8w_n}{5}h_n^6f''(X_i)f(X_i)^{-1} \nonumber\\
    & \quad + (w_n^2-2w_n)h_n^6\{f'(X_i)^2f(X_i)^{-2}-\frac{2}{5}f''(X_i)f(X_i)^{-1}\} + O_p(h_n^{L+6}) \nonumber
\end{align}
then, rearranging the right hand side for $(w_n-1)$, 
\begin{align}
    \mathbb{E}_{X_{1:n}}[\widetilde{II}_{ijklm}|X_i] 
    &= \{2f''(X_i)f(X_i)^{-1}-f'(X_i)^2f(X_i)^{-2}\}+ h_n^{L+6}\rho_{2,1}(X_i) \nonumber\\
    & \quad + (w_n-1)h_n^6\rho_{2,2}(X_i) + o_p\{h_n^{L+6}+(w_n-1)h_n^6\} \nonumber
\end{align}
In the following, we provide the result of computation of  $\mathbb{E}_{X_{1:n}}[II_{\sigma(1)\sigma(2)\sigma(3)\sigma(4)\sigma(5)}|X_i]$ for each permutation separately. The computation process of them is same as of $\mathbb{E}_{X_{1:n}}[I_{\sigma(1)\sigma(2)\sigma(3)}|X_i]$ and straightforward but much longer and more tedious, so we provide only the results, for each permutation separately, for some function $\rho_{2,1,1}, \rho_{2,1,2}, \rho_{2,1,3}$ and $\rho_{2,1,4}$ of $X_i$.
\begin{itemize}
    \item $\mathbb{E}_{X_{1:n}}[II_{i\sigma(1)\sigma(2)\sigma(3)\sigma(4)}|X_i]$
    \begin{align*}
        \mathbb{E}_{X_{1:n}}[II_{ijklm}|X_i] 
        &= \mathbb{E}_{X_{1:n}}[\{2w_nK''_{ij}K_{ik}K_{il}K_{im}+(w_n^2-2w_n)K'_{ij}K'_{ik}K_{il}K_{im}\}f(X_i)^{-4}|X_i] \\
        & = 2w_nf(X_i)^{-4}\mathbb{E}[K''_{ij}|X_i]\mathbb{E}[K_{ij}|X_i]^3 \\
        & \quad + (w_n^2-2w_n)f(X_i)^{-4}\mathbb{E}[K'_{ij}|X_i]^2\mathbb{E}[K_{ij}|X_i]^2 \\
        & = 2w_nh_n^6f''(X_i)f(X_i)^{-1} \\
        & \quad + (w_n^2-2w_n)h_n^6f'(X_i)^2f(X_i)^{-2}+ h_n^{L+6}\rho_{2,1,1}(X_i) + o_p(h_n^{L+6})
    \end{align*}
    \item $\mathbb{E}_{X_{1:n}}[II_{\sigma(1)i\sigma(2)\sigma(3)\sigma(4)}|X_i]$
    \begin{align*}
        \mathbb{E}_{X_{1:n}}[II_{jiklm}|X_i] 
        &= \mathbb{E}_{X_{1:n}}[\{2w_nK''_{ji}K_{jk}K_{jl}K_{jm}+(w_n^2-2w_n)K'_{ji}K'_{jk}K_{jl}K_{jm}\}f(X_j)^{-4}|X_i] \\
        & = (w_n^2-2w_n)h_n^6\{-f''(X_i)f(X_i)^{-1}+f'(X_i)^2f(X_i)^{-2}\}\\
        & \quad +h_n^{L+6}\rho_{2,1,2}(X_i) + o_p(h_n^{L+6})
    \end{align*}
    \item $\mathbb{E}_{X_{1:n}}[II_{\sigma(1)\sigma(2)i\sigma(3)\sigma(4)}|X_i]$
    \begin{align*}
        \mathbb{E}_{X_{1:n}}[II_{jkilm}|X_i] 
        &= \mathbb{E}_{X_{1:n}}[\{2w_nK''_{jk}K_{ji}K_{jl}K_{jm}+(w_n^2-2w_n)K'_{jk}K'_{ji}K_{jl}K_{jm}\}f(X_j)^{-4}|X_i] \\
        & = 2w_nh_n^6 f''(X_i)f(X_i)^{-1} + \\
        & \quad + (w_n^2-2w_n)h_n^6\{-f''(X_i)f(X_i)^{-1}+f'(X_i)^2f(X_i)^{-2}\} \\
        & \quad + h_n^{L+6}\rho_{2,1,3}(X_i) + o_p(h_n^{L+6})
    \end{align*}
    \item $\mathbb{E}_{X_{1:n}}[II_{\sigma(1)\sigma(2)\sigma(3)i\sigma(4)}|X_i]$ and $\mathbb{E}_{X_{1:n}}[II_{\sigma(1)\sigma(2)\sigma(3)\sigma(4)i}|X_i]$
    \begin{align*}
        \mathbb{E}_{X_{1:n}}[II_{jklim}|X_i] 
        &= \mathbb{E}_{X_{1:n}}[\{2w_nK''_{jk}K_{jl}K_{ji}K_{jm}+(w_n^2-2w_n)K'_{jk}K'_{jl}K_{ji}K_{jm}\}f(X_j)^{-4}|X_i] \\
        & = 2w_nh_n^6f''(X_i)f(X_i)^{-1} \\
        & \quad + (w_n^2-2w_n)h_n^6f'(X_i)^2f(X_i)^{-2} + h_n^{L+6}\rho_{2,1,4}(X_i) + o_p(h_n^{L+6})
    \end{align*}
\end{itemize}
\end{proof}
\end{lem}

\subsubsection{Lemmas for the Quadratic Projection Terms of Empirical Hyv\"arinen Socre}
\begin{lem} \label{lem:quad_1} Under Assumption \ref{ass:DGP} -  \ref{ass:fdiffrentiability1}, \ref{ass:Kernel}, \ref{ass:integrationbyparts}, \ref{ass:model}, it holds that
\begin{align}
    \mathbb{E}_{X_{1:n}}[\widetilde{I}_{ijk}|X_i,X_j] = O_p(h_n^{3/2}). \nonumber
\end{align}
\begin{proof}
The proof is straightforward but too lengthy, so we will omit it.
\end{proof}
\end{lem}
\begin{lem} \label{lem:quad_2} Under Assumption \ref{ass:DGP} -  \ref{ass:fdiffrentiability1}, \ref{ass:Kernel}, \ref{ass:integrationbyparts}, \ref{ass:model}, it holds that
\begin{align}
    \mathbb{E}_{X_{1:n}}[\widetilde{II}_{ijklm}|X_i,X_j] = O_p(h_n^{7/2}). \nonumber
\end{align}
\begin{proof}
The proof is straightforward but too lengthy, so we will omit it.
\end{proof}
\end{lem}

\subsubsection{Lemmas for Bounding the Remainder Terms of Empirical Hyv\"arinen Socre}

\begin{lem} \label{Lemma:third_order_U_evaluation}
Letting $p(X_i,X_j,X_k,h_n)$ be a summand of a third order U-statistics, under the assumption that $\mathbb{E}_{X_{1:n}}[|p(X_i,X_j,X_k,h_n)|^2]=O(a_n^2)$, at least, the first projection term converges at the rate of $O_p(n^{-1/2}a_n)$, the quadratic term $O_p(n^{-1}a_n)$ and the cubic term $O_p(n^{-3/2}a_n)$.
\begin{proof}
Define $r(X_i,h_n)$ as follows,
\begin{equation*}
    r(X_i,h_n)=\mathbb{E}_{X_{1:n}}[p(X_i,X_j,X_k,h_n)|X_i]-\mathbb{E}_{X_{1:n}}[p(X_i,X_j,X_k,,h_n)],
\end{equation*}
then the first projection term of a third order U-statistics $(U.III_p)$ and its squared moment are
\begin{align*}
    &(U.III_p)=\frac{3}{n}\sum_{i=1}^n r(X_i,h_n),\\
    &\mathbb{E}_{X_{1:n}}[\|(U.III_p)\|^2]=\frac{9}{n^2}\sum_{i=1}^n\sum_{j=1}^n\mathbb{E}_{X_{1:n}}[r(X_i,h_n)r(X_j,h_n)].
\end{align*}
 Since $\mathbb{E}_{X_{1:n}}[r(X_i,h)r(X_j,h_n)]=0$ for $i\neq j$, from the standard property of U-statistics,
\begin{equation*}
    \mathbb{E}_{X_{1:n}}[\|(U.III_p)\|^2]=\frac{9}{n^2}\sum_{i=1}^n\mathbb{E}_{X_{1:n}}[\|r(X_i,h_n)\|^2]
\end{equation*}
From the assumption, $O(\mathbb{E}_{X_{1:n}}[\|r(X_i,X_j,h_n)\|^2])=O(\mathbb{E}_{X_{1:n}}[\|p(X_i,X_j,h_n)\|^2])=o(a_n^2)$
this implies
\begin{align*}
    \mathbb{E}_{X_{1:n}}[\|(U.III_p)\|^2]=O(n^{-2}) \cdot O(na_n^2)=O(n^{-1}a_n^2)\Longrightarrow  (U.III_p)=O_p(n^{-\frac{1}{2}}a_n)
\end{align*}
The proof for the quadratic term and for cubic term are similar.
\end{proof}
\end{lem}

\begin{lem} \label{Lemma:fifth_order_U_evaluation}
Letting $p(X_i,X_j,X_k,X_l,X_m,h_n)$ be a summand of a fifth order U-statistics, under the assumption that $\mathbb{E}_{X_{1:n}}[|p(X_i,X_j,X_k,X_l,X_m,h_n)|^2]=O(a_n^2)$, at least, the projection term converges at the rate of $O_p(n^{-1/2}a_n)$, the quadratic term $O_p(n^{-1}a_n)$ and the cubic term $O_p(n^{-3/2}a_n)$.
\begin{proof}
The proof is similar to Lemma \ref{Lemma:third_order_U_evaluation}. 
\end{proof}
\end{lem}
}

\subsection{Lemmas for MISE}
\begin{lem} \label{lem:flogf}
Under Assumption \ref{ass:DGP}, \ref{ass:interiorpoint}, \ref{ass:fdiffrentiability1}, \ref{ass:Kernel}, \ref{ass:model} it holds that
\begin{align*}
    & \mathbb{E}_{X_{1:n}} \left[ \hat{f}_{h_n}(x) \log \hat{f}_{h_n}(x) \right] \\
    & \quad = f(x) \log  f(x) + \kappa_L f^{(L)}(x) \left\{ \log f(x) + 1 \right\}h_n^L + o(h^L)  + O\left( \frac{1}{nh_n} \right)
\end{align*}
\end{lem}
\begin{proof}
\begin{align*}
     & \mathbb{E}_{X_{1:n}} \left[ \hat{f}_{h_n}(x) \log \hat{f}_{h_n}(x) \right] \\
     & = \mathbb{E}_{X_{1:n}} \Bigl[ \mathbb{E}_{X_{1:n}}\left[ \hat{f}_{h_n}(x) \right] \log\left( \mathbb{E}_{X_{1:n}}\left[ \hat{f}_{h_n}(x) \right]  \right)  \Bigr] \\
     & \quad + \mathbb{E}_{X_{1:n}} \left[ \left( \hat{f}_{h_n}(x) - \mathbb{E}_{X_{1:n}}\left[ \hat{f}_{h_n}(x) \right]  \right) \left\{ \log\left( \mathbb{E}_{X_{1:n}}\left[ \hat{f}_{h_n}(x) \right]  \right) +  1 \right\} \right] + O\left( \frac{1}{nh_n} \right) \\
     & = \mathbb{E}_{X_{1:n}}\left[ \hat{f}_{h_n}(x) \right] \log\left( \mathbb{E}_{X_{1:n}}\left[ \hat{f}_{h_n}(x) \right]  \right)   + O\left( \frac{1}{nh_n} \right) \\
     & = \big\{ f(x) + \kappa_L f^{(L)}(x) h_n^L + o(h^L) \big\} \log  \big\{ f(x) + \kappa_L f^{(L)}(x) h_n^L + o(h^L) \big\} + O\left( \frac{1}{nh_n} \right) \\
     & = f(x) \log  f(x) + \kappa_L f^{(L)}(x) \left\{ \log f(x) + 1 \right\}h_n^L + o(h^L)  + O\left( \frac{1}{nh_n} \right)
\end{align*}
where the first equality expands  the left hand side around $\hat{f}_{h_n}(x) = \mathbb{E}[\hat{f}_{h_n}(x)]$, and the remainder of the second equality follows from the next Lemma.
\end{proof}

\begin{lem}
Under Assumption \ref{ass:DGP}, \ref{ass:interiorpoint}, \ref{ass:Kernel}, \ref{ass:model} it holds that
\begin{align*}
      \hat{f}_{h_n}(x) \log \hat{f}_{h_n}(x)   - \mathbb{E}_{X_{1:n}} \left[ \hat{f}_{h_n}(x) \log \hat{f}_{h_n}(x) \right] = O_p\left( \frac{1}{ \sqrt{nh_n}} \right).
\end{align*}
\begin{proof}
 Since the expansion of $\log \hat{f}_{h_n}(x)$ around $ \hat{f}_{h_n}(x) = \mathbb{E}_{X_{1:n}}[ \hat{f}_{h_n}(x)  ]$ yields
\begin{align*}
    \log \hat{f}_{h_n}(x)  
    &= \log\left( \mathbb{E}_{X_{1:n}}\left[ \hat{f}_{h_n}(x) \right]  \right)   + O_p\left\{ \hat{f}_{h_n}(x)  - \mathbb{E}_{X_{1:n}}\left[ \hat{f}_{h_n}(x) \right] \right\}   = \log  f(x) + o_p(1),
\end{align*}
it holds that
\begin{align*}
    \mathbb{V}_{X_{1:n}}\left[ \hat{f}_{h_n}(x) \log \hat{f}_{h_n}(x) \right] \lesssim \mathbb{V}_{X_{1:n}}\left[ \hat{f}_{h_n}(x) \right] =  O_p\left( \frac{1}{ nh_n } \right)
\end{align*}
This result and Markov's  inequality imply the lemma.
\end{proof}
\end{lem}

\begin{lem} \label{lem:V_intf}
Under Assumption \ref{ass:DGP}, \ref{ass:interiorpoint}, \ref{ass:fdiffrentiability1}, \ref{ass:Kernel}, \ref{ass:model} it holds that
\begin{align*}
    \mathbb{V}_{X_{1:n}}\left[ \int \hat{f}_{h_n}(x) dx \right] =  o\{(nh_n)^{-1}\}.
\end{align*}
\begin{proof}
\begin{align*}
    \mathbb{V}_{X_{1:n}}\left[ \int \hat{f}_{h_n}(x) dx \right] 
    & =  \mathbb{V}_{X_{1:n}}\left[ \int \frac{1}{nh_n} \sum_{i=1}^n K\left( \frac{X_i - x}{h_n} \right) dx \right] \\
    & = \frac{1}{nh_n^2} \mathbb{V}_{X_{1:n}}\left[ \int  K\left( \frac{X_i - x}{h_n} \right) dx \right] \\
    & = \frac{1}{nh_n^2} \mathbb{E}_{X_{1:n}}\left[ \left(\int  K\left( \frac{X_i - x}{h_n} \right) dx\right)^2 \right] 
    - \frac{1}{nh_n^2} \mathbb{E}_{X_{1:n}}\left[ \int  K\left( \frac{X_i - x}{h_n} \right) dx \right]^2\\
    & = \frac{1}{nh_n^2} \int \left( \int  K\left( \frac{v-x}{h_n} \right) dx \right)^2 f(v)dv - \frac{1}{nh_n^2} \left(\int\int K\left(\frac{v-x}{h_n}\right)f(v)dxdv \right)^2  \\
    & = \frac{1}{n} \int \left( \int K(u)du \right)^2  f(v)dv - \frac{1}{n} \left(\int\int K(w)f(x+uh_n)dxdu \right)\\
    & =  o\{(nh_n)^{-1}\}.
\end{align*}
where the first term in the fifth equality follows form the change of variables of $x=v-uh_n$.
\end{proof}
\end{lem}

\begin{lem} \label{lem:cov_fintf}
Under Assumption \ref{ass:DGP}, \ref{ass:interiorpoint}, \ref{ass:fdiffrentiability1}, \ref{ass:Kernel}, \ref{ass:model} it holds that
\begin{align*}
    Cov\left(\hat{f}_{h_n}(x), \int \hat{f}_{h_n}(x) dx \right)  =  o\{(nh_n)^{-1}\}.
\end{align*}
\begin{proof}
\begin{align*}
    Cov\left(\hat{f}_{h_n}(x), \int \hat{f}_{h_n}(x) dx \right) 
    & = Cov\left( \frac{1}{nh_n} \sum_{i=1}^n K\left(  \frac{X_i- x}{h_n} \right), \int \frac{1}{nh_n} \sum_{i=1}^n K\left(  \frac{X_i- x}{h_n} \right) dx \right) \\
    & = \frac{1}{nh_n^2} Cov\left( K\left(  \frac{X_i- x}{h_n} \right), \int K\left(  \frac{X_i- x}{h_n} \right) dx \right) \\
    & =\frac{1}{nh_n^2} \mathbb{E}_{X_i}\left[K\left(  \frac{X_i- x}{h_n} \right) \int K\left(  \frac{X_i- x}{h_n} \right) dx \right] \\
    & \quad - \frac{1}{nh_n^2} \mathbb{E}_{X_i}\left[ K\left(  \frac{X_i- x}{h_n} \right)  \right] \mathbb{E}_{X_i}\left[\int K\left(  \frac{X_i- x}{h_n} \right) dx \right] \\
    & = \frac{1}{nh_n^2} \int K\left(  \frac{v- x}{h_n} \right) \left\{ \int K\left(  \frac{v- x}{h_n} \right) dx \right\}f(v)dv \\
    & \quad - \frac{1}{nh_n^2} \left( \int  K\left(  \frac{v- x}{h_n} \right) f(v)dv \right) \left( \int\int  K\left(  \frac{v- x}{h_n} \right) f(v)dxdv \right)\\
    & = \frac{1}{n} \int  K\left( y \right) \left\{\int K(u)du\right\}  f(x+yh_n)dy \\
    & \quad - \frac{1}{n} \left( \int K(u)f(x+uh_n)du \right)\left( \int\int K(u)f(x+uh_n)dudv \right) \\
    & = o\{(nh_n)^{-1}\}
\end{align*}
where the fifth equality uses change of variables $(v-x)/h_n = u$ and $(v-x)/h_n = y$.
\end{proof}
\end{lem}

\end{document}